%% file: zoo-techrep.tex
\newif\ifspringer
\ifspringer
\documentclass{svjour3}
\else
\documentclass[a4paper]{article}
\fi

\usepackage{amsmath}
\usepackage{amsfonts}
\usepackage{amssymb}         %additional math symbols (see amsguide.tex)
\usepackage{amsopn}
\usepackage{theorem}          %additional LaTeX2e package (see pctex\latex2e)
\usepackage{latexsym}
\usepackage{graphicx}        %graphic and .eps files
\usepackage{mathrsfs}		%calligraphic fonts
\usepackage{psfrag}
\usepackage{algorithmic}
\usepackage{algorithm}
\usepackage{color}
\usepackage{verbatim}
\usepackage{url}
\usepackage{tikz}
\usepackage{rotating}

\tikzstyle{startstop} = [rectangle, rounded corners, text centered, draw=black,
  fill=red!30] 
\tikzstyle{io} = [trapezium, trapezium left angle=70, trapezium right
  angle=110, text centered, draw=black, fill=blue!30] 
\tikzstyle{process} = [rectangle, text centered, draw=black, fill=orange!30]
\tikzstyle{decision} = [diamond, aspect=2, text centered, draw=black, fill=green!30] 
\tikzstyle{arrow} = [thick,->,>=stealth]

\ifspringer
\else
\fi

\ifspringer
\else
\fi

\newcommand{\CLA}{\bf}
\newcommand{\revision}[1]{#1}
\newtheorem{result}{\ }[section]
\theoremstyle{changebreak}                % (see LATEX2E\THEOREM.DTX)

\newtheorem{lem}[result]{Lemma}

\newtheorem{prop}[result]{Proposition}
\newtheorem{example}[result]{Example}

\ifspringer
\renewcommand{\qed}{\hfill \ensuremath{\Box}}
\else
\newenvironment{proof}
 {{\sl Proof.}\hspace*{1 ex}}%
 {{\nopagebreak\hspace*{\fill}$\Box$\par\vspace{12pt}}}
\newcommand{\qed}{\hfill \ensuremath{\Box}}
\fi

\newcommand{\iDGP}{{\it i}\,DGP}
\newcommand{\kDMDGP}{${}^{\sf K}$DMDGP}
\begin{document}

\ifspringer

\title{New error measures and methods for realizing protein graphs from distance data} 
\titlerunning{Error measures and methods for realizing protein graphs}

\author{C.~D'Ambrosio \and Ky Vu\and C.~Lavor \and L.~Liberti\and N.~Maculan}
\institute{C.~D'Ambrosio\and Ky Vu\and L.~Liberti\at CNRS LIX, Ecole Polytechnique, 91128 Palaiseau, France\\\email{\{dambrosio,vu,liberti\}@lix.polytechnique.fr} \and C.~Lavor\at IMECC, University of Campinas, SP 13081-970, Brazil\\\email{clavor@ime.unicamp.br} \and N.~Maculan\at COPPE, Federal University of Rio de Janeiro, RJ 21941-972, Brazil\\\email{maculan@cos.ufrj.br}}
\authorrunning{C.~D'Ambrosio {\it et al.}}

\maketitle

\else
\thispagestyle{empty}
\begin{center} 

{\LARGE New error measures and methods for realizing protein graphs from distance data}   
\par \bigskip
{\sc Claudia D'Ambrosio${}^1$, Vu Khac Ky${}^1$, Carlile Lavor${}^2$,
  Leo Liberti${}^{1}$, Nelson Maculan${}^3$}   
\par \bigskip
\begin{minipage}{15cm}
\begin{flushleft}
{\small
\begin{itemize}
\item[${}^1$] {\it CNRS, UMR 7161 LIX, \'Ecole Polytechnique, F-91128 Palaiseau CEDEX,
France} \\ Email:\url{{dambrosio,vu,liberti}@lix.polytechnique.fr}
\item[${}^2$] {\it IMECC, University of Campinas, SP 13081-970, Brazil}
  \\ Email:\url{clavor@ime.unicamp.br} 
\item[${}^3$] {\it COPPE, Federal University of Rio de Janeiro, RJ 21941-972, Brazil} 
  \\ Email:\url{maculan@cos.ufrj.br}
\end{itemize}
}
\end{flushleft}
\end{minipage}
\par \medskip \today
\end{center}
\par \bigskip
\fi

% insert abstract
\begin{abstract} 
The interval Distance Geometry Problem (\iDGP) consists in finding a realization in $\mathbb{R}^K$ of a simple undirected graph $G=(V,E)$ with nonnegative intervals assigned to the edges in such a way that, for each edge, the Euclidean distance between the realization of the adjacent vertices is within the edge interval bounds. \revision{In this paper, we focus on the application to the conformation of proteins in space, which is a basic step in determining protein function: given interval estimations of some of the inter-atomic distances, find their shape. Among different families of methods for accomplishing this task, we look at mathematical programming based methods, which are well suited for dealing with intervals. The basic question we want to answer is: what is the {\it best} such method for the problem? The most meaningful error measure for evaluating solution quality is the coordinate root mean square deviation. We first introduce a new error measure which addresses a particular feature of protein backbones, i.e.~many partial reflections also yield acceptable backbones.} We then present a set of new and existing quadratic and semidefinite programming formulations of this problem, and a set of new and existing methods for solving these formulations. Finally, we perform a computational evaluation of all the feasible solver$+$formulation combinations according to new and existing error measures, finding that the best methodology is a new heuristic method based on multiplicative weights updates. \\ \ifspringer
 \else {\bf Keywords}: distance geometry, protein
conformation, mathematical programming.  \fi
\end{abstract}

% insert paper
%\tableofcontents

\section{Introduction}
\label{s:intro}
The Distance Geometry Problem (DGP) is defined formally as follows:
given an integer $K>0$, a simple undirected graph $G=(V,E)$, and an
edge weight function $U:E\to\mathbb{R}_+$, establish or deny the
existence of a vertex realization function $x:V\to\mathbb{R}^K$ such
that:
\begin{equation}
  \forall\{u,v\}\in E\qquad \|x_u-x_v\|_2 = U_{uv};\label{eq:dgp}
\end{equation}
realizations satisfying \eqref{eq:dgp} are called {\it valid
  realizations}. The DGP arises in many important applications:
determination of protein conformation from distance data
\cite{muchbook}, localization of mobile sensors in communication
networks \cite{snl}, synchronization of clocks from phase information
\cite{singer4}, control of unmanned submarine fleets \cite{bahr},
spatial logic \cite{splogic}, and more \cite{dgp-sirev}. It is {\bf
  NP}-complete when $K=1$ and {\bf NP}-hard for larger values of $K$
\cite{saxe79}.  Notationwise, we let $n=|V|$ and $m=|E|$.

\revision{The aim of this paper is to find the quality-wise best and pratically fastest method for solving a DGP variant arising in finding the shape of proteins using incomplete and imprecise distance data. We achieve this through an extensive computational benchmark of many (new and existing) heuristic methods and many instances constructed from Protein Data Bank (PDB) data \cite{pdb}. First, however, we make a theoretical contribution related to a new solution quality measure which is specially suited to evaluate the solution quality of protein isomers (i.e.~proteins which have the same chemical composition but a different shape). This is necessary to evaluating the computationally obtained solutions, since the symmetry group of protein backbones contains partial reflections \cite{bppolybook} (these are visible in most molecules, which may occur in nature in their left handed or right handed conformation).}

\subsection{The number of solutions}
\label{s:numofsols}
Let $\tilde{X}$ be the set of valid realizations of $G$. If
$x\in \tilde{X}$, any congruence (translation, rotation,
reflection) of $x$ yields another valid realization of $G$. We
therefore focus on the quotient set $X=\tilde{X}/\!\sim$, where $x\sim
y$ whenever there is a congruence mapping $x$ to $y$.

We have that $X=\varnothing$ if the corresponding DGP instance has no
solutions; $G$ is {\it rigid} if $|X|$ is finite; $G$ is \revision{{\it globally
rigid}} if $|X|=1$; and $G$ is {\it flexible} if $|X|$ is uncountable.
We note that $|X|$ cannot be countably infinite. By Milnor's theorem
on the Betti numbers of real algebraic varieties \cite{milnor}, the
number of connected components of $X$ is bounded above by $2\times
3^{{nK}-1}$. Suppose that $|X|$ is countably infinite: then it cannot
be flexible. This implies that incongruent elements of $X$ are on
distinct connected components of the manifold containing $X$. Milnor's
theorem shows that there are only finitely many such connected
components, which implies that $|X|$ is finite. \revision{This result also
follows by the cylindrical decomposition theorem of semi-algebraic
sets \cite{pollack,benedetti}.}

\subsection{Proteins and the Branch-and-Prune algorithm}
\label{s:proteins}
Our motivating application is finding the shape of protein proteins in
space (thus we fix $K=3$) knowing interval estimations of some of the
inter-atomic distances \cite{bipbip}. The protein backbone graph $G$
belongs to a specific subclass of Henneberg type I
graphs \cite{tay-whiteley}, namely there is an order $<$ on $V$ such
that, for each $v>3$, $v$ is adjacent to
$v-1,v-2,v-3$ \cite{dmdgp}. The backbone itself provides such an order
on the atoms, although other orders, which may be more convenient to
algorithmic efficiency, have been defined \cite{dvop,orders-dam}. DGP
instances with this property form a problem called {\sc Discretizable
Molecular Distance Geometry Problem} (DMDGP), which is also {\bf
NP}-hard \cite{dmdgp}. In \cite{lln5}, we proposed a fast and accurate
mixed-combinatorial algorithm for solving the DMDGP, called
Branch-and-Prune (BP). Unsurprisingly, the BP has exponential
complexity in the worst case, but the DMDGP has many interesting
properties which hold almost surely:
\begin{itemize}
\item $G$ is rigid, so $|X|$ is finite; \cite{lln5}
\item in particular, $|X|$ is a power of two; \cite{powerof2}
\item the BP algorithm is Fixed-Parameter Tractable (FPT) on the DMDGP
  \cite{bppolybook}, and in all the protein instances we tested, the
  parameter was always fixed at the same constant, yielding polytime
  behaviour.
\end{itemize}
By ``almost surely'' we mean that the set of weighted input graphs for
which the above properties may not hold has Lebesgue measure zero in
the set of all weighted input graphs (assuming the weights to be real
numbers). The BP algorithm relies on the given distances being
precise; unfortunately, however, inter-atomic distance data measured
through Nuclear Magnetic Resonance (NMR) are subject to experimental
errors, modelled as real intervals $[L,U]$ assigned to all edges
$\{u,v\}$ whenever $v-u\ge 3$ in the vertex order. To overcome this
difficulty, two research directions have been pursued: (i) the
discretization of the uncertainty intervals \cite{bpinterval}; (ii)
the analytical description, using Clifford algebra, of the locus of
vertex $v$ when the edge $\{v,v-3\}$ is weighted by an interval
\cite{cliffordalgebra}. The formulation study in this paper moves a
first step towards a third direction: the integration of purely
continuous techniques within mixed-combinatorial algorithms such as
BP. To this end, in this paper we pursue a computational study of some
these techniques.

\subsection{The interval DGP}
\label{s:idgp}
This brings us to the {\it interval} {\sc Distance Geometry Problem}
(\iDGP), which is a variant of the DGP defined as follows: the edge
function is an interval function $[L,U]:E\to\mathbb{IR}_+$, where
$L,U$ are two nonnegative functions from $E\to\mathbb{R}_+$ such that
$L_{uv}\le U_{uv}$ for each $\{u,v\}\in E$, $\mathbb{IR}_+$ is the
set of nonnegative real intervals, and Eq.~\eqref{eq:dgp} is replaced
by:
\begin{equation}
  \forall\{u,v\}\in E\qquad L_{uv}\le \|x_u-x_v\|_2\le
  U_{uv}.\label{eq:Idgp}
\end{equation}
Note that Eq.~\eqref{eq:Idgp} is often written as:
\begin{equation}
  \forall\{u,v\}\in E\qquad L_{uv}^2\le \|x_u-x_v\|_2^2\le
  U^2_{uv}.\label{eq:Idgp2}
\end{equation}
As explained later, Eq.~\eqref{eq:Idgp2} minimizes the chances that
numerical solvers, which rely on the floating-point representation of
real numbers, might stumble upon a negative representation of zero,
thereby raising a ``not a number'' ({\tt NaN}) error upon calculating
the square root. \revision{Note that the \iDGP\ contains (and hence generalizes)
the DGP, since the latter corresponds to the case $L=U$.}

\subsection{Aim of this paper}
\label{s:paperaim}
Most solution techniques for solving \iDGP\ instances require a
continuous search in Euclidean space, even if the given graph is
rigid. The most direct approach is to formulate the \iDGP\ as a
Mathematical Program (MP), which can then be solved by a MP
solver. The aim of this paper is to determine the best
solver$+$formulation combination for the \iDGP.  \revision{To this end, we need
to know: (a) how to evaluate the quality of the solutions computed by
the solvers; (b) which formulations to employ; (c) which solvers to
employ. We therefore introduce new and existing error measures,
formulations and solvers, before proceeding to evaluate them all
computationally. Since we want our algorithms to be fast and scale
well, we focus on heuristic approaches. This means that we forsake a
proof of exactness, so evaluating these algorithms require test sets
with given (trusted) solutions. Such test sets can be put together
using the PDB.}

\subsection{Solution quality evaluation}
\label{s:solqual}
\revision{The simplest measures used for evaluation DGP solution quality are based on computing the average or maximum relative error of the realization with respect to the given distance value on the edges. The drawback of these simple edge-based measures is that even a small error might correspond (in sufficiently large proteins) to a wrong protein shape. Even worse, plotting the DGP solution versus the trusted solution usually yields nothing to the human eye, since the alignment is likely to be completely off.}

\revision{A more meaningful measure is provided by {\it Procrustes analysis} \cite{goodall}, also called {\it coordinate root mean square deviation} (cRMSD) \cite{crippen-rmsd}. Informally, this is the error derived by the best alignment, via translations and rotations, of a DGP solution to the trusted solution. It provides a visual tool for a human to evaluate the error, so even when the error is non-zero the visualization helps determine whether the error is due to floating point issues or structural differences.}

\revision{Unfortunately, for protein backbones there is an added difficulty: their symmetry group includes at least one partial reflection (starting from the fourth atom along the backbone), and may include many more \cite{powerof2,bppolybook,liberti-gsi13}: in general, the partial reflection group structure is a cartesian product of cyclic groups of order two, yielding an exponential number of elements. All of these symmetric solutions are isomers. They are equivalent from the point of view of the simple, edge-based error measures, but they may have very different cRMSD values with respect to the trusted solution. Again, visualizing a trusted solution and a DGP solution from a heuristic method with a low cRMSD might yield structures which look nothing like each other.}

\revision{The first contribution of this paper is the definition of a modified error measure that extends the cRMSD in that it aligns two structures in the best possible way using translations, rotations and partial reflections, and which allows us to properly evaluate the protein backbone solutions proposed by DGP heuristics. Our new measure could be described as a ``cRMSD modulo isomers''.}

\subsection{Innovations and outcomes}
\label{s:innov}
\revision{To sum up, the innovations introduced in this paper are:} (i) the new
cRMSD modulo isomers; (ii) some new MP formulations for the iDGP;
(iii) the concept of ``pointwise formulation'' to be used in
alternating-type algorithms; (iv) an adaptation of the Multiplicative
Weights Update (MWU) algorithm to the iDGP. We conclude that the MWU
algorithm with its pointwise formulation is the best combination, and
that the new ``square factoring'' MP formulation, used within either a
pure MultiStart (MS) or a Variable Neighbourhood Search (VNS)
heuristic, is second best.

\subsection{Structure of the paper}
\label{s:paperstruc}
The rest of the paper is organized as follows. In
Sect.~\ref{s:benchmarking}, we define error measures to meaningfully
compare protein backbones found algorithmically with those stored in
PDB files \cite{pdb}, and introduce a new cRMSD type measure modulo
certain partial reflection isomers. In Sect.~\ref{s:formulations}, we
list several formulations, relaxations and variants for the \iDGP,
some of which are new. In Sect.~\ref{s:mwu}, we propose a new
algorithm for solving the \iDGP: namely, an adaptation of the
Multiplicative Weights Update method
\cite{kale}. In Sect.~\ref{s:compres}, we discuss comparative
computational results, which show that, on average, our newly proposed
algorithm provides the best quality solutions.

\section{Error measures for realizations of protein graphs}
\label{s:benchmarking}
Since we aim at ascertaining which formulation(s) can provide the best
and/or fastest bound, we need a method to benchmark quality and speed
with respect to any solution algorithm. We benchmark speed by simply
measuring CPU time.

\revision{Benchmarking solution quality is more complicated. In the Turing
Machine (TM) model, decision problems are in {\bf NP} whenever
feasible instances can be certified feasible in polynomial
time. Although the DGP and \iDGP\ are {\bf NP}-hard decision problems,
they are not known to be in {\bf NP}: feasible instances of the DGP
and \iDGP\ can in general yield realizations with irrational
components, for which polynomially-sized representations are not
generally available (some simple ideas have been tried
in \cite{dgpinnp} but failed to prove membership of the DGP to {\bf
NP}). The methods employed in this paper replace irrational numbers by
floating point numbers, and, as such, do not provide a valid
certificate. On the other hand, this is the situation with all real
number computations that need to be carry out efficiently over medium
to large-scale problems. Instead, we compute feasibility errors for
the floating point solutions we obtain.}

\subsection{The edge error}
\label{s:edgerror}
Given a realization $x^\ast:V\to\mathbb{R}^K$, we can measure the
error of $x^\ast$ with respect to a given \iDGP\ instance by assigning
an $\ell_2$-norm error to each edge $\{u,v\}$ of the graph $G=(V,E)$, given
by \cite{mdgpsurvey}:
\begin{equation}
  \alpha_{uv}(x^\ast) = \max\left(0,L_{uv} -
      \|x^\ast_u-x^\ast_v\|_2
%     \sqrt{\sum_{k\le K} (x^\ast_{uk}-x^\ast_{vk})^2}
    \right) 
  + \max\left(0,
      \|x^\ast_u-x^\ast_v\|_2
%     \sqrt{\sum_{k\le K} (x^\ast_{uk}-x^\ast_{vk})^2} 
  - U_{uv}\right).
\label{eq:alpha}
\end{equation}
We remark that the corresponding error for non-interval DGP instances is:
\begin{equation*}
  \beta_{uv}(x^\ast) = \left|
       \|x^\ast_u-x^\ast_v\|_2
%      \sqrt{\sum_{k\le K} (x^\ast_{uk}-x^\ast_{vk})^2} 
               - U_{uv}\right|.
\end{equation*}
Accordingly, we define the edge error as follows:
\begin{equation*}
  \eta_{uv}(x^\ast) = \left\{\begin{array}{rl} 
     \alpha_{uv}(x^\ast) & \mbox{if the instance is \iDGP} \\
     \beta_{uv}(x^\ast) & \mbox{if the instance is DGP.}
  \end{array}\right.
\end{equation*}
We can now define the {\it average error} associated to the instance
graph $G$ and a realization $x^\ast$ as:
\begin{equation}
  \Phi(x^\ast,G) = \frac{1}{|E|}\sum_{\{u,v\}\in E} \eta_{uv}(x^\ast),
  \label{eq:avgerr}
\end{equation}
and the {\it maximum error} as:
\begin{equation}
  \Psi(x^\ast,G) = \max_{\{u,v\}\in E} \eta_{uv}(x^\ast).
  \label{eq:maxerr}
\end{equation}

\revision{The above are absolute edge error measures. Relative error measures also exist, where each term $L_{uv}-\|x_u-x_v\|_2$ is replaced by $\frac{L_{uv}-\|x_u-x_v\|_2}{|L_{uv}|}$ (and similary for $\|x_u-x_v\|_2-U_{uv}$. Whether one or the other is used depends on the application at hand, and how poorly scaled the input data $L,U$ are. In the case of proteins, bounds are generall well scaled, as they are often between $1$ and $6${\AA}; so absolute error measures are more appropriate.}

\subsection{The coordinate root mean square deviation}
\label{s:rmsd}
The edge errors go a long way in determining when a realization
$x^\ast$ is not valid. In many applications, however, we know {\it a
priori} that a problem instance should feasible. Take e.g.~the
reconstruction of protein conformations from inter-atomic distances:
the protein certainly exists (this is also the case when localizing
sensors in wireless networks: the network is being measured, so it
exists). Furthermore, we might have a given (precise or approximate) realization
$\bar{x}$. In this setting, we want to evaluate the error with respect
to the given realization $\bar{x}$.

An obvious way to adapt the edge error to this situation is to compute
the average, over edges in $E$, of an absolute $\ell_2$-norm distance
difference:
\begin{equation}
  \Delta(x^\ast,\bar{x}) = \frac{1}{|E|} \sum_{\{u,v\}\in E} 
    \left|\,
       \|x^\ast_u-x^\ast_v\|_2
%      \sqrt{\sum_{k\le K} (x^\ast_{uk}-x^\ast_{vk})^2} 
     - \|\bar{x}_u-\bar{x}_v\|_2
%      \sqrt{\sum_{k\le K} (\bar{x}_{uk}-\bar{x}_{vk})^2}
    \,\right|.
  \label{eq:rlzerr}
\end{equation}
Unfortunately this approach is wrong, since different congruent
realizations yield different error values, making the comparison
impossible.

To this end, the cRMSD is often used
instead: i.e.,~translate both $x^\ast$ and $\bar{x}$ so that their
centroids $\gamma(x^\ast)=\gamma(\bar{x})=0$, where the {\it centroid}
is the vector $\gamma(x)\in\mathbb{R}^K$ defined as:
\begin{equation}
  \gamma(x) = \sum_{v\le K} x_{v},
  \label{eq:centroid}
\end{equation}
and then find the congruence $\rho$ (consisting of a rotation composed
with at most one reflection) such that $\|x^\ast-\rho(\bar{x})\|$ is
minimum. Note that the norm $\|\cdot\|$ on $\mathbb{R}^{Kn}$ is
induced by the $\ell_2$-norm in $\mathbb{R}^K$:
\begin{equation}
  \|x^\ast-\bar{x}\|=\sum_{v\in V} \|x_v^\ast-\bar{x}_v\|_2. 
  \label{eq:pcnorm}
\end{equation}
The cRMSD between $x^\ast$ and $\bar{x}$ is defined as
$\min_{\rho}\|x^\ast-\rho(\bar{x})\|$.

\subsection{Distance error modulo isometries}
\label{s:demi}
Although the cRMSD is widely used in computational geometry, it still
falls short in one of the properties of molecules, namely
isomers, which are molecules having the same chemical formula but
different 3D structure.

If we consider protein backbones only, their graphs $G=(V,E)$ possess
a further structural property. They have an order $<$ on $V$ such
that:
\begin{enumerate}
  \setlength{\parskip}{-0.2em}
  \item the first $K$ vertices in the order form a clique in
    $G$ ({\it clique property});
  \item each vertex $v>K$ is adjacent to
    $v-1,\ldots,v-K$ ({\it contiguous trilateration order property}). 
\end{enumerate}
Although protein backbones have $K=3$, we develop the theory
for general $K$. DGP instances having these properties are also
collectively known as \kDMDGP, which are a subclass of Henneberg type
I graphs \cite{henneberg1911}. Contiguous trilateration orders are
also known as cTOP or \kDMDGP\ orders \cite{orders-dam}. The edges
induced by these properties in a \kDMDGP\ graph are called {\it
  discretization edges}, and the edges which are not discretization
edges are called {\it pruning edges}.

Many mathematical aspects of the \kDMDGP\ have been investigated in
the past (see \cite{powerof2,bppolybook,liberti-gsi13}). The problem
itself is {\bf NP}-hard. The automorphism group of $X$ generally
contains a subgroup $\mathscr{G}_P$ consisting of partial reflections
$g_v$, called the {\it pruning group}, such that the action of $g_v$
over a realization $x\in X$ is:
\begin{equation}
      g_v(x) = (x_1,\ldots,x_{v-1},R_x^v(x_v),\ldots,R_x^v(x_n)),
  \label{eq:prefl}
\end{equation}
where $R_x^v$ is the reflection with respect to the affine subspace
spanned by $x_{v-1}$, $\ldots$, $x_{v-K}$, and where $v$ ranges over a
vertex set 
\begin{equation*}
  Z = V\smallsetminus (\{1,\ldots,K\}\cup \bigcup_{\{u,w\}\in E\atop
  u+K<w} \{u+K+1,\ldots,w\}),
\end{equation*}
or, in other words, $v$ must not be ``covered'' by any pruning edge.

\begin{example}
\revision{Consider the DGP instance with $V=\{1,2,3,4\}$,
\[E=\{\{1,2\},\{1,3\},\{2,3\},\{2,4\},\{3,4\}\}\]
consisting of two triangles on $\{1,2,3\}$ and $\{2,3,4\}$, and $K=2$. There is a partial reflection $\rho_1$ fixing $1,2$ and reflecting $3,4$ across the line through $1,2$, and another partial reflection $\rho_2$ fixing $1,2,3$ and reflecting $4$ across the line through $1,2,3$. The range of the pruning edge $\{1,4\}$ is $\{1+K+1,\ldots,4\}=\{4\}$. Therefore, if we add $\{1,4\}$ to $E$, $Z=\{3\}$, which means that the pruning group of this instance has the single generator $\rho_1$.}
\end{example}

The protein backbone isomers of a valid realization $\bar{x}$ are
given by the orbit $\mathscr{G}_Px=\{g_v(x)\;|\;v\in Z\}$. It turns
out that all backbone isomers in $\mathscr{G}_Px$ are valid
realizations of the given DGP instance $G$. So we might obtain a
realization $x^\ast$ which is a valid isomer (and hence has zero edge
errors), but has a large cRMSD with the given (different) isomer
$\bar{x}$.

A serious issue arises when considering \iDGP\ instances, however: if
the cRMSD between $x^\ast$ and $\bar{x}$ is positive, is it due to the
``slack'' induced by the interval edge weights, or is it due to the
fact that $x^\ast$ and $\bar{x}$ are different isomers of essentially
the same backbone (a similar issue was described in
\cite{crippen-rmsd})? This motivates us to define the following
problem:
\begin{quote}
  {\sc Distance Error Modulo Isometries} (DEMI). Given integers $n,K$
  with $n\ge K$, two $n$-point realizations $x,y\in\mathbb{R}^{Kn}$
  such that the centroids $\gamma(x)=\gamma(y)=0$, and a description
  of a pruning group $\mathscr{G}_P$, find the rotation $\rho$ and a
  partial reflection composition $g\in\mathscr{G}_P$ such that
  $\|x-g\rho(y)\|$ is minimum.
\end{quote}
Note that groups can be described by listing their elements, or by a set of generators (and possibly relations) which, when multiplied together up to closure, are guaranteed to generate the whole group. The latter description is usually much shorter than the former. 

We let $\partial(x,y)$ be the minimum value of $\|x-g\rho(y)\|$ which
solves the DEMI. We note that $\partial$ is {\it not} a semimetric
(hence not even a metric), since $\partial(x,y)$ can be zero even
though $x\not=y$ (just take $y$ as a partial reflection of $x$).

\subsubsection{Complexity of DEMI}
\label{s:demicomplexity}
The computational complexity class of DEMI depends on the description
of the pruning group. If it is given explicitly, by listing all the
partial reflection compositions in $\mathscr{G}_P$, then the trivial
Algorithm \ref{alg:trivial} solves the problem in polynomial time for
fixed $K$. For a realization $x\in\mathbb{R}^{Kn}$ and an integer
$h\le n$, let $x[h]$ be the partial realization $(x_i\;|\;1\le i\le
h)$.
\begin{algorithm}[!ht]
  \begin{algorithmic}[1]
\STATE Find a congruence $\rho$ minimizing
  $\|x[K]-y[K]\|$\label{trivial1} 
\STATE Let $\partial(x,y)=\min\{\|x-g\rho(y)\|\;|\;g\in\mathscr{G}_P\}$
\label{trivial3}  
  \end{algorithmic}
  \caption{{\sf SolveDEMI}$(x,y{\CLA ,\mathscr{G}_P})$}
  \label{alg:trivial}
\end{algorithm}
Step \ref{trivial1} takes a polynomial amount of time for fixed $K$
(an $O(n^{K-2}\log n)$ algorithm was described in \cite{altmehl}), but
more efficient methods exist for $K=3$, see \cite{atkinson,coutsias}.
Step \ref{trivial3} depends linearly on the order of the pruning
group, which was shown in \cite{liberti-gsi13} to be $2^{|Z|}$. Since
$Z$ is usually small in practice (see Sect.~\ref{s:demisizeofZ}) and
on average (see Sect.~\ref{s:varZ}), assuming the input to DEMI to be
the explicit list of all partial reflection compositions is not out of
place.

We have not been able to prove that DEMI can be solved in polynomial
time (for fixed $K$) \revision{if its input is $x$, $n$, and the compact group
generators description $Z$,} nor that DEMI is {\bf NP}-hard under the
same conditions. We leave this as an open question.

%% Let us assume now that the input description of $\mathscr{G}_P$ is by
%% a list of generators $\{g_v\;|\;v\in Z\}$. Then the algorithm above
%% has worst-case exponential complexity for obvious reasons; 
%% but unless
%% $\mbox{\bf P}=\mbox{\bf NP}$, finding a polynomial algorithm is
%% impossible, as the next result shows.
%% \begin{thm}
%% DEMI is {\bf NP}-hard.
%% \end{thm}
%% \begin{proof}
%% We reduce {\sc Partition} to DEMI. Let $(a_1,\ldots,a_m)$ be an
%% instance of {\sc Partition}. {\bf TODO} 
%% \end{proof}
%% The above proof is based on an idea of Yemini \cite[p.~2]{yemini},
%% which is in turn inspired by the classic result by Saxe that the DGP
%% is {\bf NP}-complete even when $K=1$ \cite{saxe79}.

\subsubsection{Empirical observations on the size of $Z$}
\label{s:demisizeofZ}
In this section we exhibit empirical evidence to the effect that $|Z|$
is rarely large. First, we note that $|Z|\ge 1$: this follows by the
definition of $Z=\{v>K\;|\nexists\{u,w\}\in E\;(u+K<v\le w)\}$,
since $v=K+1$ is obviously always in $Z$ (this can also be shown by
other means \cite[Sect.~2.1]{dmdgp}).

\begin{figure}[!ht]
\begin{center}
\hspace*{-0.5cm}\begin{tabular}{rcc} 
 $s/K$  & $2$ & $3$ \\ [-3em] 
\begin{minipage}{0.05\textwidth}$0.1$\end{minipage} &
\begin{minipage}{0.45\textwidth}\includegraphics[trim=100 130 100 130, clip, width=1.0\textwidth]{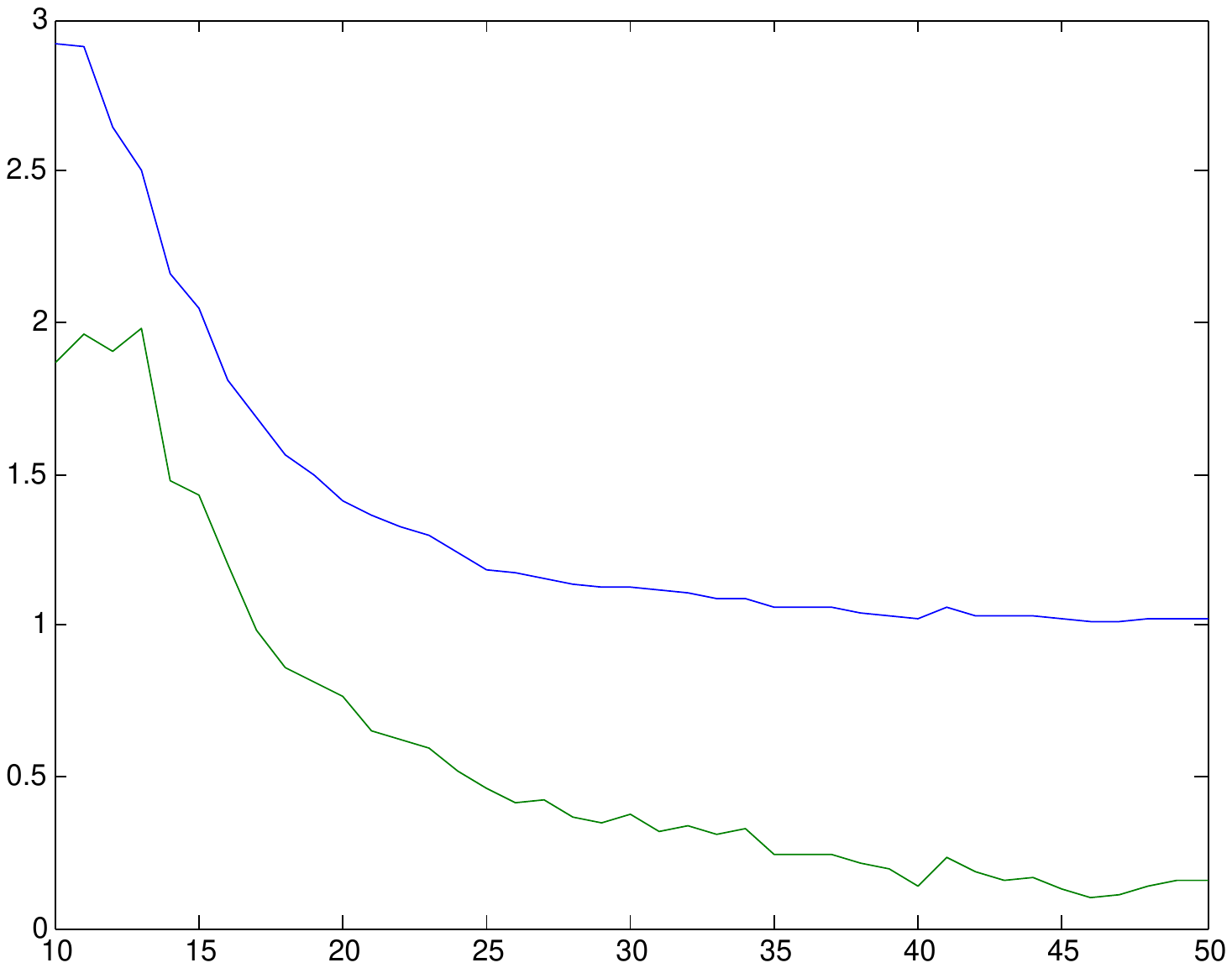}\end{minipage} &
\begin{minipage}{0.45\textwidth}\includegraphics[trim=100 130 100 130, clip, width=1.0\textwidth]{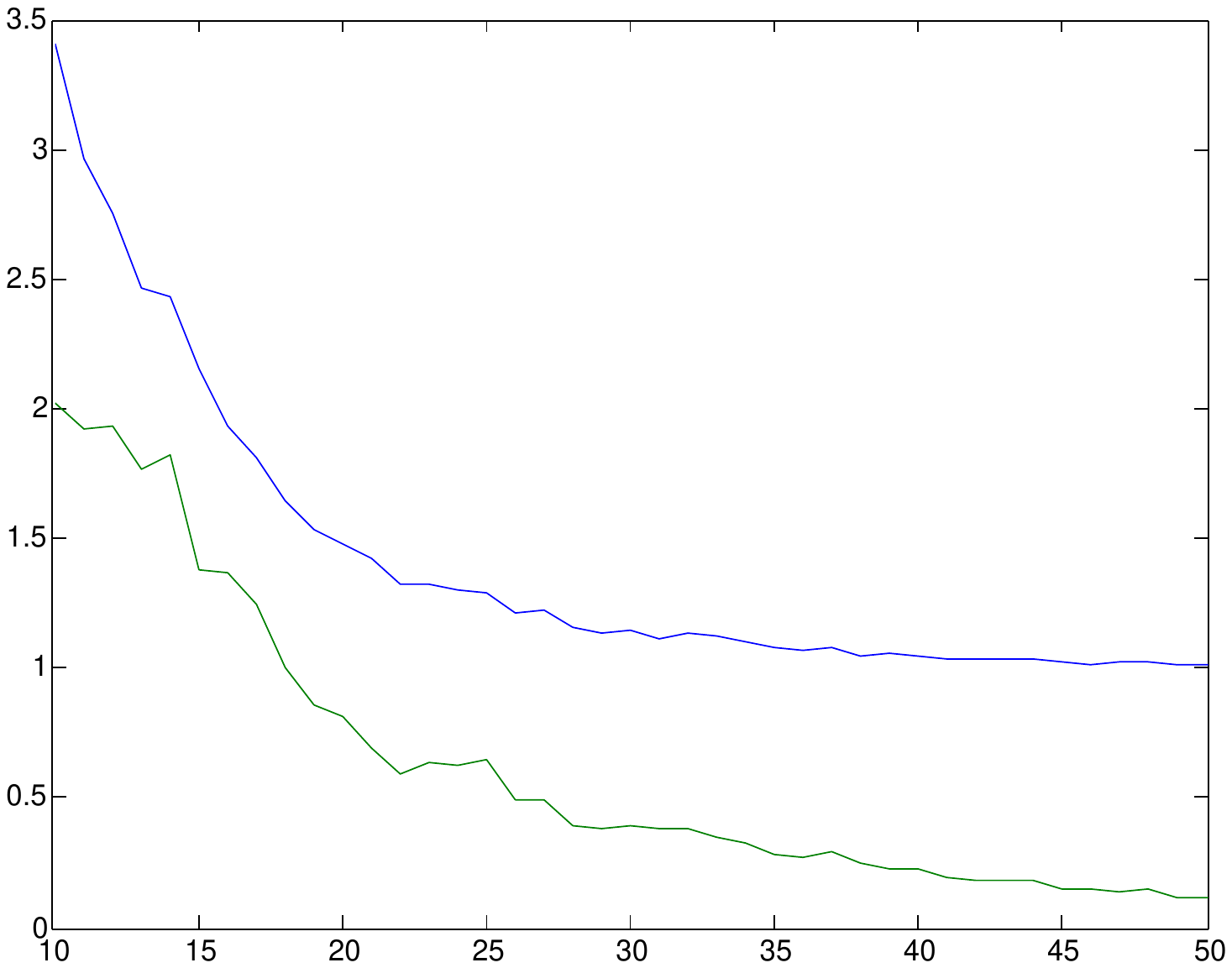}\end{minipage} \\ [-7em]
\begin{minipage}{0.05\textwidth}$0.2$\end{minipage} &
\begin{minipage}{0.45\textwidth}\includegraphics[trim=100 130 100 130, clip, width=1.0\textwidth]{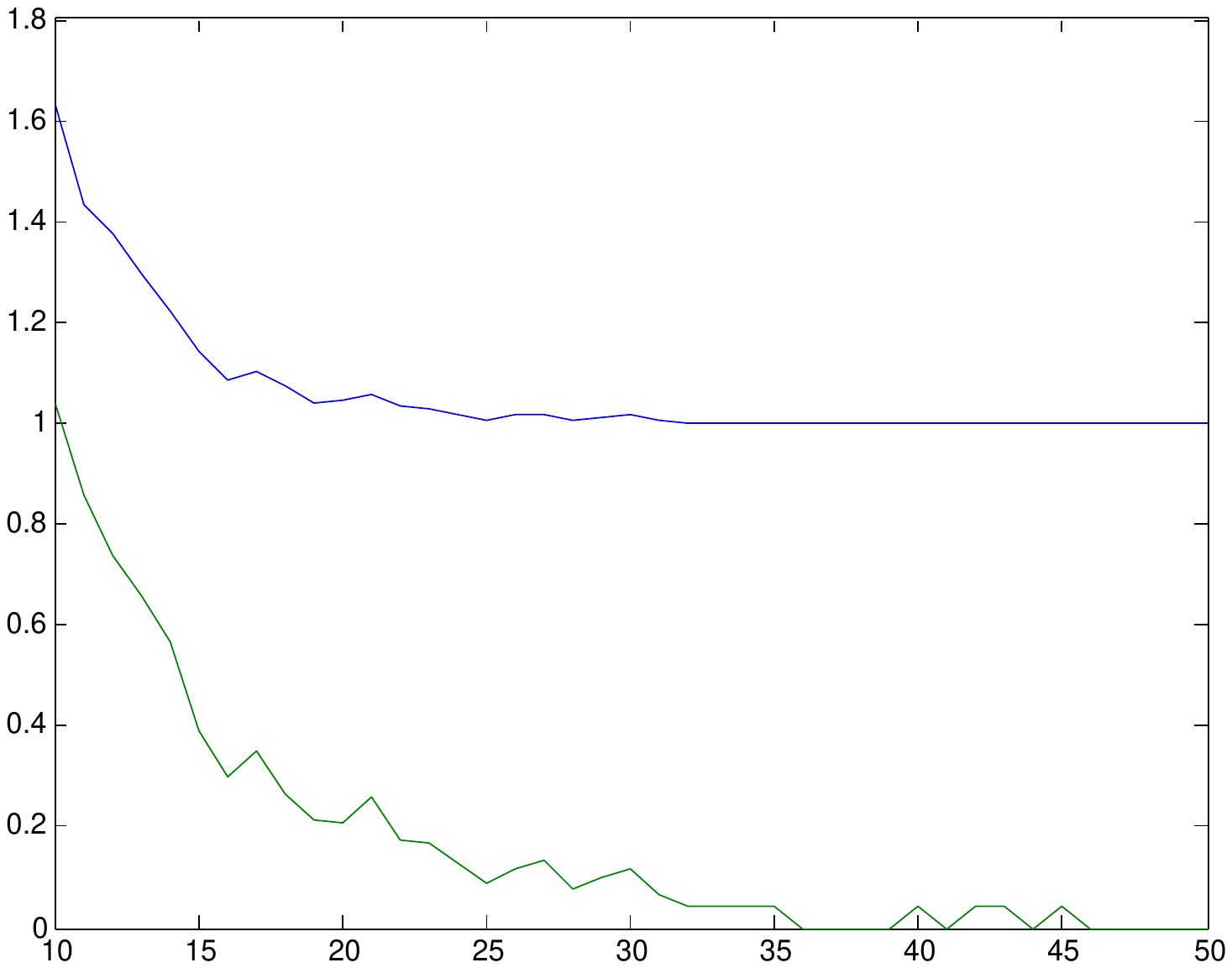}\end{minipage} &
\begin{minipage}{0.45\textwidth}\includegraphics[trim=100 130 100 130, clip, width=1.0\textwidth]{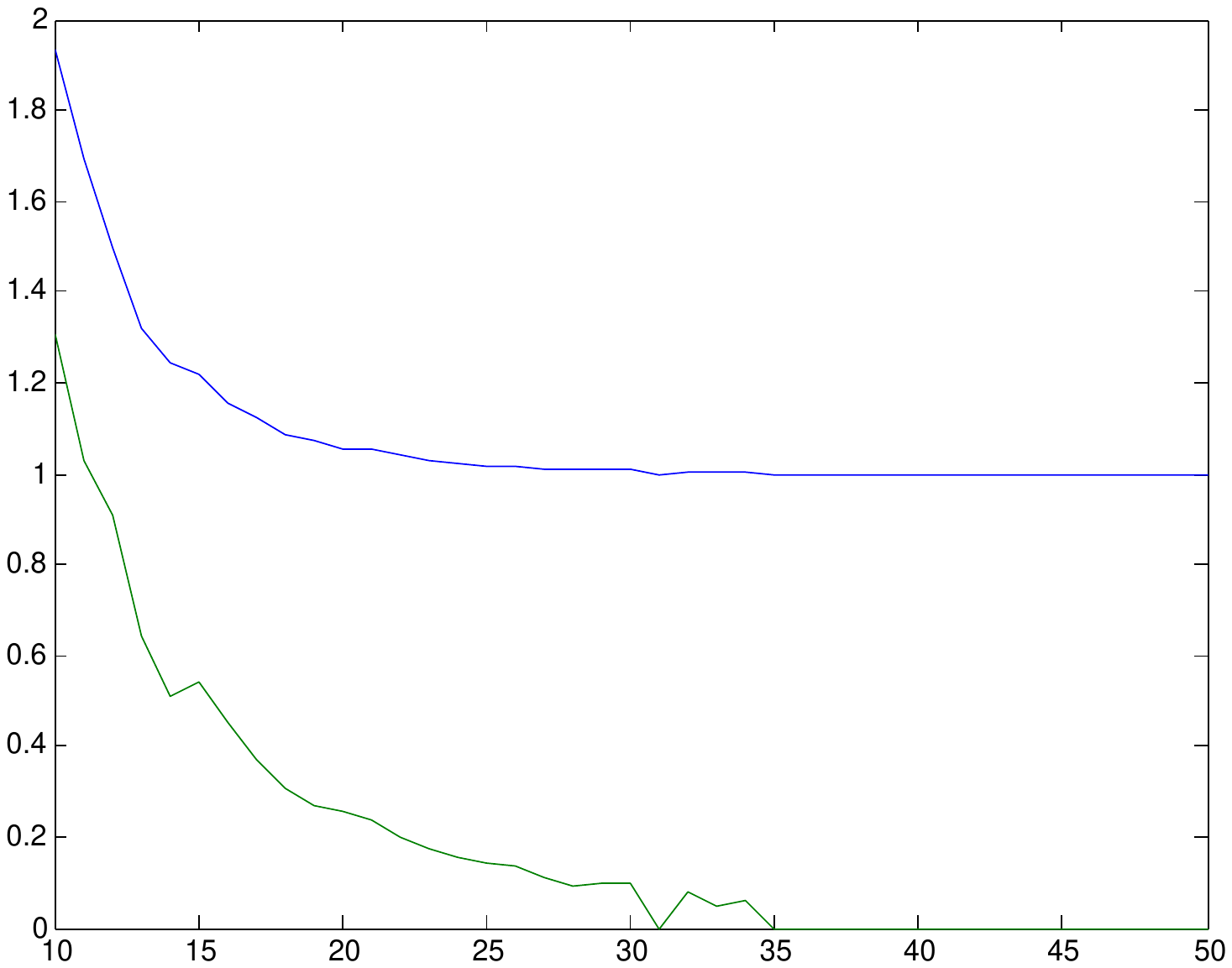}\end{minipage} \\ [-7em]
\begin{minipage}{0.05\textwidth}$0.3$\end{minipage} &
\begin{minipage}{0.45\textwidth}\includegraphics[trim=100 130 100 130, clip, width=1.0\textwidth]{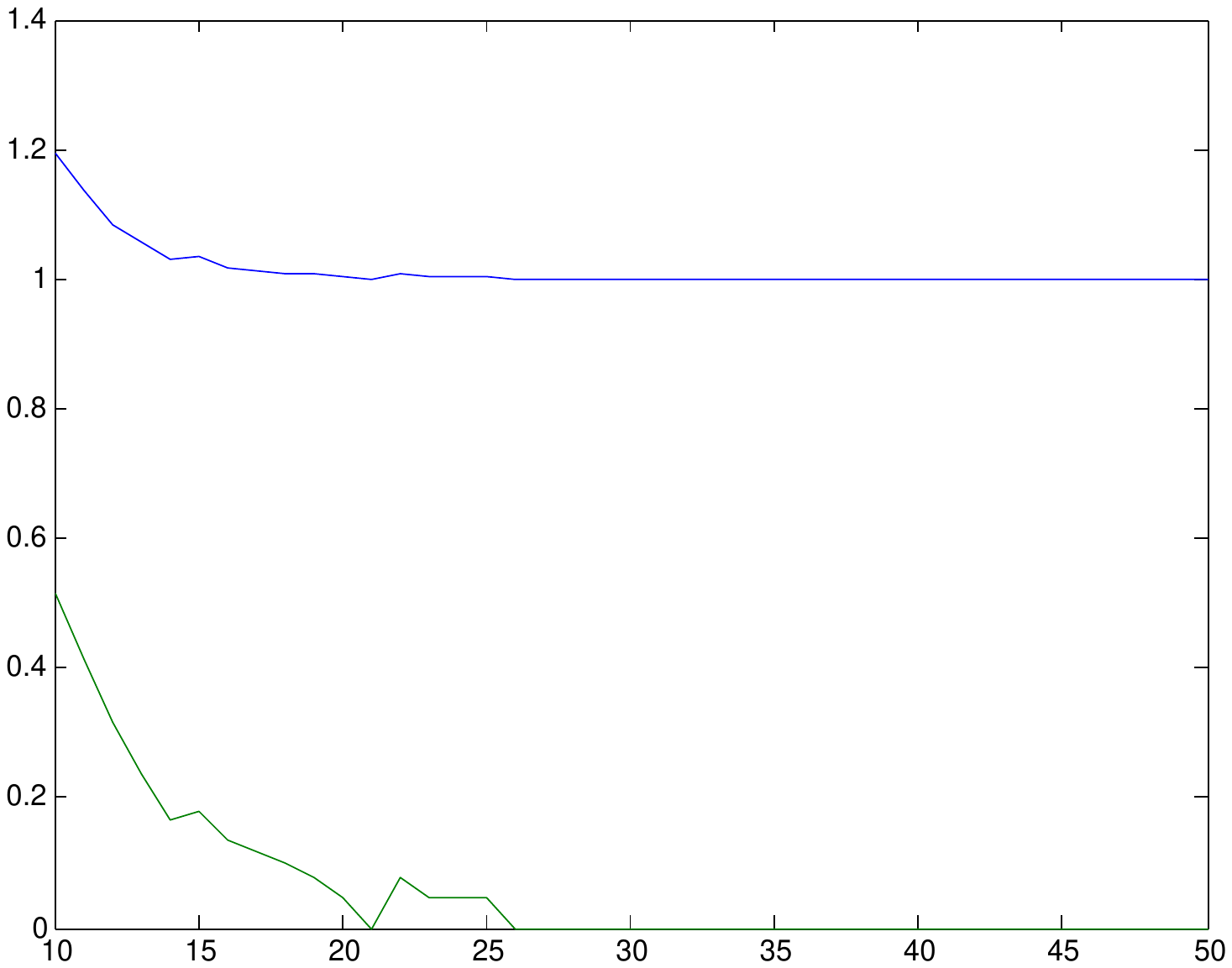}\end{minipage} &
\begin{minipage}{0.45\textwidth}\includegraphics[trim=100 130 100 130, clip, width=1.0\textwidth]{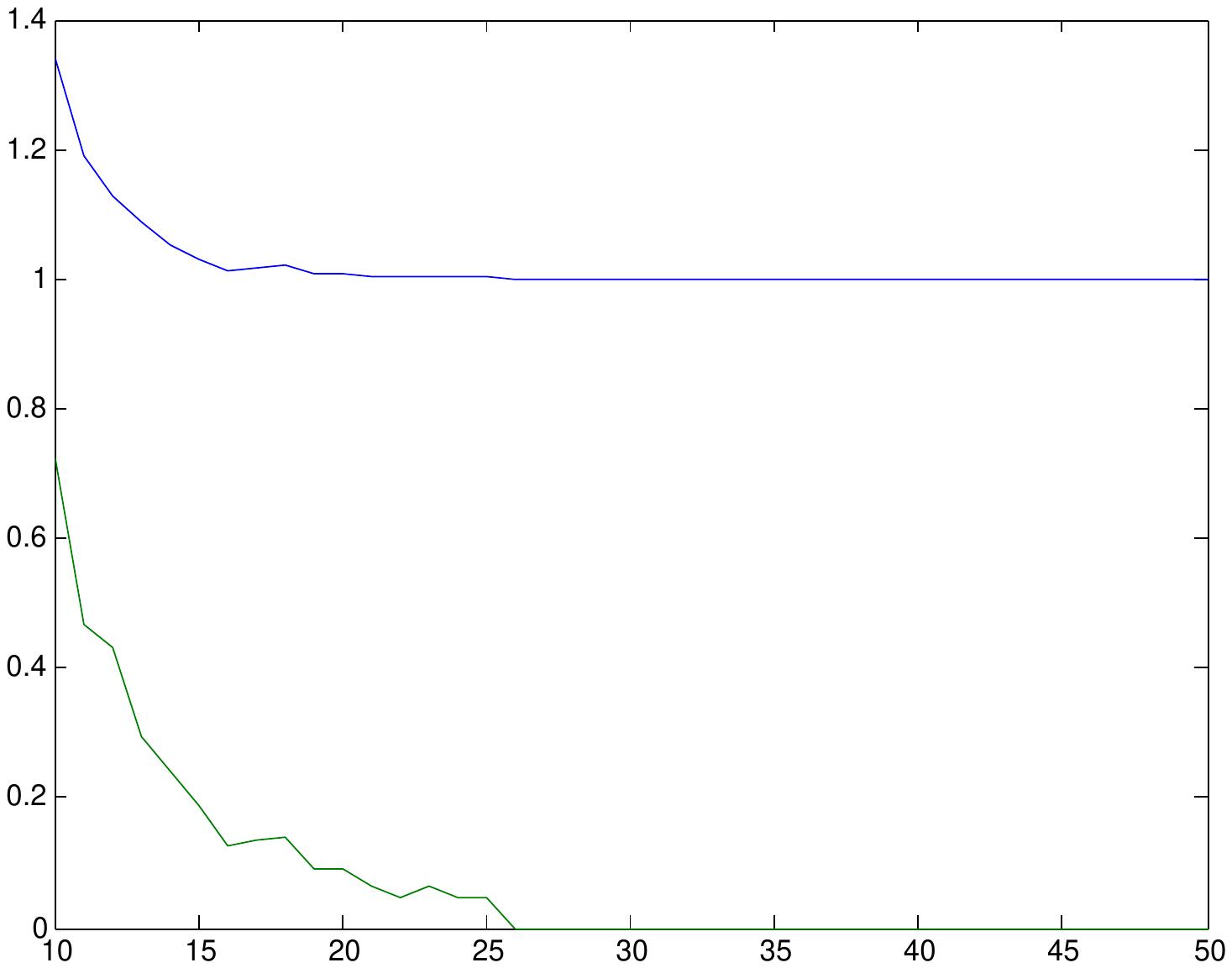}\end{minipage} 
\end{tabular}
%% \hspace*{-1.0cm}\begin{tabular}{c@{\hspace*{-0.4cm}}c@{\hspace*{-0.4cm}}c} 
%%   $s=0.1$ & $s=0.2$ & $s=0.3$ \\ %[-0.5cm]
%% \includegraphics[width=5.0cm]{pgrpavgsize-2_x_01_500-10_50} &
%% \includegraphics[width=5.0cm]{pgrpavgsize-2_x_02_500-10_50} &
%% \includegraphics[width=5.0cm]{pgrpavgsize-2_x_03_500-10_50} \\ %[-3cm]
%% \includegraphics[width=5.0cm]{pgrpavgsize-3_x_01_500-10_50} &
%% \includegraphics[width=5.0cm]{pgrpavgsize-3_x_02_500-10_50} &
%% \includegraphics[width=5.0cm]{pgrpavgsize-3_x_03_500-10_50} 
%% \end{tabular}
\vspace*{-1.5cm}
\end{center}
\caption{\revision{In each picture: mean (top curve) and standard deviation (bottom curve) of the pruning group size as a function of $n$ for fixed values of $K=2$ (left column), $K=3$ (right column), and the edge sparsity $s$ (values in $\{0.1,0.2,0.3\}$ in top, middle and bottom rows).}}
\label{f:pgrpsize_n}
\end{figure}

\begin{figure}[!ht]
\begin{center}
\hspace*{-0.5cm}\begin{tabular}{rcc} 
 $n/K$  & $2$ & $3$ \\ [-3em] 
\begin{minipage}{0.05\textwidth}$10$\end{minipage} &
\begin{minipage}{0.45\textwidth}\includegraphics[trim=100 130 100 130, clip, width=1.0\textwidth]{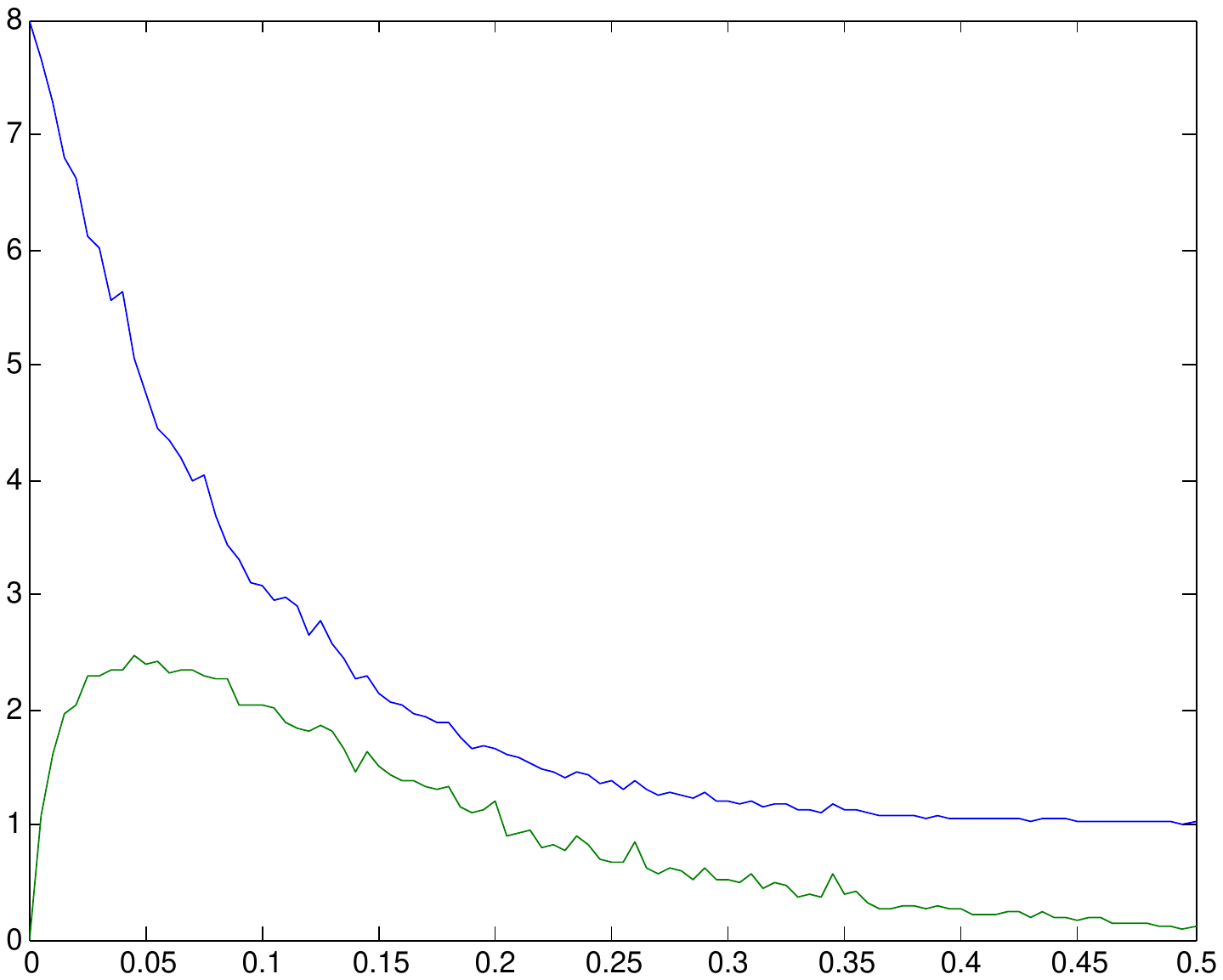}\end{minipage} &
\begin{minipage}{0.45\textwidth}\includegraphics[trim=100 130 100 130, clip, width=1.0\textwidth]{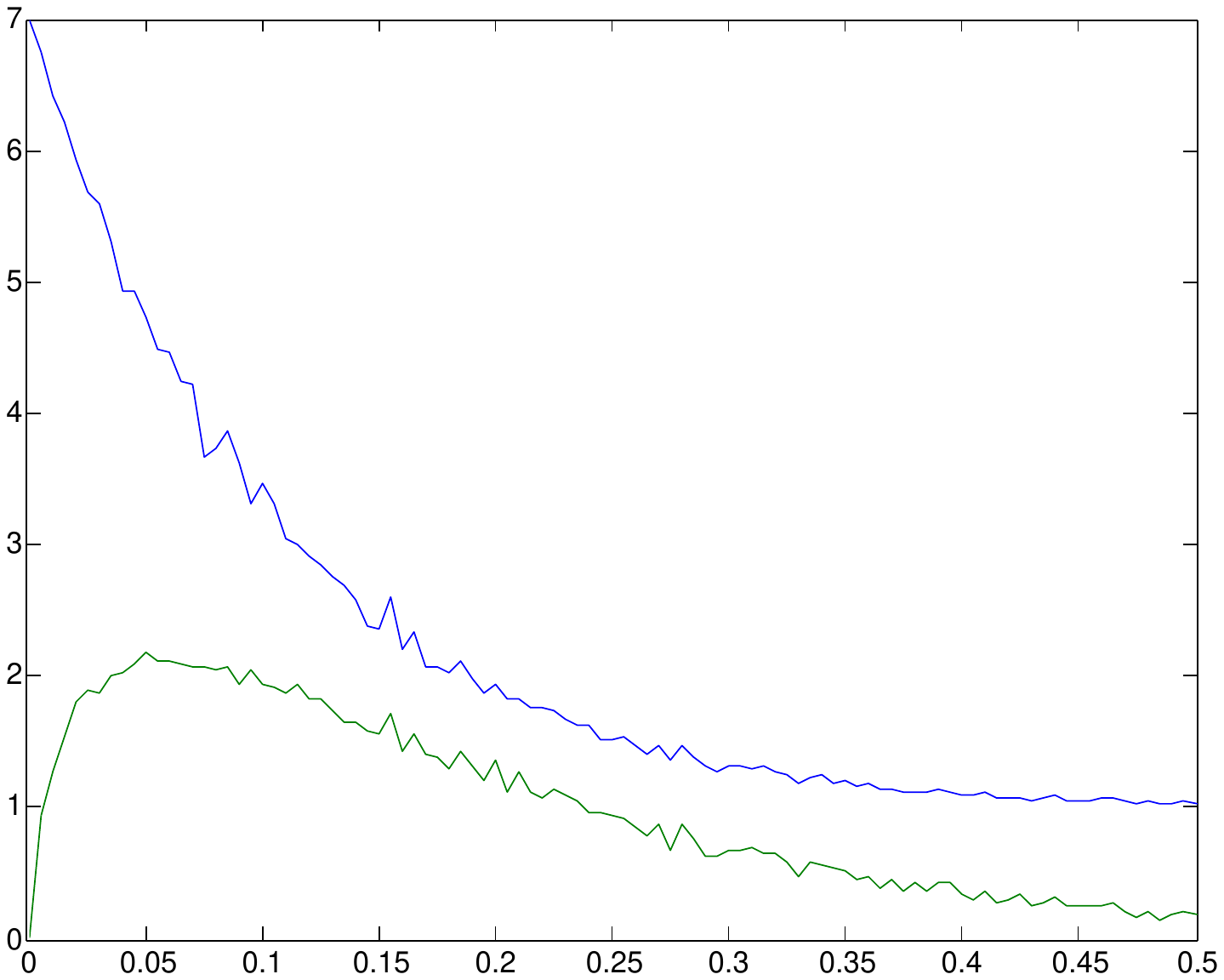}\end{minipage} \\ [-7em]
\begin{minipage}{0.05\textwidth}$15$\end{minipage} &
\begin{minipage}{0.45\textwidth}\includegraphics[trim=100 130 100 130, clip, width=1.0\textwidth]{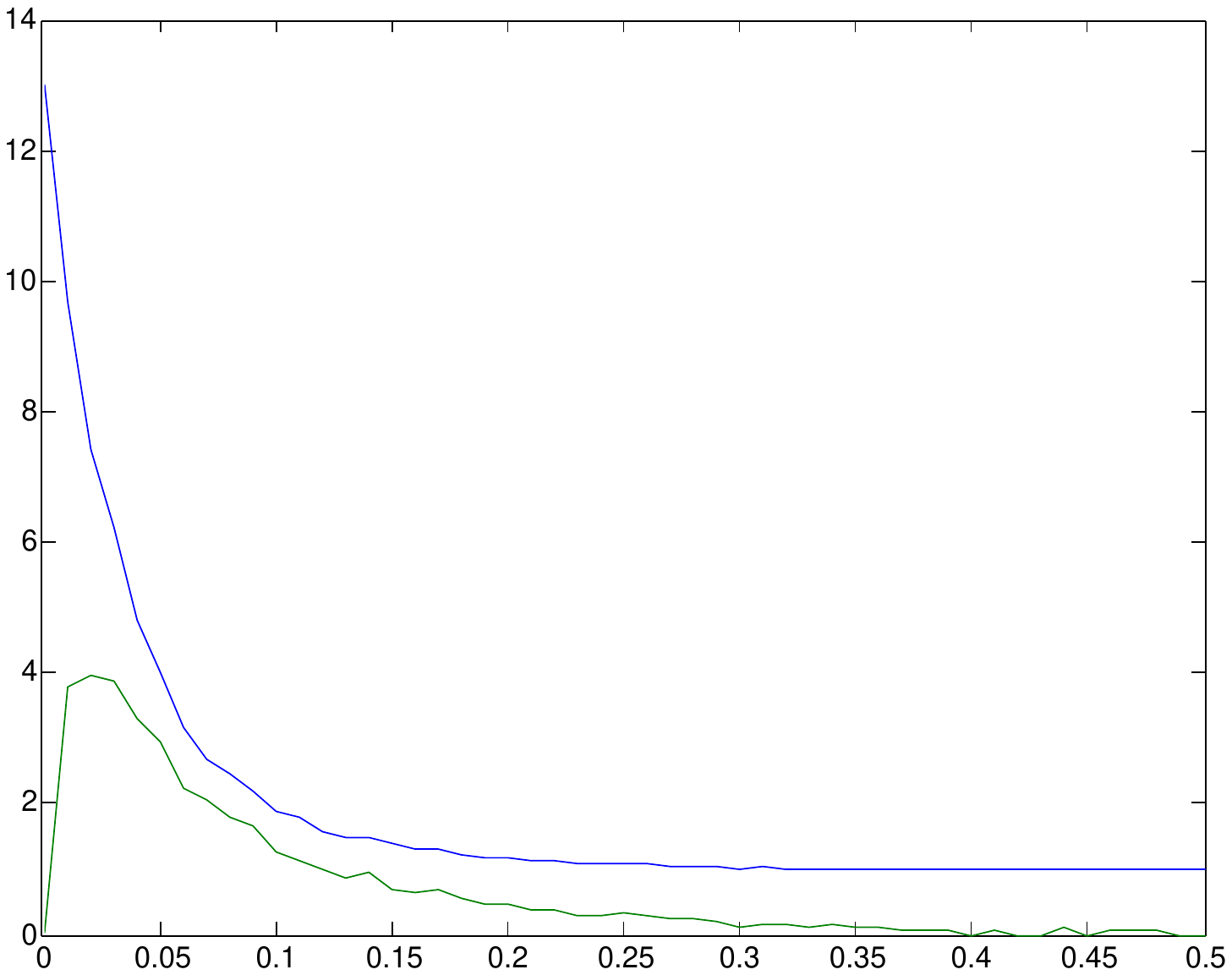}\end{minipage} &
\begin{minipage}{0.45\textwidth}\includegraphics[trim=100 130 100 130, clip, width=1.0\textwidth]{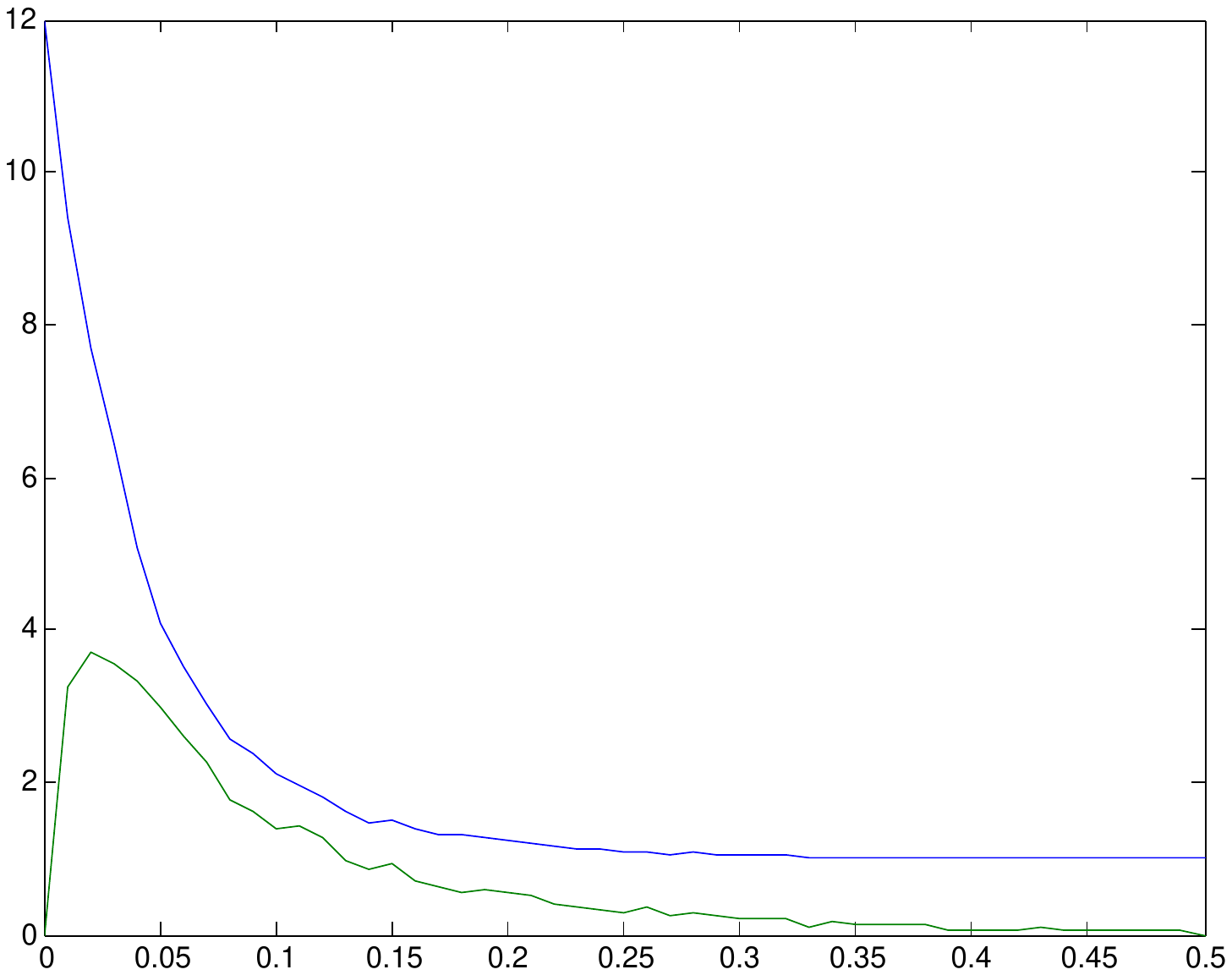}\end{minipage} \\ [-7em]
\begin{minipage}{0.05\textwidth}$20$\end{minipage} &
\begin{minipage}{0.45\textwidth}\includegraphics[trim=100 130 100 130, clip, width=1.0\textwidth]{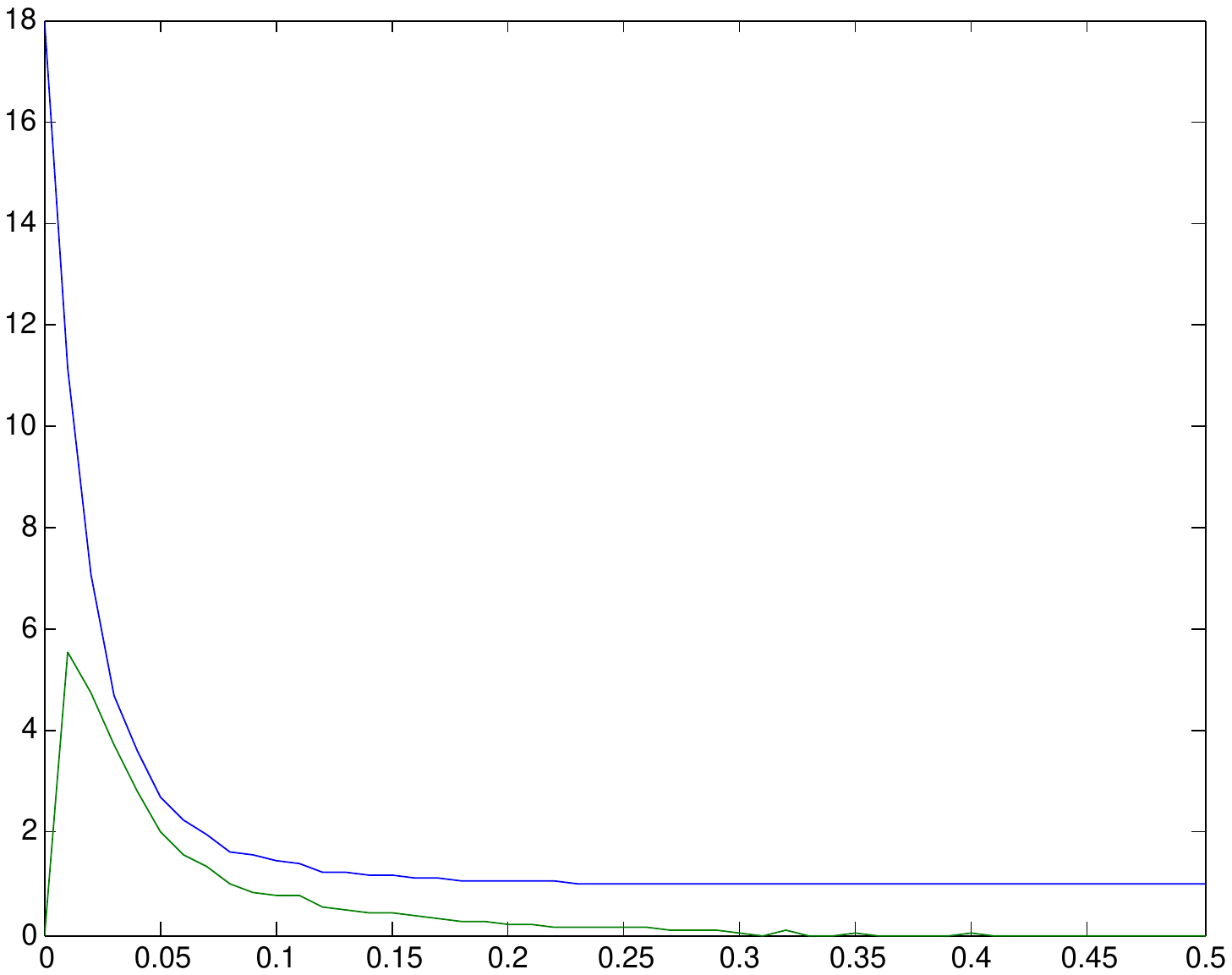}\end{minipage} &
\begin{minipage}{0.45\textwidth}\includegraphics[trim=100 130 100 130, clip, width=1.0\textwidth]{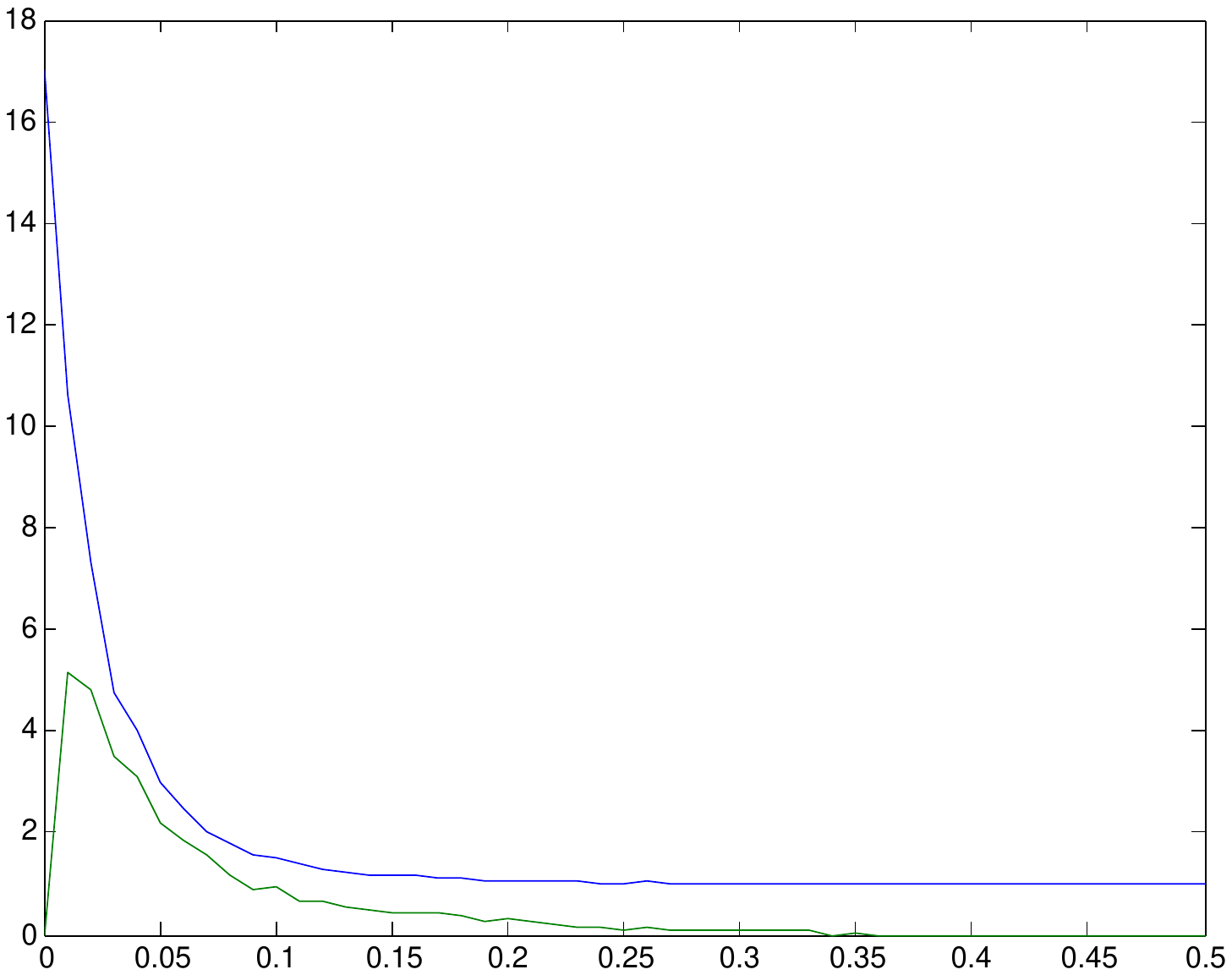}\end{minipage} 
\end{tabular}
%% \hspace*{-1.0cm}\begin{tabular}{c@{\hspace*{-0.4cm}}c@{\hspace*{-0.4cm}}c}
%%  $n=10$ & $n=15$ & $n=20$ \\ %[-0.5cm]
%% \includegraphics[width=5.0cm]{pgrpavgsize-2_10_x_500-00_05} &
%% \includegraphics[width=5.0cm]{pgrpavgsize-2_15_x_500-00_05} &
%% \includegraphics[width=5.0cm]{pgrpavgsize-2_20_x_500-00_05} \\ %[-3cm]
%% \includegraphics[width=5.0cm]{pgrpavgsize-3_10_x_500-00_05} &
%% \includegraphics[width=5.0cm]{pgrpavgsize-3_15_x_500-00_05} &
%% \includegraphics[width=5.0cm]{pgrpavgsize-3_20_x_500-00_05}
%% \end{tabular}
\vspace*{-1.5cm}
\end{center}
\caption{\revision{In each picture: mean (top curve) and standard deviation (bottom curve) of the pruning group size as a function of the edge sparsity $s$ for fixed values of $K=2$ (left column), $K=3$ (right column), and $n$ (values in $\{10,15,20\}$ in top, middle and bottom rows).}}
\label{f:pgrpsize_s}
\end{figure}

Figures \ref{f:pgrpsize_n}-\ref{f:pgrpsize_s} show the mean and
standard deviations of $|Z|$ relative to samples of 500 randomly
generated \kDMDGP\ instances for each value of $K\in\{2,3\}$ and
various values of the edge sparsity $s$. The generation procedure is
as follows: given $n=|V|$ and $K$, we initially generate a
\kDMDGP\ instance with all the necessary discretization edges in its
edge set $E$ (there are $K(K-1)/2+(n-K)K$ of them), but no pruning
edges. Then we loop over all $\{i,j\}$ which are not discretization
edges, and with given probability $s$ we insert a pruning edge in
$E$. So $s$ is in fact the density of the pruning edges.

The exact dependency of $|Z|$ on the number of pruning edges is given
in \cite{bppolybook}, and it is used to show that the BP algorithm is
FPT. It should be clear by definition that the denser the graph, the
smaller $Z$ must be. Figures \ref{f:pgrpsize_n}-\ref{f:pgrpsize_s}
show (empirically) that $|Z|$ tends to 1 very fast and very reliably
as $n$ and $s$ increase, with $n,s$ as small as, respectively, $20$
and $0.3$. Large graphs with $|Z|>1$ are very rare.

It is interesting to note that the standard deviation of $|Z|$ as a
function of the sparsity $s$ has a maximum in $[0,0.05]$ (see
Fig.~\ref{f:pgrpsize_s}). This phenomenon is analyzed below in more
detail.

\subsubsection{Expectation and variance of $|Z|$}
\label{s:varZ}
As explained in Sect.~\ref{s:demi}, \kDMDGP\ instances consist of a
backbone subgraph (a minimal graph satisfying the clique and
contiguous trilateration order properties) and some pruning
edges. Accordingly, random \kDMDGP\ graphs $G=(V,E)$ are generated as
follows:
\begin{itemize}
\item a backbone which only depends on $K,n$ and determines the order on
  $V$;
\item for each pair $\{u,w\}$ which is not a discretization edge, we
  independently add $\{u,w\}$ as a pruning edge in $E$ with probability
  $s\in[0,1]$.
\end{itemize}
Now consider the subset $Z \subseteq V$, defined as in
Sect.~\ref{s:demi} as
\begin{equation*}
  Z = \{v > K \, | \, \nexists \{u,w\} \in E \; (u+K < v \le w) \}.
\end{equation*}

We consider $|Z|$ as a random variable depending on the edge
probability $s$ (also known as the sparsity of the \kDMDGP\ graph
$G$), and compute its expected value. In the following, $\mbox{\sf
  P}(\cdot)$ is the probability of an event, $\mbox{\sf E}(\cdot)$ is
the expectation of a random variable and $\mbox{\sf Var}(\cdot)$ is
its variance.
\ifspringer\begin{proposition}\else\begin{prop}\fi
$\mbox{\sf E}(|Z|) \le 1 + (n-K-1)(1-s)^{n-K-1}$.
\ifspringer\end{proposition}\else\end{prop}\fi
\begin{proof}
For all $v\in\{K+1,\ldots, n\}$ define $\mathcal{X}_v=0$ if $v\notin
Z$ and $1$ if $v \in Z$. Then $|Z|=\sum\limits_{v=K+1}^n
\mathcal{X}_v$, which implies:
\begin{equation*}
  \mbox{\sf E}(|Z|)=\sum\limits_{v=K+1}^n \mbox{\sf
    E}(\mathcal{X}_v)=\sum_{v=K+1}^n \mbox{\sf P}(v\in Z).
\end{equation*}
Now, for any $v\in\{K+1,\ldots,n\}$ there are $v-K-1$ choices of $u$
with $u+K<v$, and there are $n-v+1$ choices of $w$ with $v\le
w$. Therefore, there are $(v-K-1)(n-v+1)$ possible choices of the
pruning edge $\{u,w\}$ such that $u+K<v\le w$. Moreover, $v\in Z$ if
all these pairs are not added to the graph. Thus,
\[\mbox{\sf P}(v\in Z)=(1-s)^{(v-K-1)(n-v+1)},\] and hence:
\begin{equation*}
  \mbox{\sf E}(|Z|) = \sum_{v=K+1}^n  (1-s)^{(v - K - 1)(n - v + 1)}.
\end{equation*}
Finally, we remark that (a) the first term of the sum is 1, and (b)
$(1-s) < 1$, so we can replace all
the terms of the sum by the second largest one, and obtain:
\begin{equation}
  \mbox{\sf E}(|Z|) \le 1 +  (n-K-1)(1-s)^{n-K-1}, \label{eq:EZ}
\end{equation}
as claimed. \qed
\end{proof}
The RHS of Eq.~\eqref{eq:EZ} converges to 1 as $s\to 1$ with $n,K$
fixed, and as $n\to\infty$ with $s,K$ fixed, which is consistent with
the empirical results of Sect.~\ref{s:demisizeofZ}. We are therefore
justified in making the qualitative statements that, for random DMGDP,
$|Z|\approx 1$. 

\revision{We now discuss the variance. Since $\mbox{Var}(|Z|) = \mbox{Var}(\sum_{v=K+1}^n \mathcal{X}_v)$, then by a property of sum of correlated variables \cite{wikisumcorr}, we have:
\begin{align*}
\mbox{Var}(|Z|) & = \sum_{v=K+1}^n \mbox{Var}(\mathcal{X}_v) + 2\sum_{k+1\le v_1 < v_2 \le n} \mbox{Cov} (\mathcal{X}_{v_1},\mathcal{X}_{v_2}) \\
& = \sum_{v=K+2}^n \mbox{Var}(\mathcal{X}_v) + 2 \sum_{k+2\le v_1 < v_2 \le n} \mbox{Cov} (\mathcal{X}_{v_1},\mathcal{X}_{v_2}) \\
& \mbox{ (this follows from $\mbox{E}(\mathcal{X}_{K+1})  = 1$ and $ \mbox{E}(\mathcal{X}_{K+1}\mathcal{X}_v) = \mbox{E}(\mathcal{X}_v)$ for all $v$)}  \\
& = \sum_{v=K+2}^n \mbox{E}(\mathcal{X}_v) + \sum_{k+2\le v_1 < v_2 \le n} \mbox{E} (\mathcal{X}_{v_1} \mathcal{X}_{v_2}) + \\
& - \sum_{v=K+2}^n [\mbox{E}(\mathcal{X}_v)]^2  - 2\sum_{k+2\le v_1 < v_2 \le n} \mbox{E} (\mathcal{X}_{v_1}) \mbox{E}( \mathcal{X}_{v_2}).
\end{align*}}
By definition of $Z$, two vertices $v_1$ and $v_2$ are in $Z$ if all
pairs $\{u,w\}$ such that either $u+K<v_1\le w$ or $u+K<v_2\le w$ are
not edges of $G$. Assume $v_1 < v_2$, then there are:
\begin{align*}
  \ & (v_1 - K-1)(n-v_1+1) + (v_2 -K-1)(n-v_2+1) - (v_1 -K-1)(n-v_2+1) \\
 =\ & (v_1 - K-1)(n-v_1+1) + (v_2 -v_1)(n-v_2+1)
%=& (j-1)(n-K+1) + i - i^2 + ij - j^2
\end{align*}
such edges (by counting all pairs of each type and subtracting the
number of doubly counted ones). So, the probability that $v_1,v_2\in Z$
is $$(1-s)^{(v_1 - K-1)(n-v_1+1) + (v_2 -v_1)(n-v_2+1)}.$$ This implies
\begin{align*}
  \mbox{\sf Var}(|Z|)
  & = \sum_{v=K+2}^n (1-s)^{(v - K-1)(n-v+1)}
    - \sum_{v=K+2}^n (1-s)^{2(v - K-1)(n-v+1)} + \\ 
  & \quad + 2\sum_{K+2 \le v_1 < v_2 \le n}
    (1-s)^{(v_1 - K-1)(n-v_1+1) + (v_2 -v_1)(n-v_2+1)} \ - \\
  & \quad - 2\sum_{K+2 \le v_1 < v_2 \le n}  (1-s)^{(v_1 - K-1)(n-v_1+1) + (v_2 -K-1)(n-v_2+1)}.
\end{align*}
To simplify the analysis of $\mbox{\sf Var}(|Z|)$, we provide an
upper bound.
\ifspringer\begin{lemma}\else\begin{lem}\fi
  \label{lemma ZTASUA}
  For all $s\in(0,1)$ and $k \ge 1$, we have $\sum\limits_{i=1}^{k-1}
  (1-s)^{i(k-i)} < \frac{2(1-s)^{k-1}}{s}$.
\ifspringer\end{lemma}\else\end{lem}\fi
\begin{proof}
For each $1 \le i < \lfloor \frac{k}{2}\rfloor$ we have the estimate
\begin{align}\label{eq:estimate}
(i+1)(k-i-1) = ik + k -i^2 -2i -1 = i(k-i) + k -2i -1 \ge i(k-i) + 1.
\end{align}
Therefore,
\begin{align*}
  \sum_{i=1}^{k-1} (1-s)^{i(k-i)} 
  & \le 2 \;\sum_{i=1}^{\lfloor \frac{k}{2}\rfloor} (1-s)^{i(k-i)} \\
  & \le 2 ((1-s)^{k-1} + (1-s)^{k} + (1-s)^{k+1}
    + \ldots + (1-s)^{k-2+\lfloor \frac{k}{2}\rfloor}) \\
  & < 2(1-s)^{k-1} \sum_{i=0}^\infty (1-s)^i \\
  & = \frac{2(1-s)^{k-1}}{s}.
\end{align*}
The second inequality follows because of estimate \eqref{eq:estimate}. \qed
\end{proof}
We can now improve the estimate for the variance (where $n,K$ only
appear in the exponent):
\begin{align*}
  0 & < \mbox{\sf Var}(|Z|) \\
  & < \sum_{v_2=K+2}^n\!\!\!\!\! (1-s)^{(v_2 - K-1)(n-v_2+1)} 
  + 2\!\!\!\!\!\!\!\!\sum_{K+2 \le v_1 < v_2 \le n} \!\!\!\!\!\!\!\!(1-s)^{(v_1 - K-1)(n-v_1+1) + (v_2 -v_1)(n-v_2+1)} \\
  & < \sum_{i=1}^{n-K-1} (1-s)^{i(n-K-i)} 
  + 2\!\!\!\!\sum_{K+2 \le v_1 \le n} \left( (1-s)^{(v_1 - K-1)(n-v_1+1)}
    \sum_{i=1}^{n-v_1}(1-s)^{i(n-v_1-i+1)} \right) \\
  & < \frac{2}{s} (1-s)^{(n-K-1)}  
  + \frac{4}{s}\sum_{K+2 \le v_1 \le n}
  \big[ (1-s)^{(v_1 - K-1)(n-v_1+1)} (1-s)^{n-v_1} \big]
  \qquad \mbox{(Lemma \ref{lemma ZTASUA})}\\
  & = \frac{2}{s} (1-s)^{(n-K-1)} + \frac{4}{s(1-s)}\sum_{i=2}^{n-K}
  (1-s)^{i(n-K-i+1)} \\
  & < \frac{2}{s} (1-s)^{(n-K-1)} +
  \frac{4}{s(1-s)} (\frac{2}{s} -1)(1-s)^{n-K} \qquad
  \mbox{(Lemma \ref{lemma ZTASUA})}\\
  & = \left(\frac{8}{s^2} - \frac{2}{s}\right) (1-s)^{n-K-1}.
\end{align*}
For example, with $s=0.2$, $n=35$, $K=2$, the estimate yields
$\left(\frac{8}{s^2} - \frac{2}{s}\right) (1-s)^{n-K-1} = 0.15$. With
$s=0.3$, $n=25$, $K=2$, we get
$\left(\frac{8}{s^2}-\frac{2}{s}\right)x(1-s)^{n-K-1}=0.03$.

Fig.~\ref{f:pgrpsize_s} shows that the standard deviation (and hence
the variance) of $|Z|$ has a maximum when $s$ is close to zero. Fixing
$n$ and $K$, consider $\mbox{\sf Var}(|Z|)$ as a function $f(t)$ of $
1-s$, let $\tau(k)=kt^k$, and rewrite $\mbox{\sf Var}(|Z|)$ as:
\begin{align*}
\mbox{\sf Var}(|Z|) = f(t) & = \sum_{v=K+2}^n t^{(v - K-1)(n-v+1)} -
\sum_{v=K+2}^n t^{2(v - K-1)(n-v+1)} + \\ & \quad + 2\sum_{K+2 \le v_1 <
  v_2 \le n} t^{(v_1 - K-1)(n-v_1+1) + (v_2 -v_1)(n-v_2+1)} - \\ & \quad -
2\sum_{K+2 \le v_1 < v_2 \le n} t^{(v_1 - K-1)(n-v_1+1) + (v_2 -K-1)(n-v_2+1) }.
\end{align*}
Taking the derivative of $f(t)$, we have:
\begin{align*}
f'(t) & = t^{-1} \left(\sum_{v=K+2}^n \tau((v - K-1)(n-v+1)) -
  \sum_{v=K+2}^n \tau(2(v - K-1)(n-v+1)) \; +\right.\\ & \quad
  +2\sum_{K+2 \le v_1 < v_2 \le n} \tau\big((v_1 - K-1)(n-v_1+1) + (v_2
    -v_1)(n-v_2+1)\big) \; -\\ & \left.\quad - 2\sum_{K+2 \le v_1 < v_2 \le n}
    \tau((v_1 - K-1)(n-v_1+1) + (v_2 -K-1)(n-v_2+1))\right).
%\\ & =t^{-1}
%\bigg[\sum_{v=K+2}^n \tau\big[(v - K-1)(n-v+1)\big] - \tau\big [2(v -
%K-1)(n-v+1)\big] \bigg] \; +\\ & \quad +2\,t^{-1} \sum_{K+2 \le u < v
%\le n} \tau\big[(u - K-1)(n-u+1) + (v -u)(n-v+1)\big] \\& -
%\tau\big[(u - K-1)(n-u+1) + (v -K-1)(n-v+1)\big] \\ &
\end{align*}

Consider the derivative of $\tau$ with respect to $k$, $\tau'(k) =
(kt^k)' = t^k (1 + k\ln (t))$, and take for example $k \ge 20$ and $t
\le 0.95$. We have $(1 + k\ln (t)) \le 1 + 20\ln(0.95) = -0.026 <
0$. Therefore, when $t < 0.95$, $\tau(k)$ is a decreasing function on
the set $\{k\;|\; k \ge 20\}$. It means that, whenever $n-K-1 \ge 20$,
$\tau((v - K-1)(n-v+1))\ge \tau(2(v - K-1)(n-v+1))$ for each
$v\in\{K+2,\ldots,n\}$, and $\tau((v_1-K-1)(n-v_1+1)+(v_2-v_1)(n-v_2+1))\ge
\tau((v_1 - K-1)(n-v_1+1) + (v_2 -K-1)(n-v_2+1))$ for each
$v_1<v_2\in\{K+2,\ldots,n\}$, since all values under $\tau$ are at least
$20$. We therefore have that $f'(t) \ge 0$ for all $t<0.95$,
i.e.,~whenever $s\in[0.05, 1]$, $\mbox{\sf Var}(|Z|)$ decreases as $s$
increases. In other words, the maximum of $\mbox{\sf Var}(|Z|)$ can
only be attained on $[0,0.05]$.

We can generalize this example to the following result.
\ifspringer\begin{lemma}\else\begin{lem}\fi
For fixed $n,K$, the maximum of $\mbox{\sf Var}(|Z|)$ can only be
attained at $s\in[0,\frac{1}{n-K-1}]$.
\ifspringer\end{lemma}\else\end{lem}\fi
\begin{proof}
We have 
\begin{equation*}
    \tau'(k) < 0 \Leftrightarrow 1 + k\ln(t) <0 \Leftrightarrow \ln
    (\frac{1}{t}) > \frac{1}{k} \Leftrightarrow \frac{1}{t} > e^{1/k}
    \Leftrightarrow t < e^{-1/k} \Leftrightarrow s > 1 - e^{-1/k}.
\end{equation*}
Since
\begin{equation*}
  e^{-1/k} = 1 - \frac{1}{k} + \frac{1}{2!k^2} - \frac{1}{3!k^3} +
  \ldots > 1 - \frac{1}{k},
\end{equation*}
we have $1-e^{-1/k}<\frac{1}{k}$. Therefore, if $s>\frac{1}{k}$ we
have $\tau'(k)<0$. So, when $s>\frac{1}{n-K-1}$, we have $\tau'(k)<0$
for all $k \ge n-K-1$. Now the same argument as in the example above
shows that $\mbox{\sf Var}(|Z|)$ decreases on the set
$[\frac{1}{n-K-1},\infty)$. \qed
\end{proof}

\subsubsection{Computing DEMI measures in practice}
\label{s:demipractice}
We believe we made a convincing argument that we can safely use
Alg.~\ref{alg:trivial} to solve DEMI instances. There is, however, a
glitch: none of the PDB instances we consider actually comes with a
pre-defined cTOP order. For some of them, the protein backbone is a
cTOP order. For others this is not the case. The state of the art in
automatically finding cTOP orders in graphs is severely limited
\cite{orders-dam}, and certainly does not scale to hundreds of
vertices easily. Thus the {\it DEMI measure} $\partial(x,y)$ of a
realization $x$ with respect to a given realization $y$ will not be
computed for all instances we test in Sect.~\ref{s:compres}, but only
for some (see Table \ref{t:demi}).

\section{New and existing \iDGP\ formulations}
\label{s:formulations}
All formulations we consider are box-constrained to bounds
$x\in[M^L,M^U]^{Kn}$, which have to be large enough to accommodate a
worst-case realization with the given distances. One could take for
example $M^L=-\frac{1}{2}\sum_{\{u,v\}\in E} U_{uv}$ and $M^U=-M^L$,
and then tighten these bounds using some pre-processing techniques
\cite{couenne}. We do not write these bounds explicitly in the
formulations below. Notationwise, $\mathbf{M}=[M^L,M^U]^m$ and
$\mathbf{M}^+=\mathbf{M}\cap[0,+\infty]$.

Most formulations come with variants. A common variant, which we refer
to as the {\it square root variant}, is the following: replace
$\|x_u-x_v\|_2^2$ by $\|x_u-x_v\|_2$ and squared distance bounds by
distance bounds. In such variants, because of floating point issues,
$\sqrt{\alpha}$ is implemented as $\sqrt{\alpha+\delta}$, where
$\delta$ is a constant in $O(10^{-10})$.

In all of our formulations, aside from the semidefinite programming
(SDP) ones, we fix the centroid at the origin, which means that we
find solutions modulo translations. This seems to improve the overall
reliability and convergence speed of the heuristic solution algorithms
we use. It is interesting that this ceases to be the case if we also
impose no rotation by fixing the first $K$ vertices, in which case the
algorithms find much worse local optima.

\subsection{Validation}
\begin{figure}[!ht]
\begin{center}
%\vspace*{-2cm}
\begin{tabular}{c@{\hspace*{-1.5cm}}c}
\begin{minipage}{6cm}
\includegraphics[width=3cm]{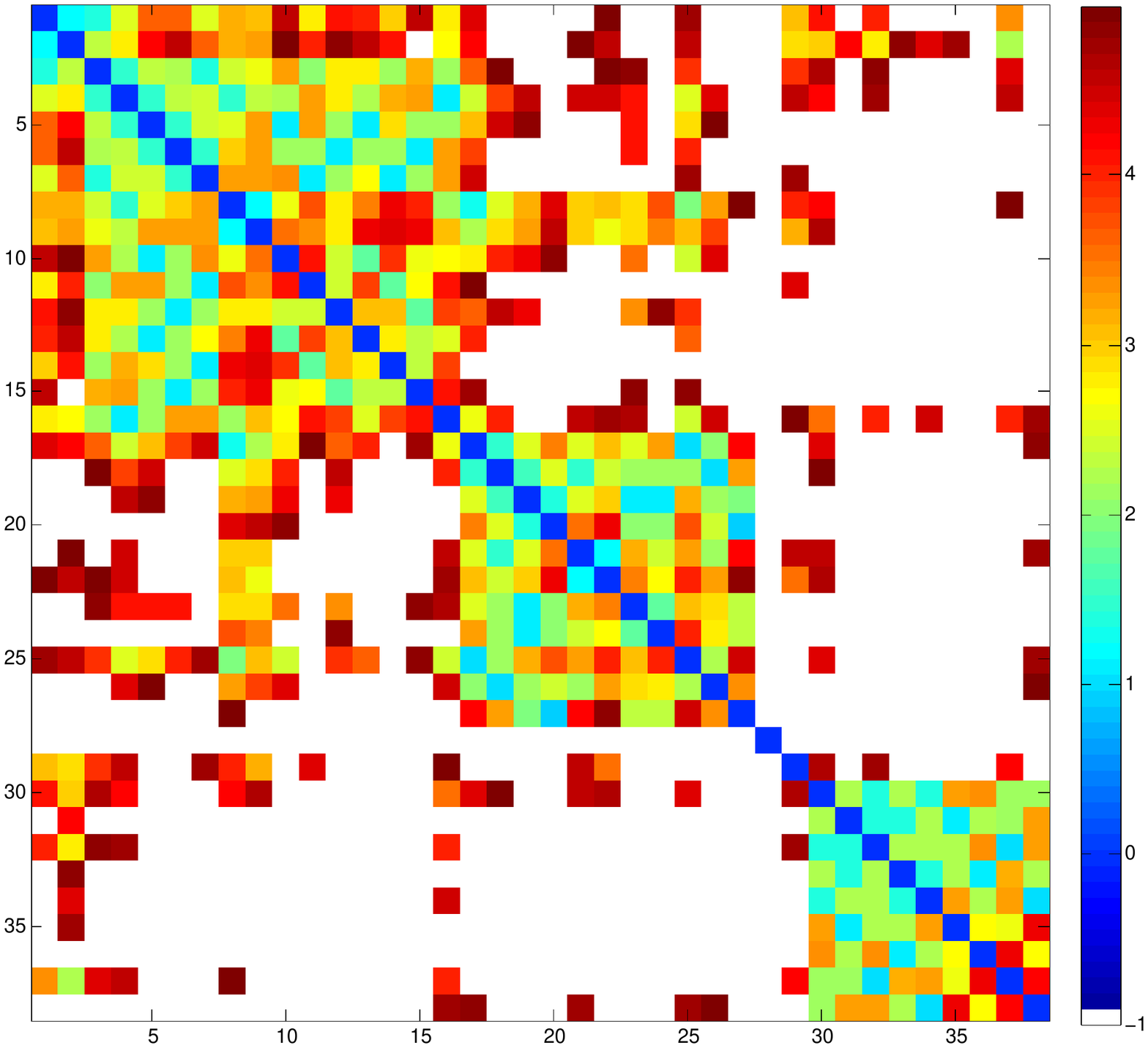} \\
\includegraphics[width=5cm]{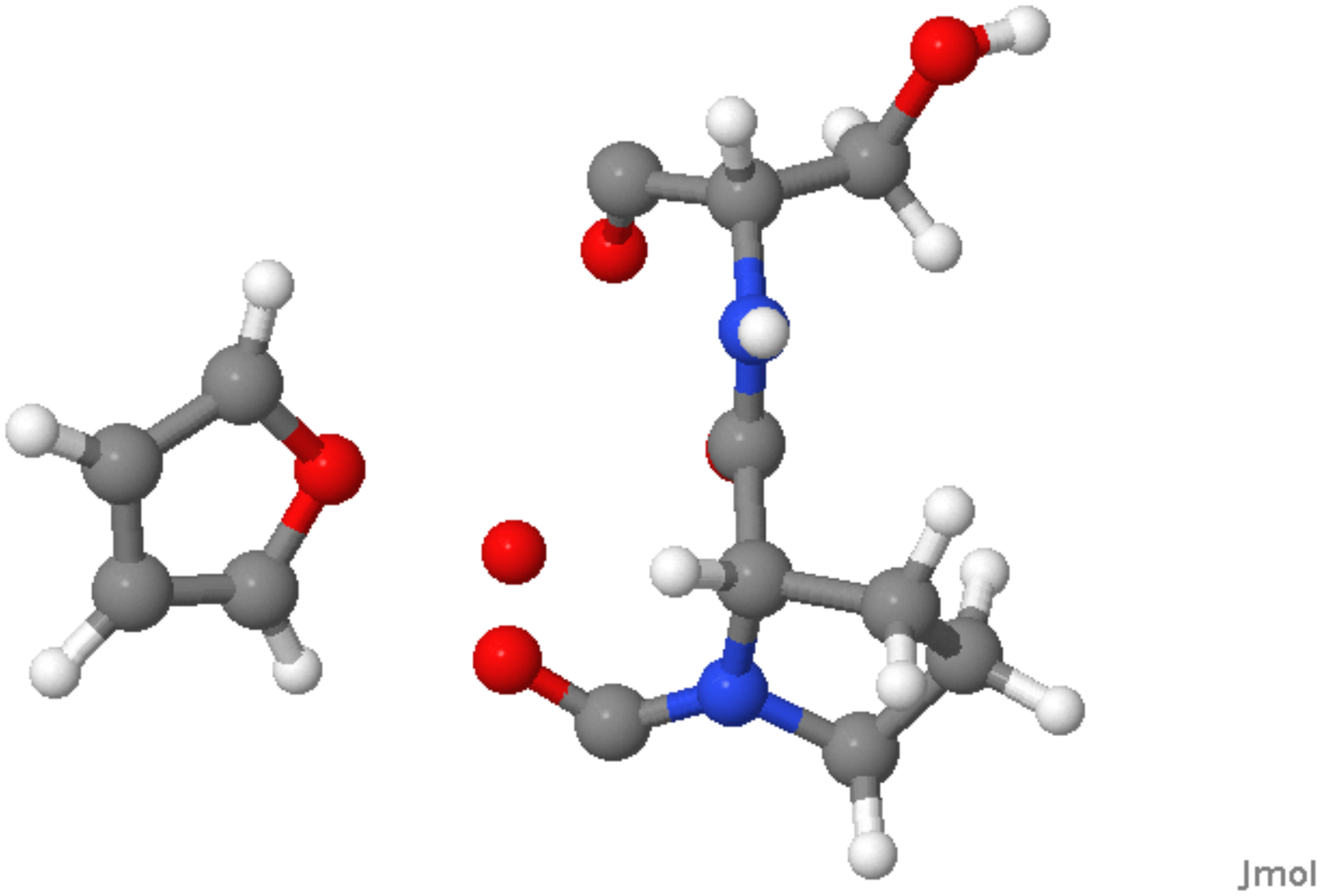} 
\end{minipage} &
\begin{minipage}{8cm}
\includegraphics[width=8cm]{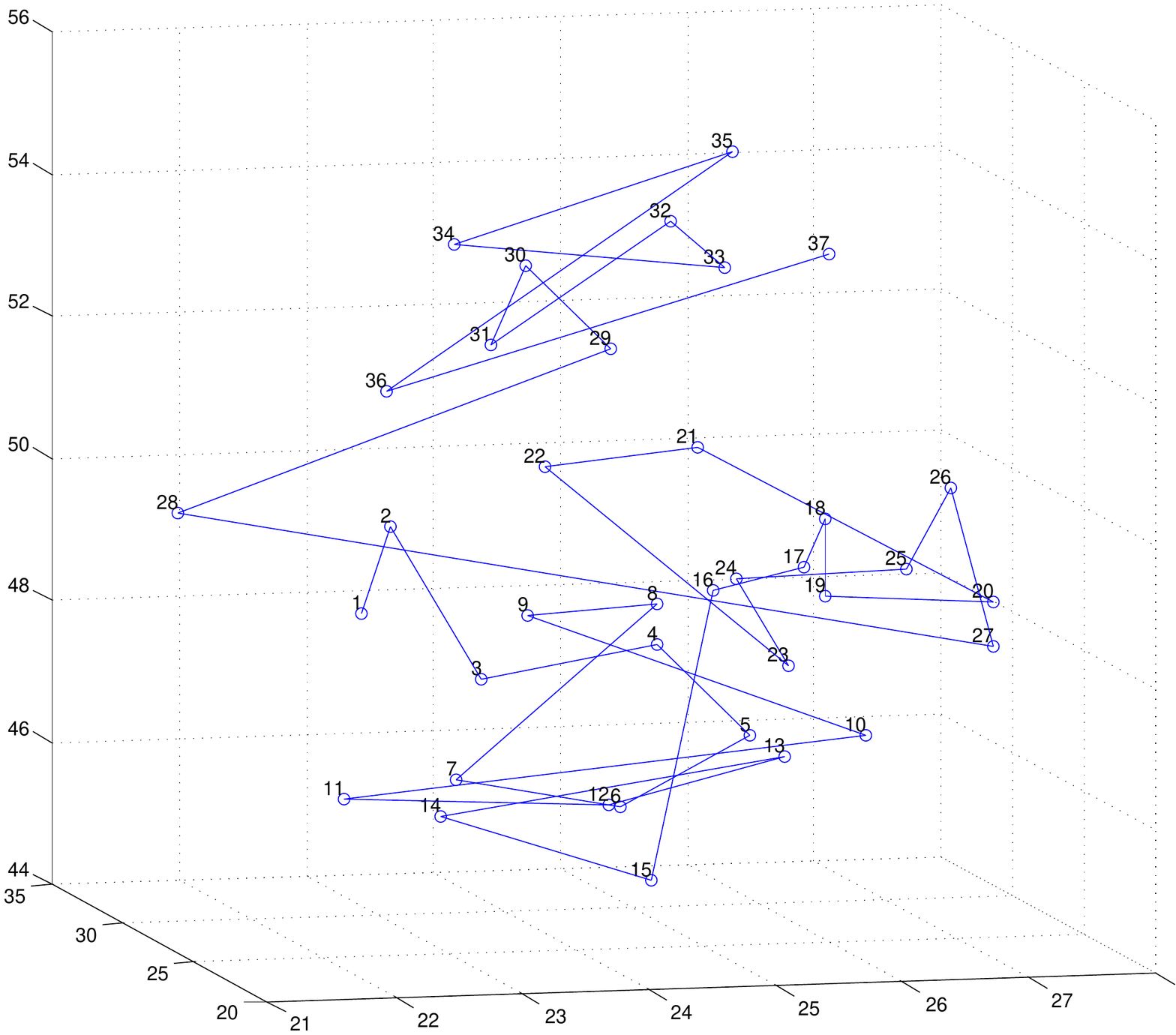}
\end{minipage}
\end{tabular}
%\vspace*{-2cm}
\end{center}
\caption{The {\tt tiny} instance: a heat map of the pEDM (upper left)
  and the correct realization in $\mathbb{R}^K$ shown by the {\sc
    Jmol} molecular visualization software (lower left) and in a
  Euclidean space plot, using the natural vertex order (right). The
  axes of the two 3D plots are not aligned to minimize overlap. The
  atom which appears disconnected in the {\sc Jmol} plot corresponds
  to vertex 28 in the Euclidean plot, and the nine atoms in a
  pentagonal arrangement correspond to vertices 29-37.}
\label{f:tiny}
\end{figure}

With each formulation, we present performances and results on a single
PDB instance called {\tt tiny}, which describes a graph
$G_{\mbox{\scriptsize\tt tiny}}=(V,E)$ with $|V|=37$, $|E|=335$ and
$K=3$. Fig.~\ref{f:tiny} shows a heat map of the partial Euclidean
Distance Matrix (pEDM) and the correct realization (found in the PDB
file) in $\mathbb{R}^K$ using two types of plots.

These validation experiments consist in solving the {\tt tiny}
instance using three different Global Optimization (GO) methods. The
first method is a deterministic GO solver based on spatial
Branch-and-Bound (sBB) \cite{couenne}, which we run for at most
900s. The second method is a stochastic matheuristic called Variable
Neighbourhood Search (VNS), described in \cite{vnssolver} with some
adaptations from \cite{recipe}. The third method is a straightforward
MultiStart (MS) algorithm, which is possibly the simplest stochastic
metaheuristic, and consists of deploying a certain number of local
descents from randomly sampled initial points. Both VNS and MS were
allowed to run for at most 20s of user CPU time (but terminated
whenever they found an optimum with average error less than
$10^{-6}$). The results report the average edge error $\Phi$ (see
Eq.~\eqref{eq:avgerr}), the maximum edge error $\Psi$ (see
Eq.~\eqref{eq:maxerr}), the DEMI measure $\partial$, and the CPU time
in seconds. All statistics referring to stochastic algorithms are
averaged over 10 runs.

These validation experiments were conducted on a single core of a
two-core Intel i7 CPU running at 2.0GHz with 8GB RAM under the Darwin
Kernel v.~13.3.0. Our sBB solver of choice is {\sc Couenne}
\cite{couenne} in its default setting. We used AMPL \cite{ampl} to
implement the VNS and MS algorithm, and {\sc Ipopt} \cite{ipopt,waechter} as a
local solver. The SDP formulations were modelled using YalMIP
\cite{yalmip} running under MATLAB \cite{matlab} and solved using
{\sc Mosek} \cite{mosek7}.

\revision{The point of these preliminary experiments is to visually show how the DEMI error measure $\partial$ impacts structural differences versus floating point errors. Each 3D plot contains two realizations (seen from the angle which best emphasizes their differences): the trusted solution found in the PDB, and the output of the corresponding algorithm. Floating point errors can be remarked when two realizations are almost aligned but not quite superimposed. Structural errors are evident when no alignment is visible.}

\subsection{Exact formulations}
\label{s:exact}
These formulations will yield a valid realization at every global optimum.

\subsubsection{Penalty minimization}
This formulation minimizes the sum of non-negative penalties $s_{uv}$
deriving from the fact that $\|x_u-x_v\|_2$ is smaller than $L_{uv}$
or larger than $U_{uv}$:
\begin{equation}
  \left. \begin{array}{rrcl}
     \min\limits_{s\in\mathbf{M}^+,x} & \sum\limits_{\{u,v\}\in
       E} s_{uv} & & \\  
     \forall\{u,v\}\in E & L^2_{uv} - \|x_u-x_v\|_2^2 &\le & s_{uv} \\
     \forall\{u,v\}\in E & \|x_u-x_v\|_2^2 - U^2_{uv} &\le & s_{uv} \\
     \forall k\le K & \sum\limits_{v\in V} x_{vk} &=& 0.
  \end{array} \right\} \label{eq:dgp1}
\end{equation}

Variants: (i) replace $\sum$ with $\max$; (ii) use different variables
$s^L,s^U$ to represent penalties w.r.t.~$L,U$; (iii) replace the
objective by any positive linear form in the penalty variables.

This formulation and its variants have the property that an optimum is
global if and only if the objective function value is identically
zero. An unconstrained and weighted version of this formulation
appeared in \cite{morewu2}. The performance of the penalty
minimization formulation and its variants on the {\tt tiny} instance
is shown in Table \ref{t:penmin}.

\begin{table}
\begin{center}
{\small
\hspace*{-0.5cm}\begin{tabular}{|l@{\hspace*{-0.2cm}}||cc@{\hspace*{-0.4cm}}c|cc@{\hspace*{-0.4cm}}c|cc@{\hspace*{-0.4cm}}c|cc@{\hspace*{-0.4cm}}c|} \hline 
& \multicolumn{3}{c|}{Original} & \multicolumn{3}{c|}{Var.~(i)} &
  \multicolumn{3}{c|}{Var.~(ii)} & \multicolumn{3}{c|}{Var.~(iii)} \\ 
{\it Solver} & $\Phi$ & $\Psi$ & {\it CPU} & $\Phi$ & $\Psi$ & {\it
  CPU} & $\Phi$ & $\Psi$ & {\it CPU} & $\Phi$ & $\Psi$ & {\it CPU}
\\ \hline \hline
{\sc Couenne} & 0 & 0 & 38.89 & 0 & 0 & 146 & 0 & 0 & 22.05 & $\infty$ &
$\infty$ & 900 \\ \hline
\begin{minipage}{1.5cm}
$x_{\mbox{\tiny\tt tiny}}$ \\and\\ $x_{\mbox{\tiny DEMI}}$
\end{minipage} &
\multicolumn{3}{c|}{% 
\begin{minipage}{2.4cm}
  \hspace*{-0.2cm}\includegraphics[width=3cm]{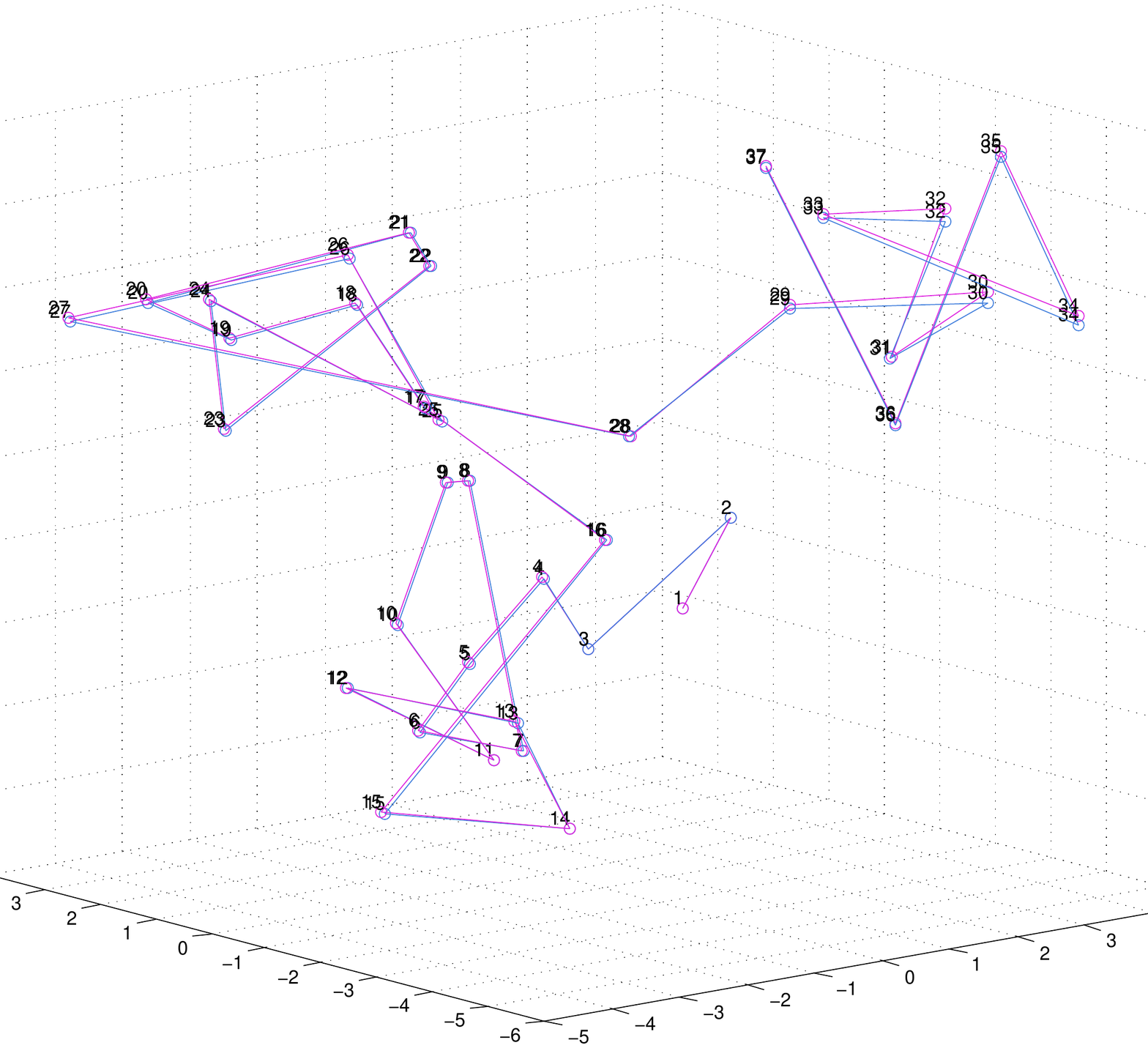} 
\end{minipage}} & 
\multicolumn{3}{c|}{%
\begin{minipage}{2.4cm}
  \hspace*{-0.2cm}\includegraphics[width=3cm]{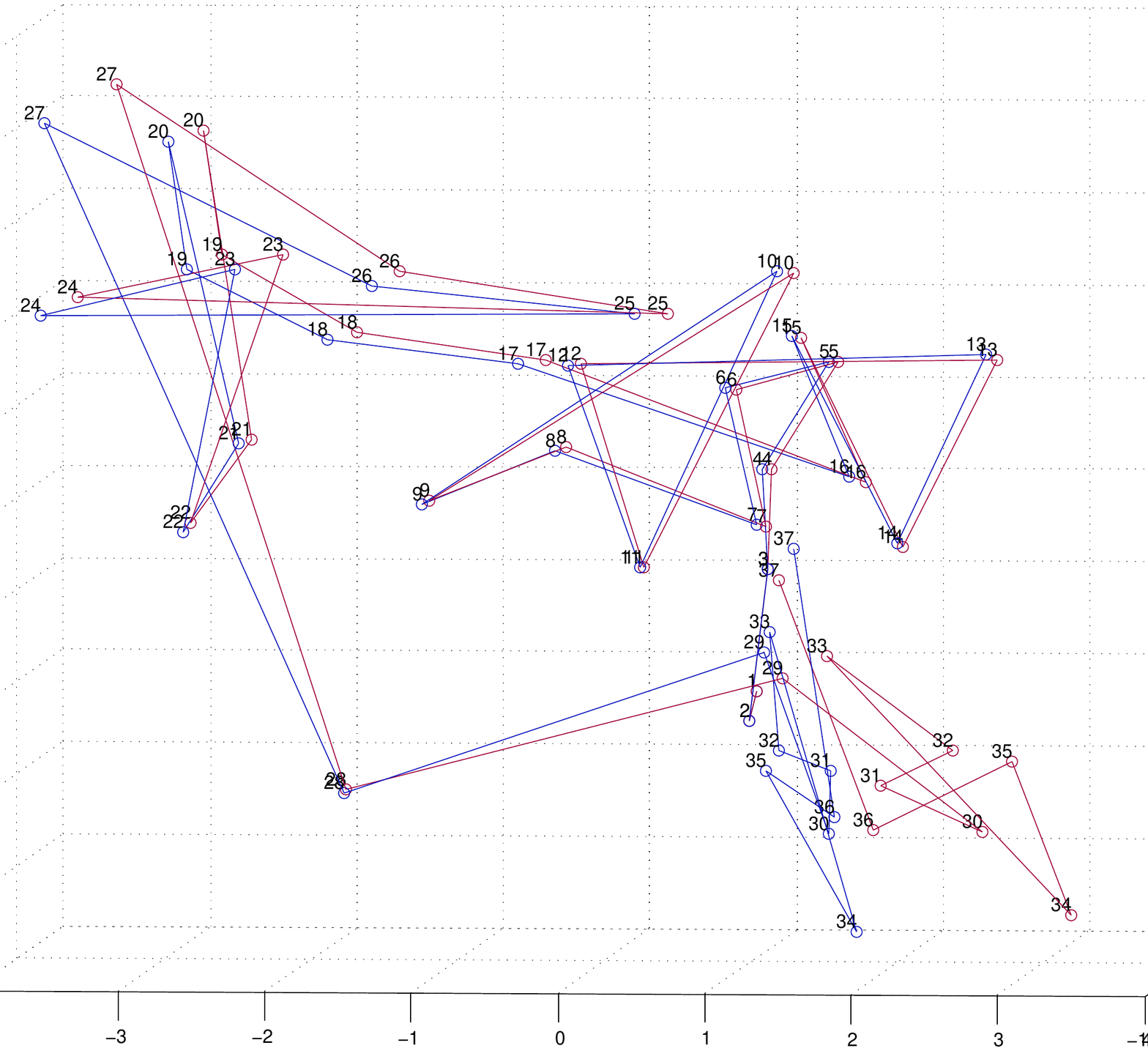} 
\end{minipage}} & 
\multicolumn{3}{c|}{%
\begin{minipage}{2.4cm}
  \hspace*{-0.2cm}\includegraphics[width=3cm]{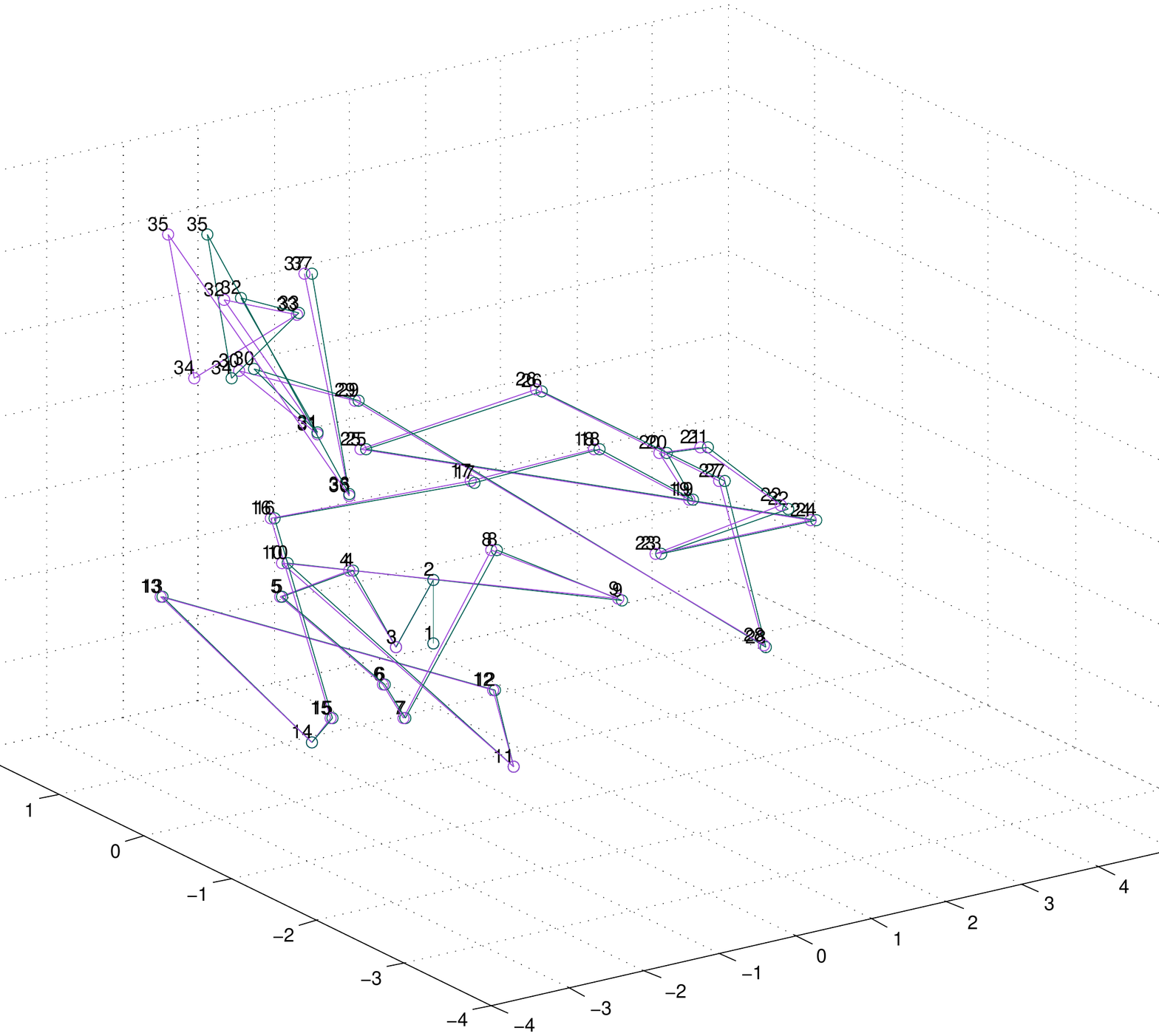} 
\end{minipage}} & 
\multicolumn{3}{c|}{% 
\begin{minipage}{2.4cm}
  (No solution)
\end{minipage}} \\ \hline
$\partial$ & \multicolumn{3}{c|}{0.3540} & \multicolumn{3}{c|}{2.9834}
& \multicolumn{3}{c|}{0.6072} & \multicolumn{3}{c|}{$\infty$}
 \\ \hline \hline
{\sc VNS} & 0 & 0 & 2.27 & 0.01 & 0.19 & 2.76 & 0.01 & 0.15 & 1.09 & 0
& 0 & 3.48\\ \hline
\begin{minipage}{1.5cm}
$x_{\mbox{\tiny\tt tiny}}$ \\ and \\ $x_{\mbox{\tiny DEMI}}$
\end{minipage}  &
\multicolumn{3}{c|}{%
\begin{minipage}{2.4cm}
  \hspace*{-0.2cm}\includegraphics[width=3cm]{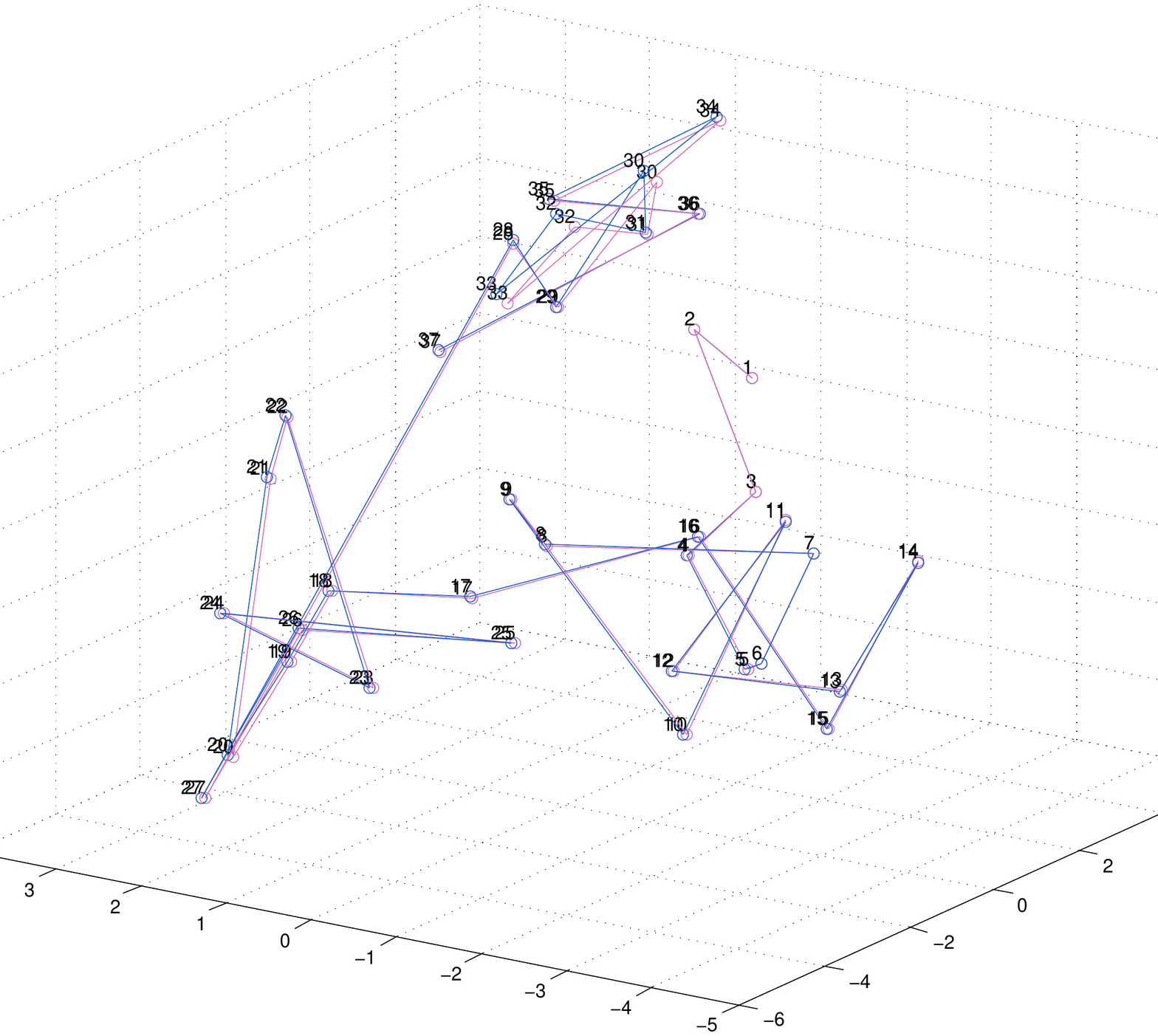} 
\end{minipage}} & 
\multicolumn{3}{c|}{%
\begin{minipage}{2.4cm}
  \hspace*{-0.2cm}\includegraphics[width=3cm]{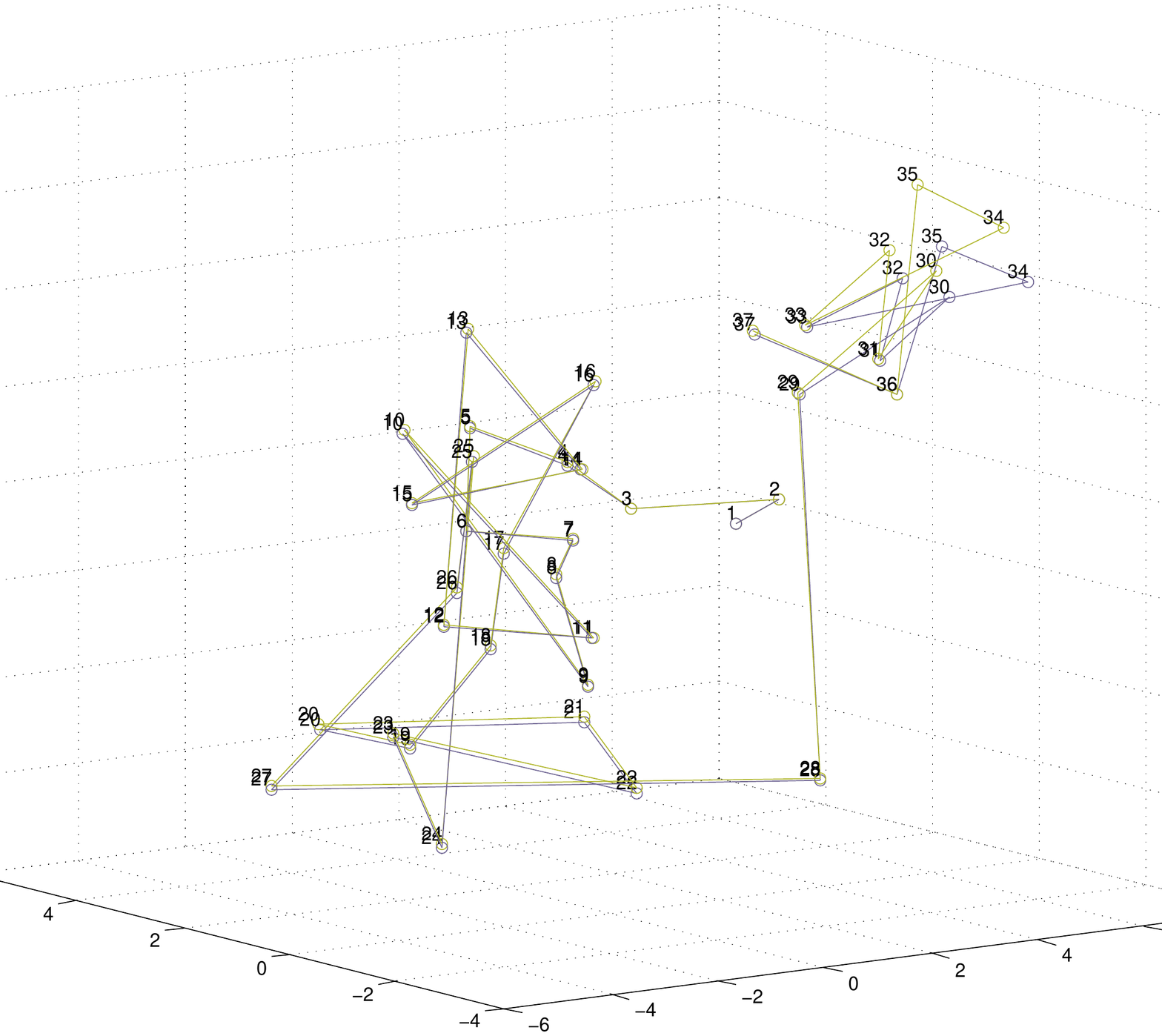} 
\end{minipage}} & 
\multicolumn{3}{c|}{%
\begin{minipage}{2.4cm}
  \hspace*{-0.2cm}\includegraphics[width=3cm]{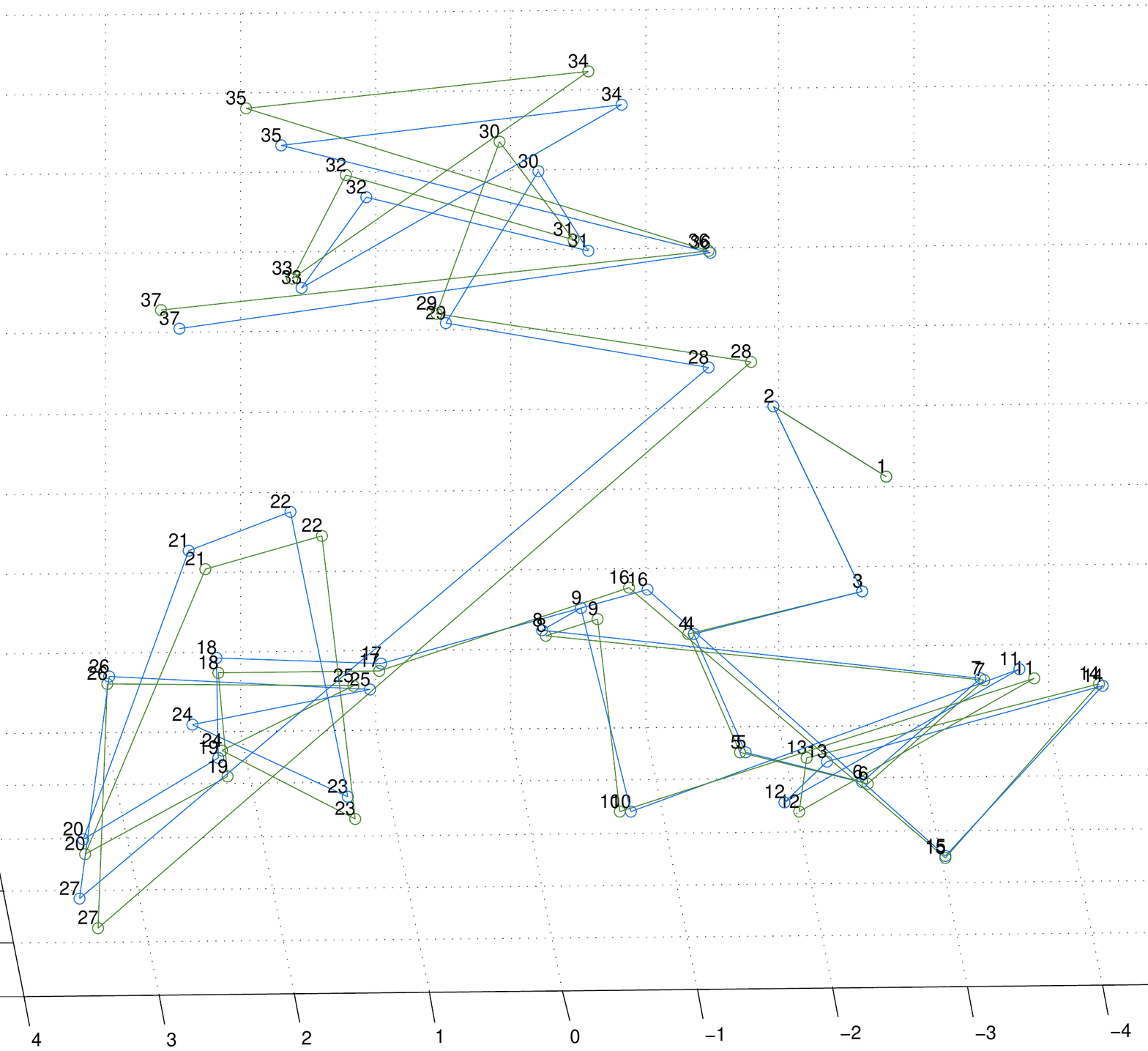} 
\end{minipage}} & 
\multicolumn{3}{c|}{%
\begin{minipage}{2.6cm}
  \hspace*{-0.2cm}\includegraphics[width=3cm]{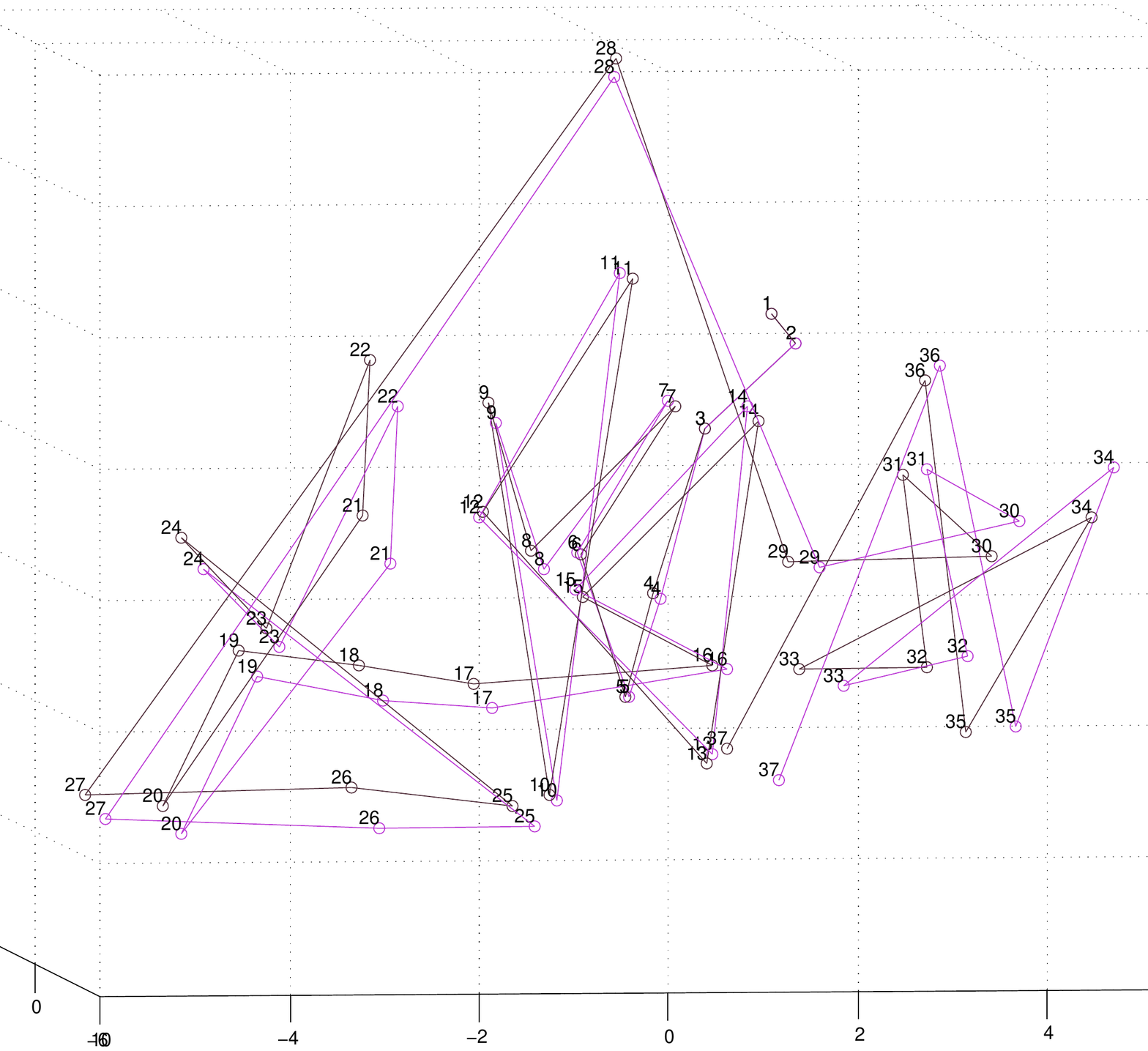} 
\end{minipage}} \\  \hline
$\partial$ & \multicolumn{3}{c|}{0.4024} & \multicolumn{3}{c|}{1.1480}
& \multicolumn{3}{c|}{1.5853} & \multicolumn{3}{c|}{1.5464}
\\ \hline \hline
{\sc MS} & 0 & 0 & 1.90 & 0 & 0 & 2.28 & 0 & 0 & 1.82 &0 &0 & 1.54\\ \hline
\begin{minipage}{1.5cm}
$x_{\mbox{\tiny\tt tiny}}$ \\ and\\  $x_{\mbox{\tiny DEMI}}$
\end{minipage} &
\multicolumn{3}{c|}{%
\begin{minipage}{2.4cm}
  \hspace*{-0.2cm}\includegraphics[width=3cm]{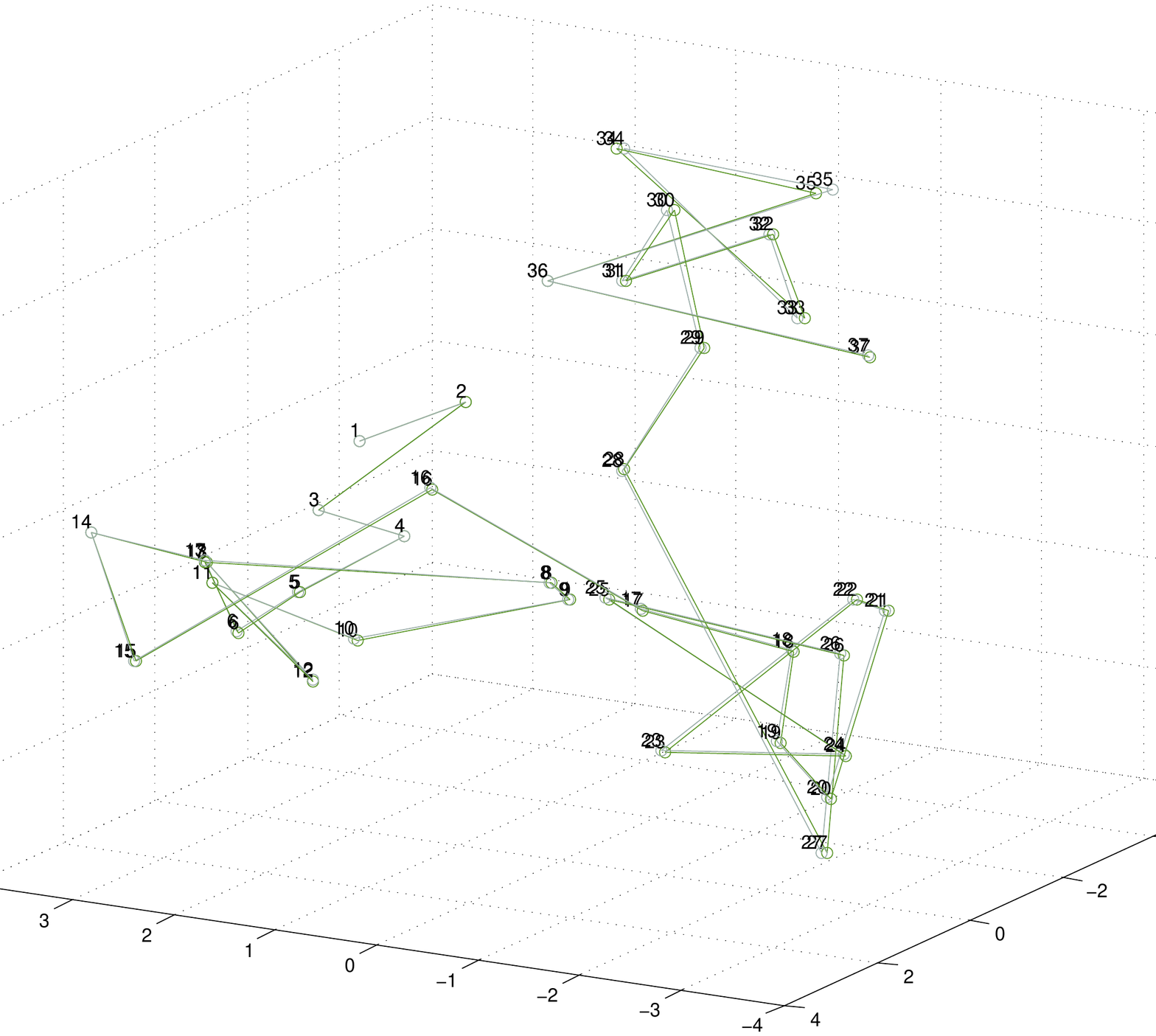} 
\end{minipage}} & 
\multicolumn{3}{c|}{%
\begin{minipage}{2.4cm}
  \hspace*{-0.2cm}\includegraphics[width=3cm]{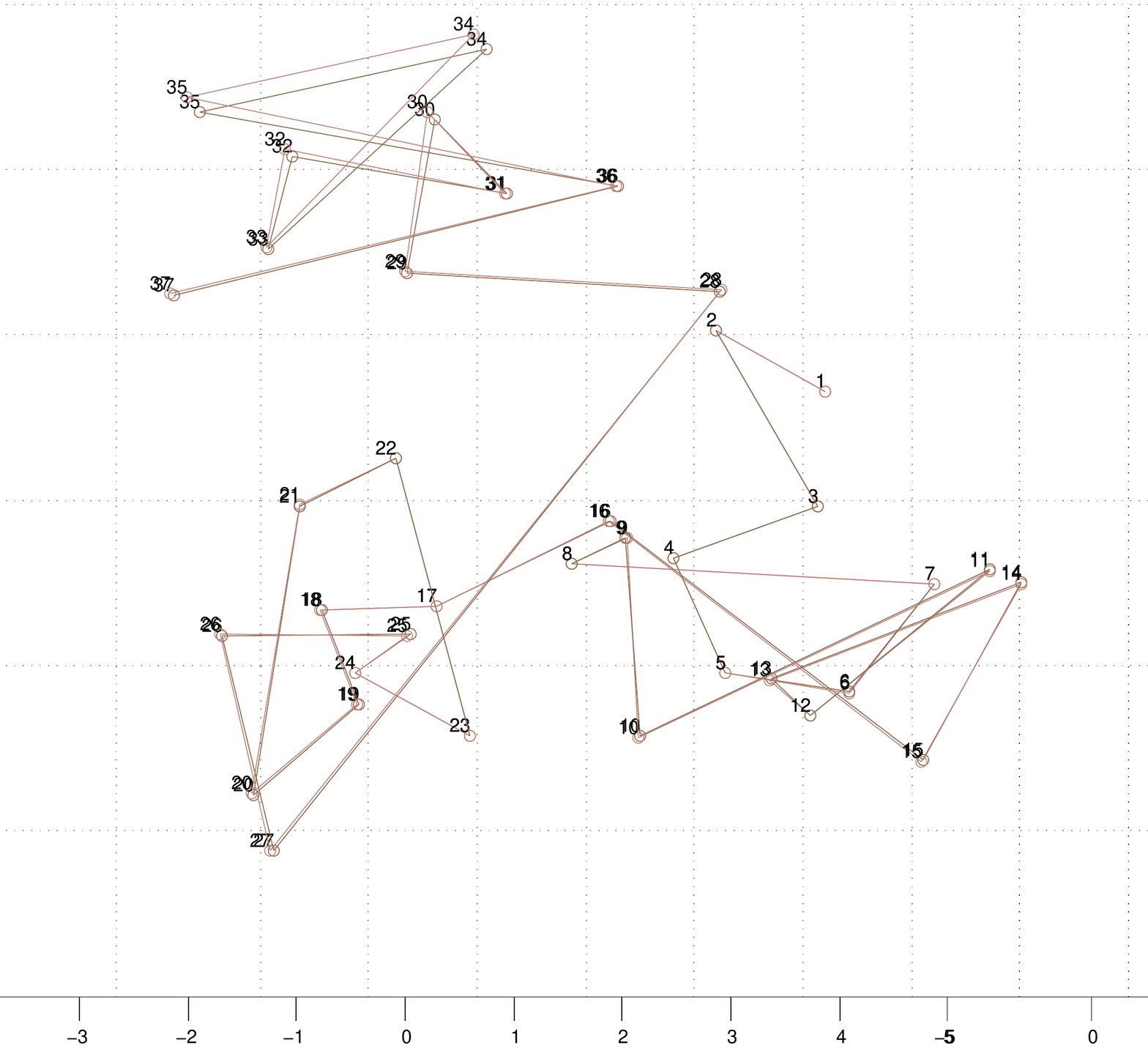} 
\end{minipage}} & 
\multicolumn{3}{c|}{%
\begin{minipage}{2.4cm}
  \hspace*{-0.2cm}\includegraphics[width=3cm]{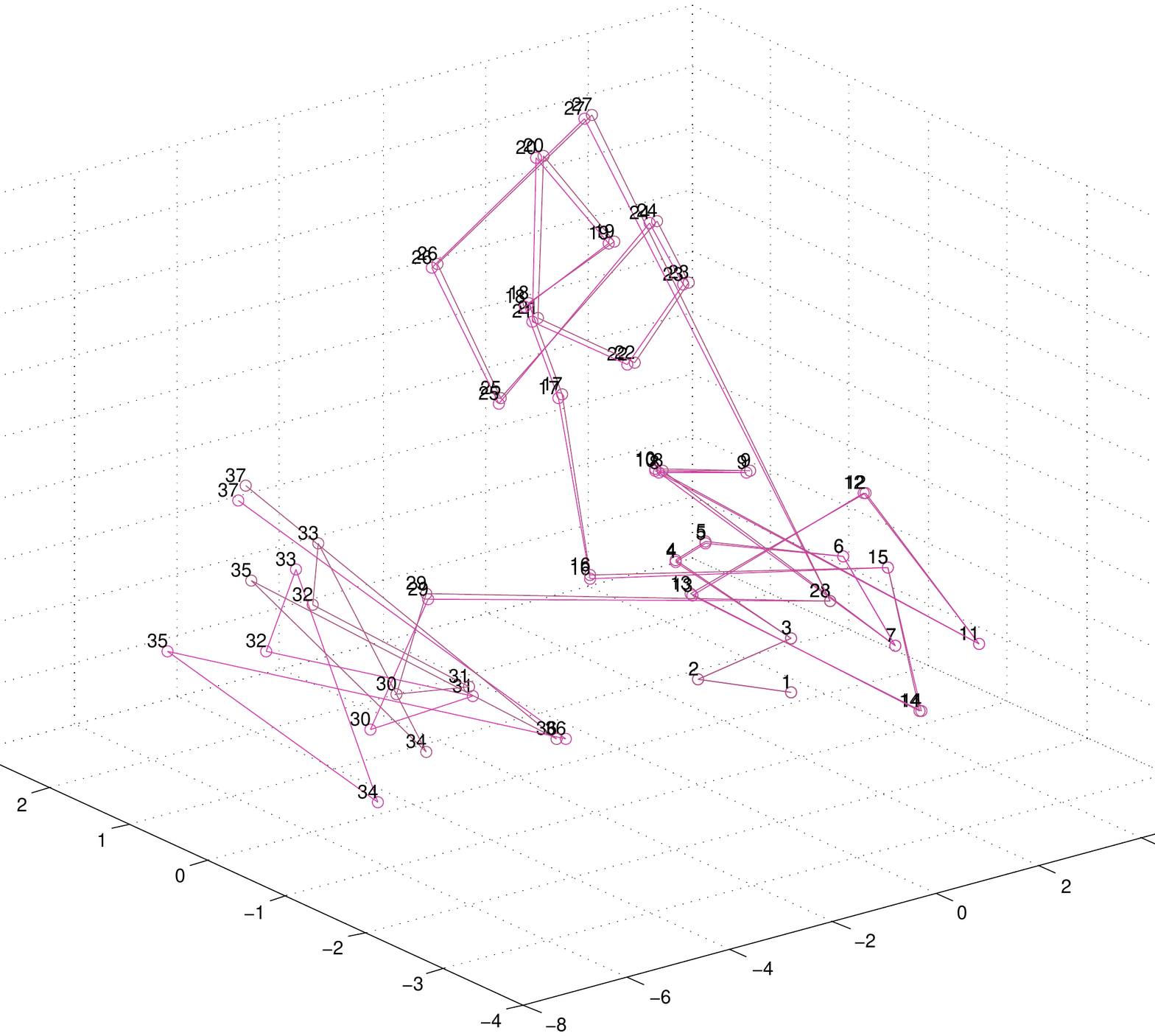} 
\end{minipage}} & 
\multicolumn{3}{c|}{%
\begin{minipage}{2.6cm}
  \hspace*{-0.2cm}\includegraphics[width=3cm]{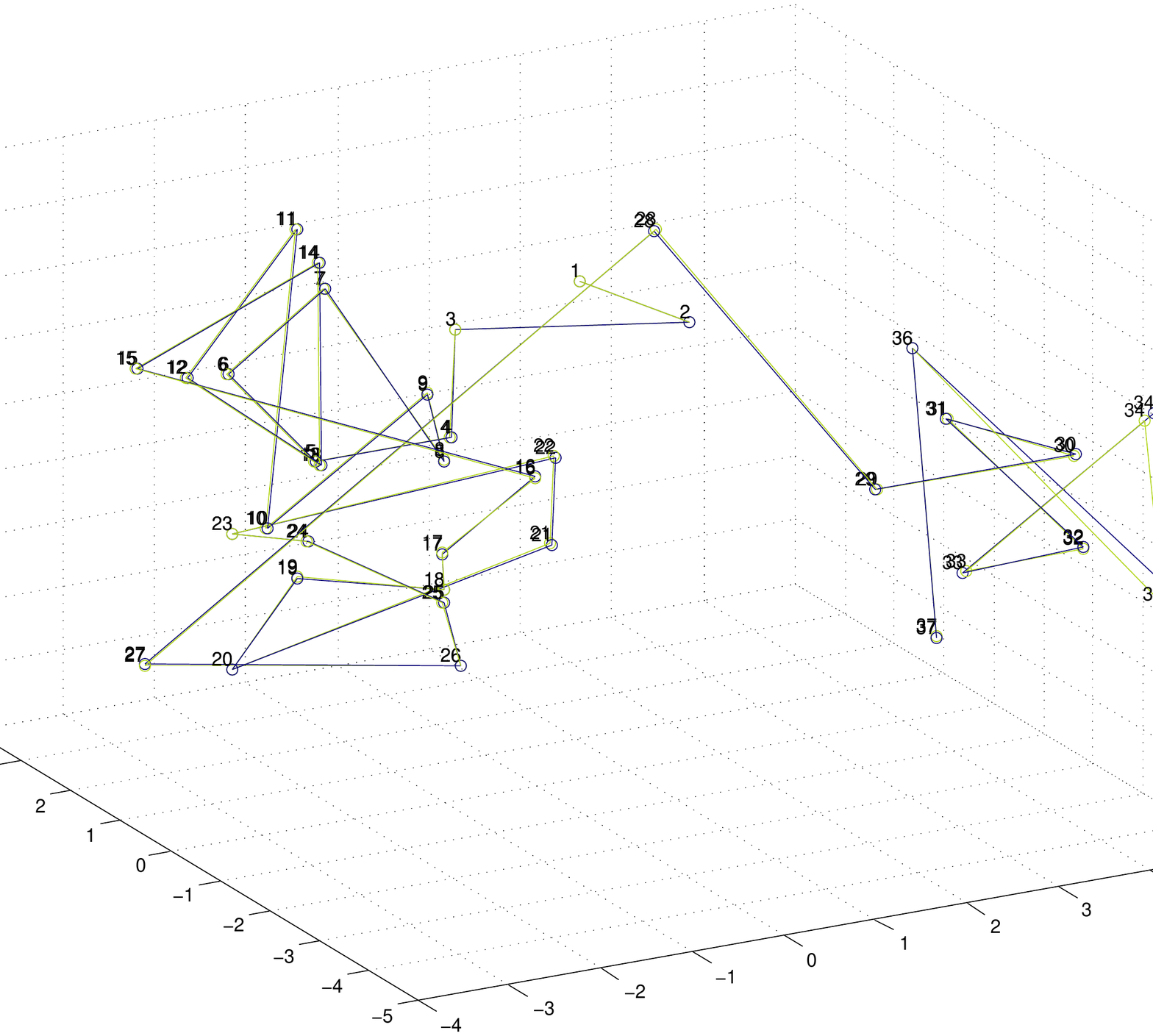} 
\end{minipage}} \\ \hline
$\partial$ & \multicolumn{3}{c|}{0.3108} & \multicolumn{3}{c|}{0.5983}
& \multicolumn{3}{c|}{3.6933} & \multicolumn{3}{c|}{0.4156}
\\ \hline \hline
\end{tabular} 
}
\end{center}
\caption{Performance of {\it penalty minimization} on {\tt tiny}. For
  each solver and formulation (variant), we report the edge errors
  $\Phi,\Psi$, the CPU time, a 3D plot of the solution
  $x_{\mbox{\tt\scriptsize tiny}}$ given in the PDB file versus the
  solution $x_{\mbox{\scriptsize DEMI}}$ found by solving the DEMI
  instance with $x=x_{\mbox{\tt\scriptsize tiny}}$ and $y$ given by
  the solution of the solver, and the corresponding DEMI measure
  $\partial(x,y)=\min\limits_{g,\rho}\|x-g\rho(y)\|$.}
\label{t:penmin}
\end{table}

\subsubsection{Square factoring}

This formulation has been adapted to the interval case from
\cite{mago14}. It exploits the identity
$\|x_u-x_v\|_2^2=(x_u-x_v)(x_u-x_v)$: 
\label{it:qdr}
\begin{equation}
  \left. \begin{array}{rrcl}
     \min\limits_{x,\sigma\in\mathbf{M}^K,\tau\in\mathbf{M}^K} &
       \sum\limits_{\{u,v\}\in E}\sum\limits_{k\le K} 
       (\sigma_{uvk}-\tau_{uvk})^2 && \\   
     \forall\{u,v\}\in E, k\le K & x_{uk}-x_{vk} &=& \sigma_{uvk} \\
     \forall\{u,v\}\in E & \sum\limits_{k\le K}\sigma_{uvk}\tau_{uvk}
       &\ge & L_{uv}^2\\
     \forall\{u,v\}\in E & \sum\limits_{k\le K}\sigma_{uvk}\tau_{uvk} 
       &\le & U_{uv}^2 \\
     \forall k\le K & \sum\limits_{v\in V} x_{vk} &=& 0.
  \end{array} \right\} \label{eq:dgp3}
\end{equation}

We propose no variants for this formulation. The performance of the
square factoring formulation on the {\tt tiny} instance is shown in
Table \ref{t:squaref}.

\begin{table}
\begin{center}
{\small
\begin{tabular}{|l||ccc|} \hline 
{\it Solver} & $\Phi$ & $\Psi$ & {\it CPU} \\ \hline \hline
{\sc Couenne} & 0 & 0 & 3.11  \\ \hline
\begin{minipage}{1cm}
$x_{\mbox{\tiny\tt tiny}}$ and $x_{\mbox{\tiny DEMI}}$
\end{minipage} &
\multicolumn{3}{c|}{% 
\begin{minipage}{4cm}
  \includegraphics[width=3.5cm]{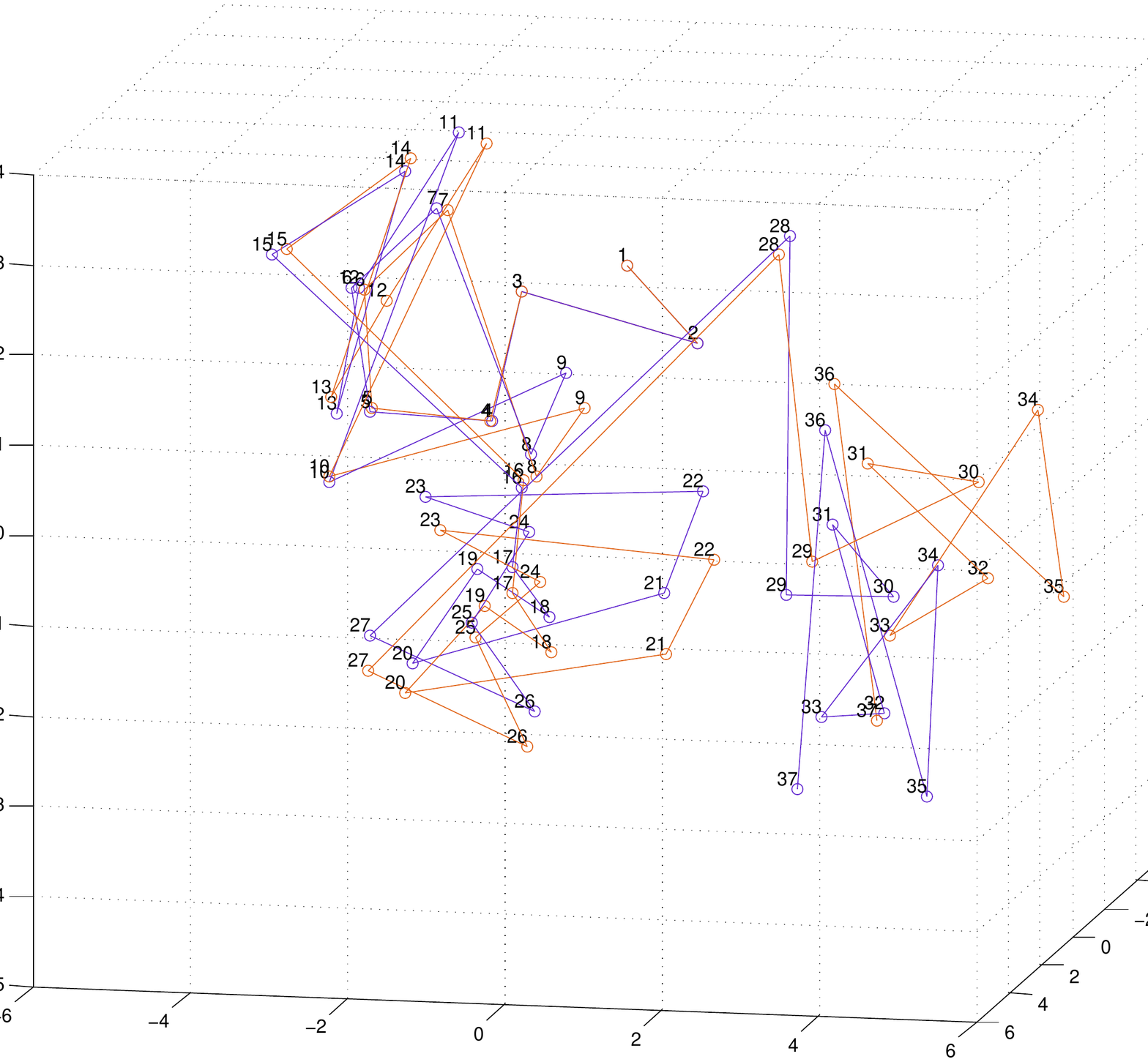} 
\end{minipage}} \\ \hline
$\partial$ &\multicolumn{3}{c|}{6.3182} \\ \hline \hline
{\sc VNS} & 0.01 & 0.02 & 4.05 \\ \hline
\begin{minipage}{1cm}
$x_{\mbox{\tiny\tt tiny}}$ and $x_{\mbox{\tiny DEMI}}$
\end{minipage} &
\multicolumn{3}{c|}{% 
\begin{minipage}{4cm}
  \includegraphics[width=3.5cm]{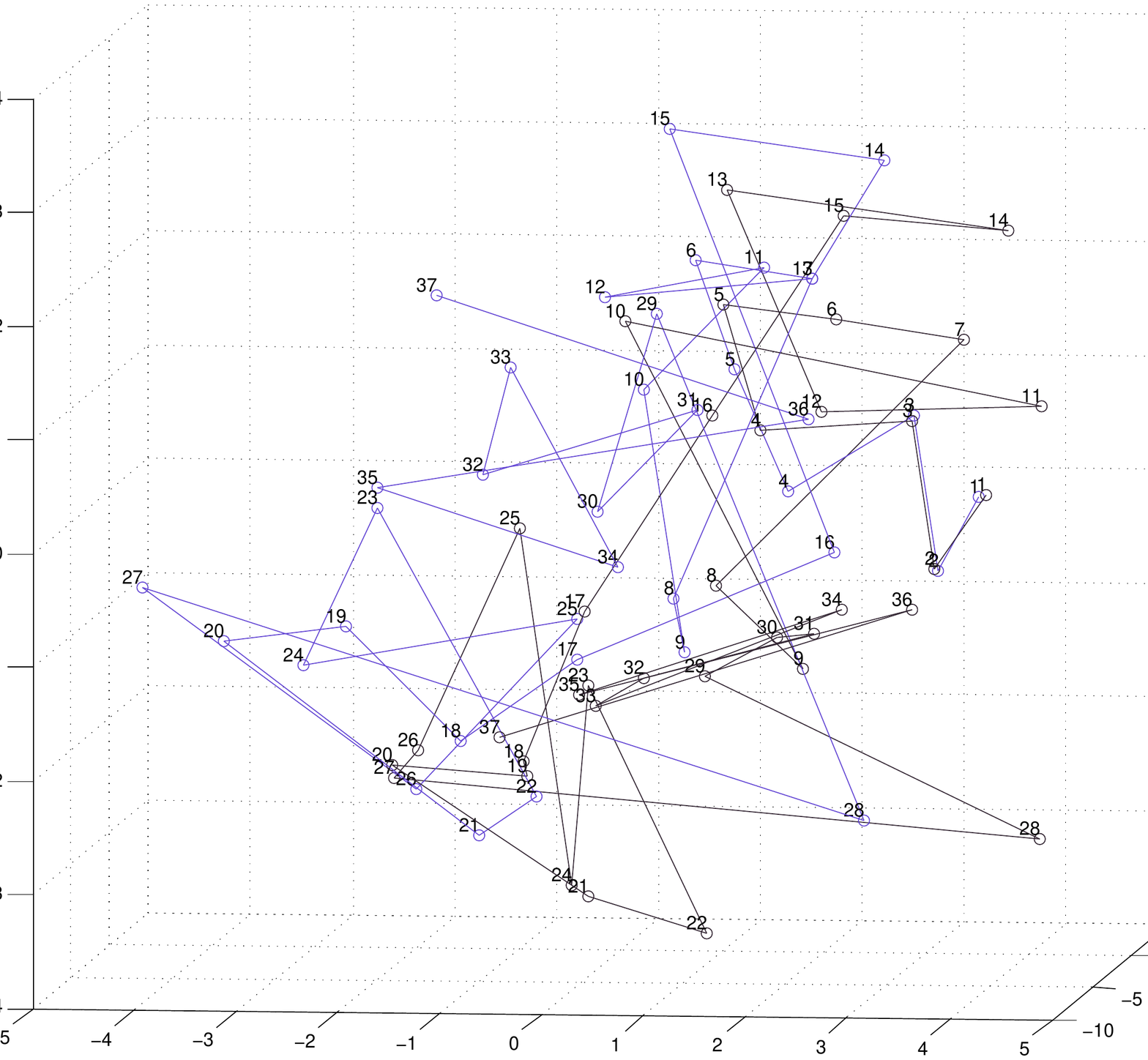} 
\end{minipage}} \\ \hline
$\partial$ &\multicolumn{3}{c|}{13.6406} \\ \hline \hline
{\sc MS} & 0 & 0 & 2.31 \\ \hline 
\begin{minipage}{1cm}
$x_{\mbox{\tiny\tt tiny}}$ and $x_{\mbox{\tiny DEMI}}$
\end{minipage} &
\multicolumn{3}{c|}{% 
\begin{minipage}{4cm}
  \includegraphics[width=3.5cm]{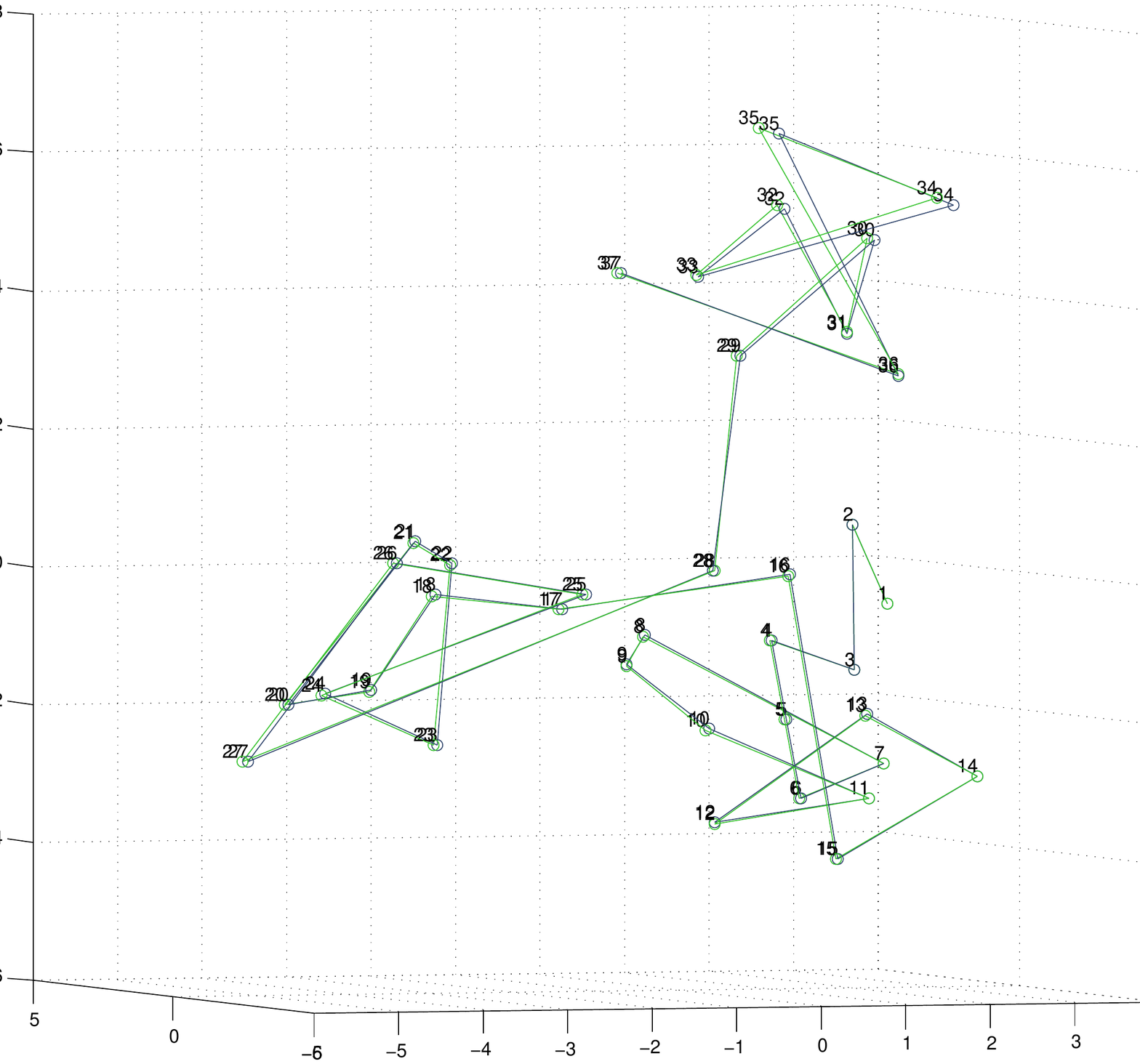} 
\end{minipage}} \\ \hline
$\partial$ &\multicolumn{3}{c|}{0.6072} \\ \hline \hline
\end{tabular}
}
\end{center}
\caption{Performance of {\it square factoring} on {\tt tiny}. For each
  solver, we report the edge errors $\Phi,\Psi$, the CPU time, a 3D
  plot of the solution $x_{\mbox{\tt\scriptsize tiny}}$ given in the
  PDB file versus the solution $x_{\mbox{\scriptsize DEMI}}$ found by
  solving the DEMI instance with $x=x_{\mbox{\tt\scriptsize tiny}}$
  and $y$ given by the solution of the solver, and the corresponding
  DEMI measure $\partial(x,y)=\min\limits_{g,\rho}\|x-g\rho(y)\|$.}
\label{t:squaref}
\end{table}

\subsection{Relaxations}
These are formulations which relax some feasibility constraints. The
obtained solution may or may not be a valid (feasible) solution to the
given instance. One should always therefore verify that the solution
satisfies \eqref{eq:Idgp}. On the other hand, if a relaxation is
infeasible, then so must be the original \iDGP\ instance.

\subsubsection{Convexity and concavity}
\label{s:convconc}
This formulation, adapted to the interval case from \cite{mago14},
exploits the convexity and concavity of the equations in
Eq.~\eqref{eq:Idgp2} separately:
\begin{equation}
  \left. \begin{array}{rrcl}
     \max\limits_{x} & \sum\limits_{\{u,v\}\in E} \|x_u-x_v\|_2^2 & & \\ 
     \forall\{u,v\}\in E & \|x_u-x_v\|_2^2 &\le & U_{uv}^2 \\
     \forall k\le K & \sum\limits_{v\in V} x_{vk} &=& 0.
  \end{array} \right\} \label{eq:dgp4}
\end{equation}
Variants: replace the objective with a positively weighted version
thereof. 

Eq.~\eqref{eq:dgp4} is an exact reformulation (in the sense of
\cite{refmathprog}) of
\begin{equation}
\min_x \sum\limits_{\{u,v\}\in E} (\|x_u-x_v\|^2-U_{uv}^2)^2, \label{eq:dgpmp}
\end{equation}
which is possibly the best known Mathematical Programming (MP)
formulation of the (non-interval) DGP so far. That
Eq.~\eqref{eq:dgpmp} and Eq.~\eqref{eq:dgp4} have the same solutions
can be intuitively visualized the edges $\{u,v\}$ of the underlying
graph $G$ as a set of interconnected cables, each of length $U_{uv}$:
the objective of Eq.~\eqref{eq:dgp4} ``pulls'' the adjacent vertices
$u,v$ apart as far as possible. As a result, all cables can be 
straightened if and only if the DGP has a valid solution. A formal
proof of this fact is given elsewhere \cite{mwa_minlp_working}.

If the given instance is an \iDGP\ one, however, Eq.~\eqref{eq:dgp4}
is a relaxation of the lower bounding constraints: by attempting to
maximize the distance between adjacent points, one hopes that
$\|x_u-x_v\|_2\ge L_{uv}$ will hold, but this need not necessarily be
the case. The performance of the convexity and concavity formulation
and its variants on the {\tt tiny} instance is shown in Table
\ref{t:concavity}.

\begin{table}
\begin{center}
{\small
\begin{tabular}{|l||ccc|ccc|} \hline 
& \multicolumn{3}{c|}{Original} & \multicolumn{3}{c|}{Var.~(i)} \\
{\it Solver} & $\Phi$ & $\Psi$ & {\it CPU} & $\Phi$ & $\Psi$ & {\it CPU} \\ \hline \hline
{\sc Couenne} & 0 & 0.03 & 1.78 & 0 & 0.03 & 1.53 \\ \hline
\begin{minipage}{1cm}
$x_{\mbox{\tiny\tt tiny}}$ and $x_{\mbox{\tiny DEMI}}$
\end{minipage} &
\multicolumn{3}{c|}{% 
\begin{minipage}{4cm}
  \includegraphics[width=3.5cm]{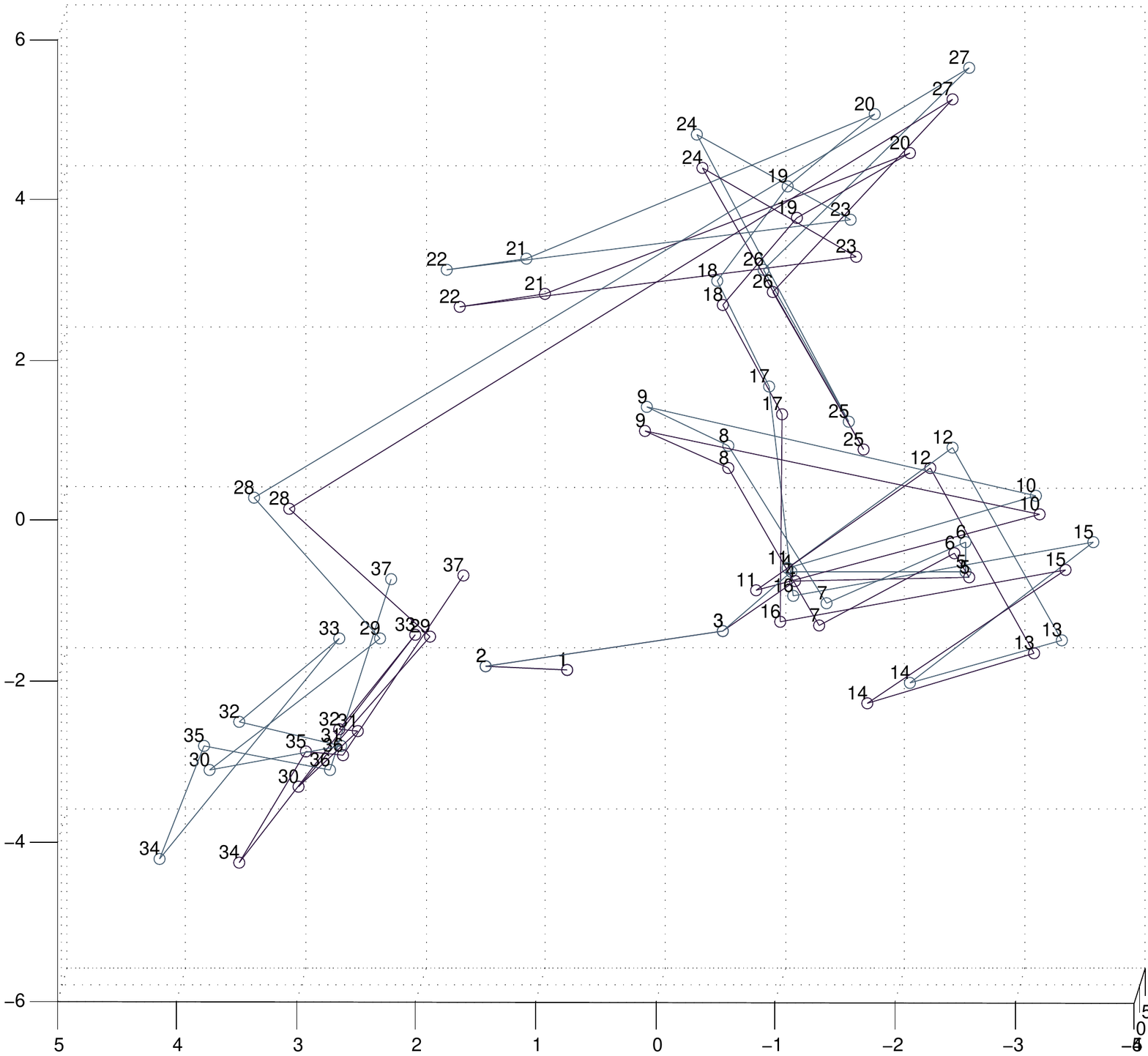} 
\end{minipage}} &
\multicolumn{3}{c|}{% 
\begin{minipage}{4cm}
  \includegraphics[width=3.5cm]{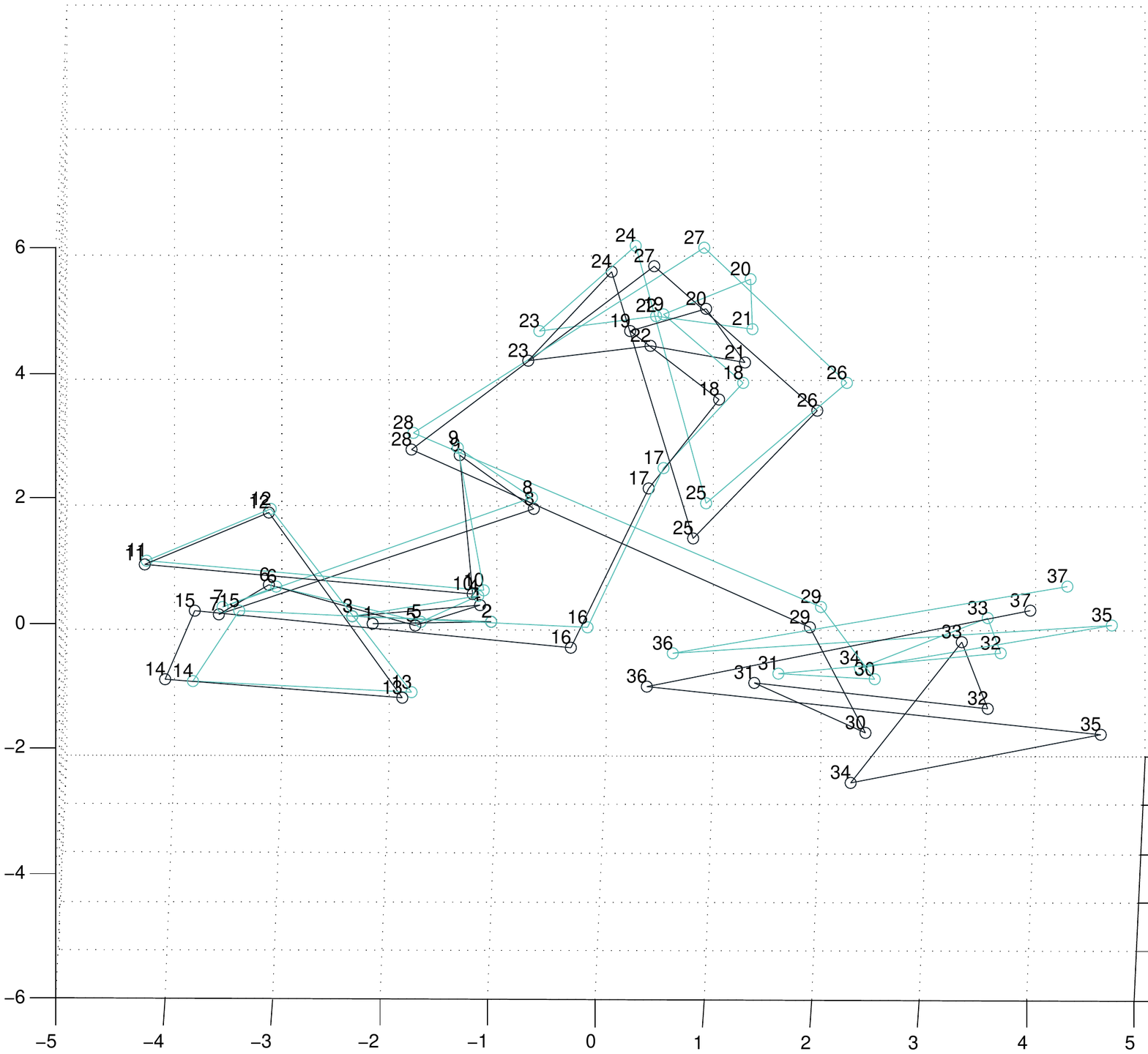} 
\end{minipage}} \\ \hline
$\partial$ &\multicolumn{3}{c|}{2.0211} &\multicolumn{3}{c|}{3.6479} \\ \hline \hline
{\sc VNS} & 0 & 0.33 & 8.80 & 0 & 0.36 & 8.94 \\ \hline
\begin{minipage}{1cm}
$x_{\mbox{\tiny\tt tiny}}$ and $x_{\mbox{\tiny DEMI}}$
\end{minipage} &
\multicolumn{3}{c|}{% 
\begin{minipage}{4cm}
  \includegraphics[width=3.5cm]{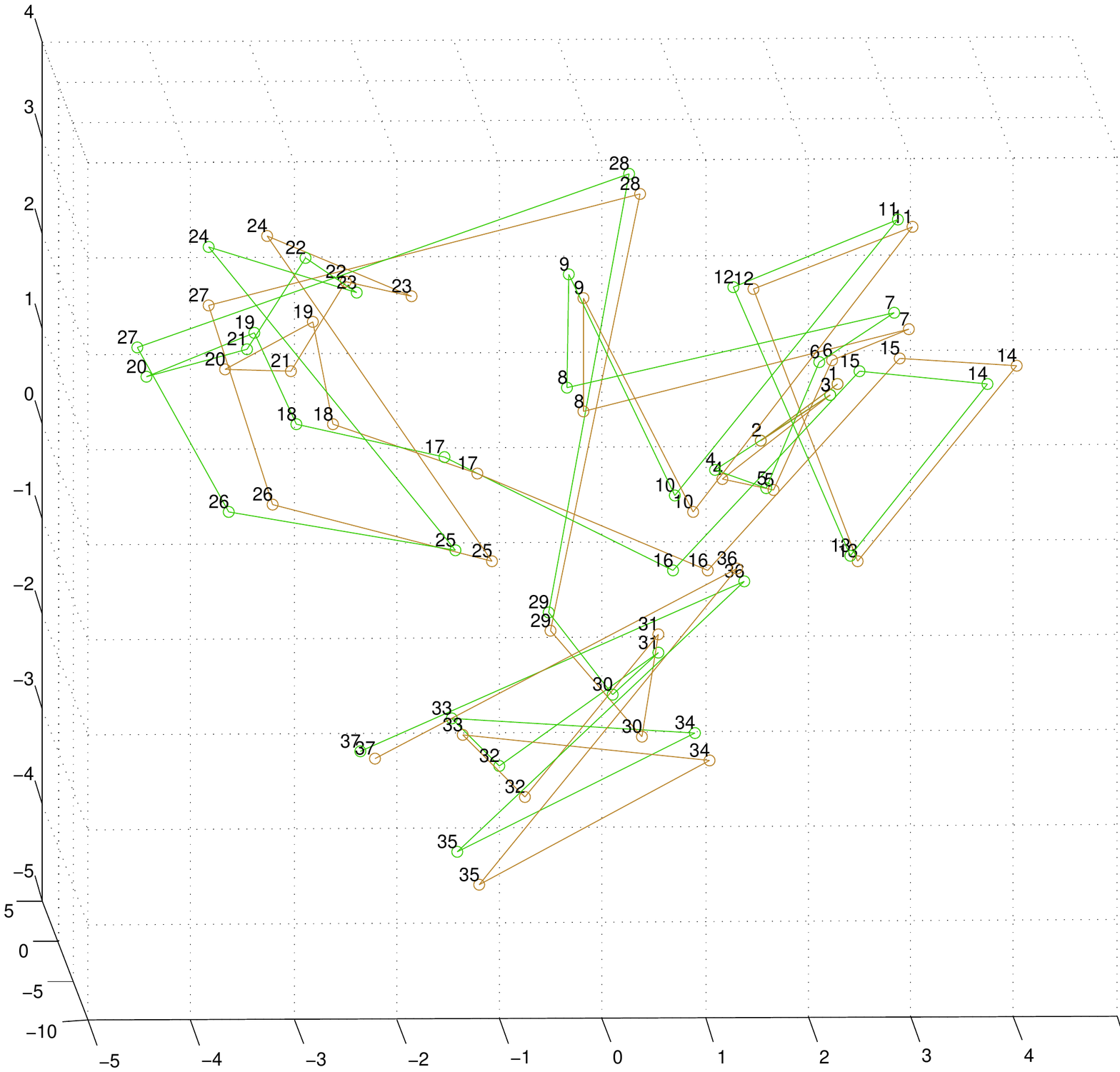} 
\end{minipage}} &
\multicolumn{3}{c|}{% 
\begin{minipage}{4cm}
  \includegraphics[width=3.5cm]{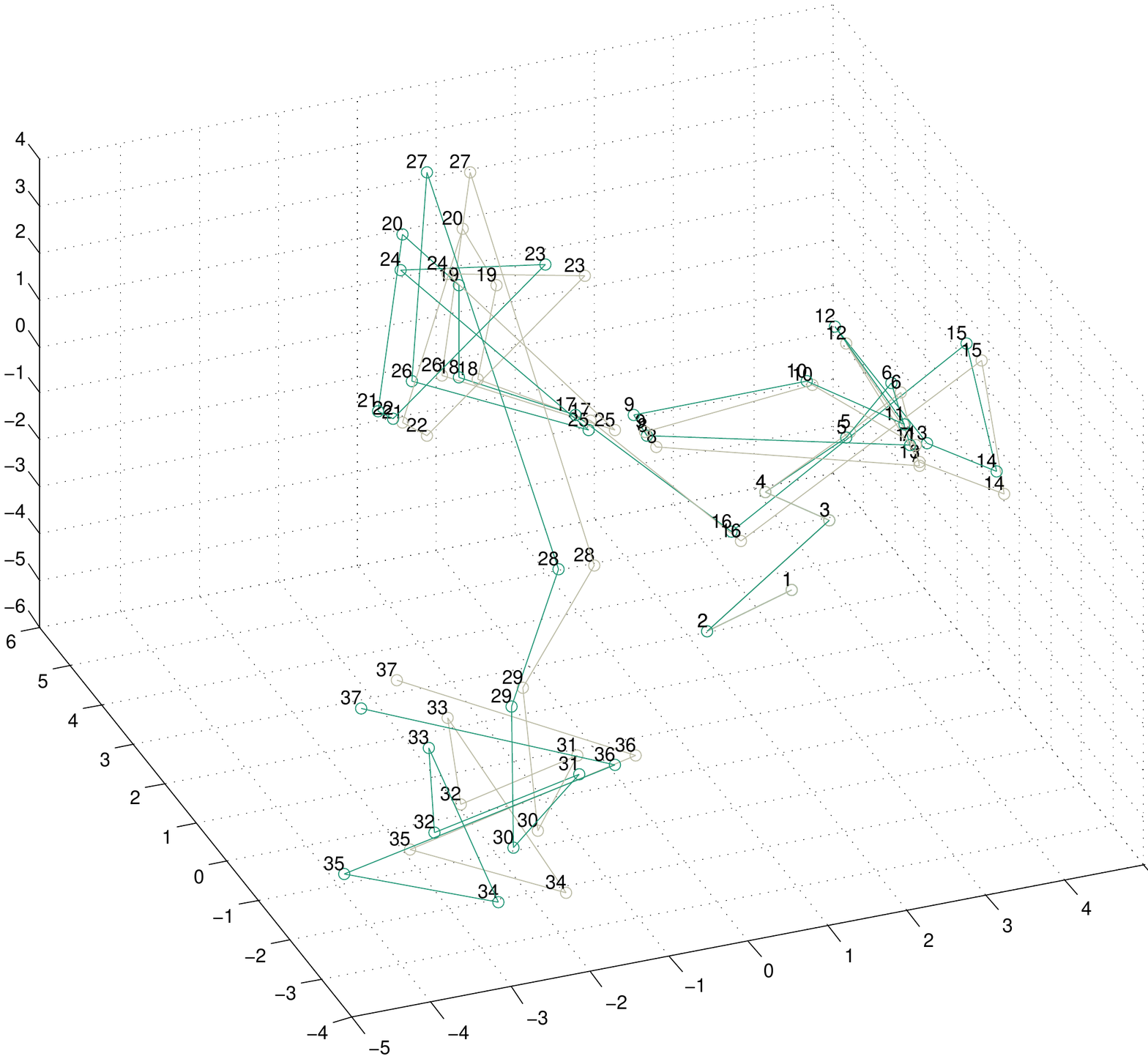} 
\end{minipage}} \\ \hline
$\partial$ &\multicolumn{3}{c|}{2.0211} &\multicolumn{3}{c|}{2.3988} \\ \hline \hline
{\sc MS} & 0 & 0.33 & 20.19 & 0 & 0.35 & 20.12 \\ \hline 
\begin{minipage}{1cm}
$x_{\mbox{\tiny\tt tiny}}$ and $x_{\mbox{\tiny DEMI}}$
\end{minipage} &
\multicolumn{3}{c|}{% 
\begin{minipage}{4cm}
  \includegraphics[width=3.5cm]{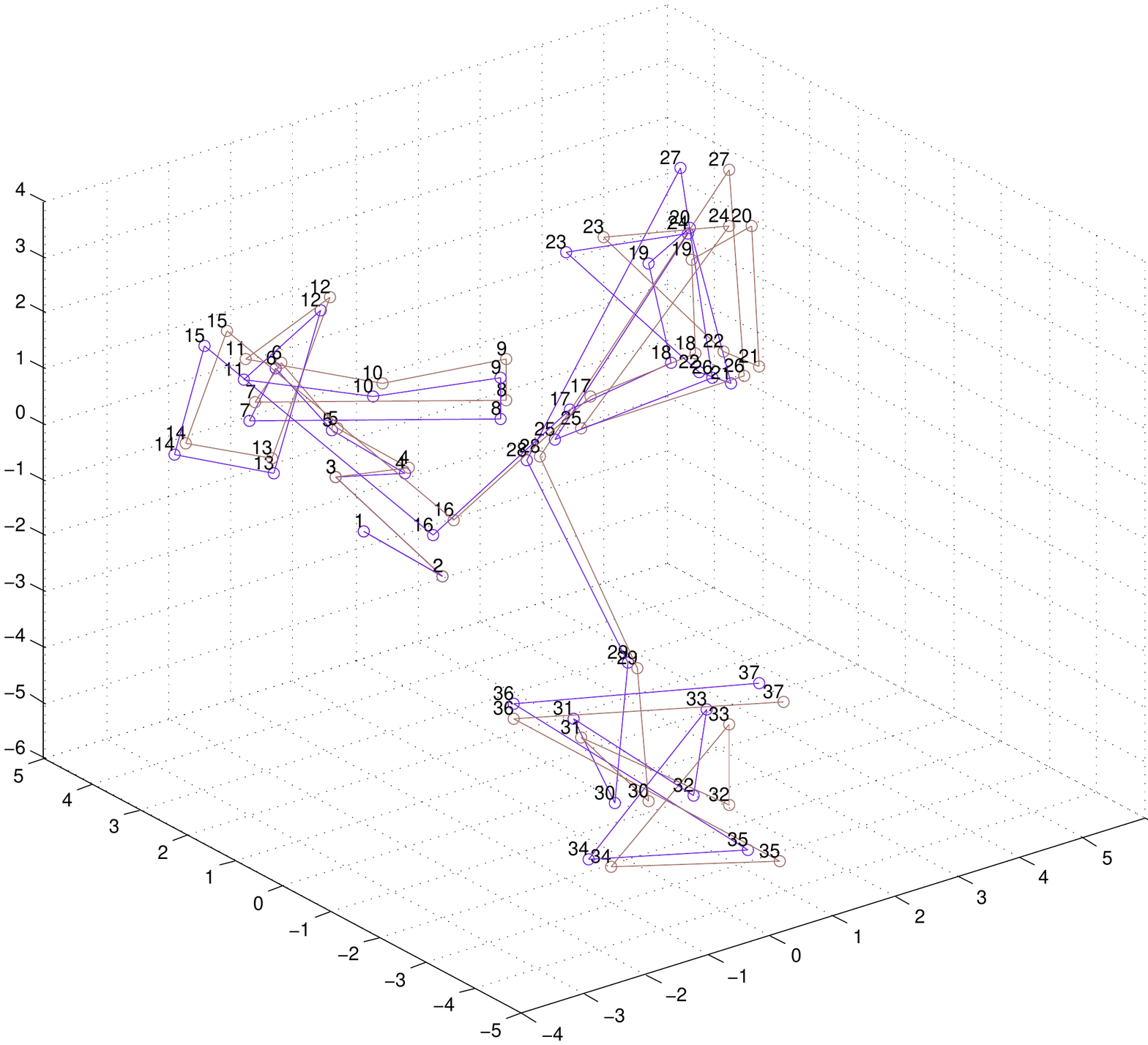} 
\end{minipage}} &
\multicolumn{3}{c|}{% 
\begin{minipage}{4cm}
  \includegraphics[width=3.5cm]{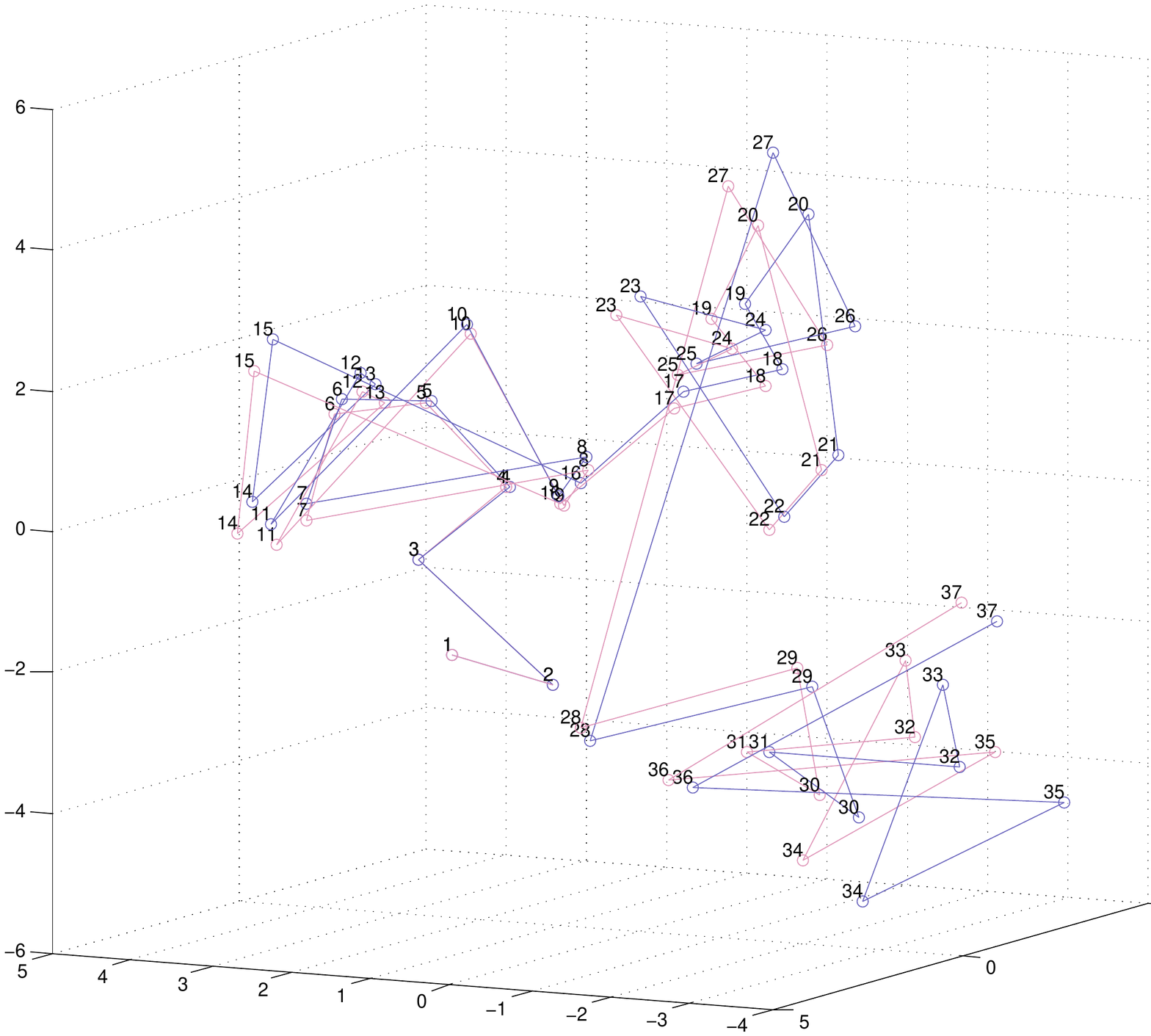} 
\end{minipage}} \\ \hline
$\partial$ &\multicolumn{3}{c|}{2.0211} &\multicolumn{3}{c|}{2.8338} \\ \hline \hline
\end{tabular}
}
\end{center}
\caption{Performance of {\it convexity and concavity} on {\tt
    tiny}. For each solver and formulation (variant), we report the
  edge errors $\Phi,\Psi$, the CPU time, a 3D plot of the solution
  $x_{\mbox{\tt\scriptsize tiny}}$ given in the PDB file versus the
  solution $x_{\mbox{\scriptsize DEMI}}$ found by solving the DEMI
  instance with $x=x_{\mbox{\tt\scriptsize tiny}}$ and $y$ given by
  the solution of the solver, and the corresponding DEMI measure
  $\partial(x,y)=\min\limits_{g,\rho}\|x-g\rho(y)\|$.}
\label{t:concavity}
\end{table}

\subsubsection{Semidefinite programming relaxation}
This is a natural SDP relaxation, similar
to many which already appeared in the literature, where
$\|x_u-x_v\|_2^2$ is linearized to $X_{uu}+X_{vv}-2X_{uv}$:
\begin{equation}
  \left. \begin{array}{rrcl}
    \max\limits_{X\succeq 0} & \sum\limits_{\{u,v\}\in E}
       (X_{uu}+X_{vv}-2X_{uv}) && \\ 
    \forall \{u,v\}\in E & X_{uu} + X_{vv} - 2X_{uv} &\ge& L^2_{uv} \\
    \forall \{u,v\}\in E & X_{uu} + X_{vv} - 2X_{uv} &\le& U^2_{uv},
  \end{array} \right\} \label{eq:dgp5}
\end{equation}
where $X\succeq 0$ means that $X$ is required to be positive
semidefinite.  Several SDP formulations for the DGP have been proposed
in the literature over the years, \revision{see
e.g.~\cite{yajima,wolkowicz,biswasacm,biswas2006ieee}.} Our formulation, which
addresses the \iDGP, is directly inspired by those in
\cite{biswasphd}, since it employs a linearization of the constraints
in Eq.~\eqref{eq:Idgp2}. As objective function, we employ a
linearization of $\sum_{\{u,v\}\in E}\|x_u-x_v\|^2$, which is
unusual. We observed empirically that this yields a good performance
on datasets arising from protein conformation.

Variants: replace the objective with $\min\mbox{Tr}(X)$ as a proxy to
rank minimization \cite{phaselift}. The performance of the SDP
relaxation and its variant on the {\tt tiny} instance is shown in
Table \ref{t:sdp}.

\begin{table}
\begin{center}
{\small
\begin{tabular}{|l||ccc|ccc|} \hline 
& \multicolumn{3}{c|}{Original} & \multicolumn{3}{c|}{Var.~(i)} \\
{\it Solver} & $\Phi$ & $\Psi$ & {\it CPU} & $\Phi$ & $\Psi$ & {\it CPU} \\ \hline \hline
{\sc Mosek} & 0.0153 & 0.3140 & 1.37 & 0.4704 & 2.4810 & 1.35 \\ \hline
\begin{minipage}{1cm}
$x_{\mbox{\tiny\tt tiny}}$ and $x_{\mbox{\tiny DEMI}}$
\end{minipage} &
\multicolumn{3}{c|}{% 
\begin{minipage}{4cm}
  \includegraphics[width=3.5cm]{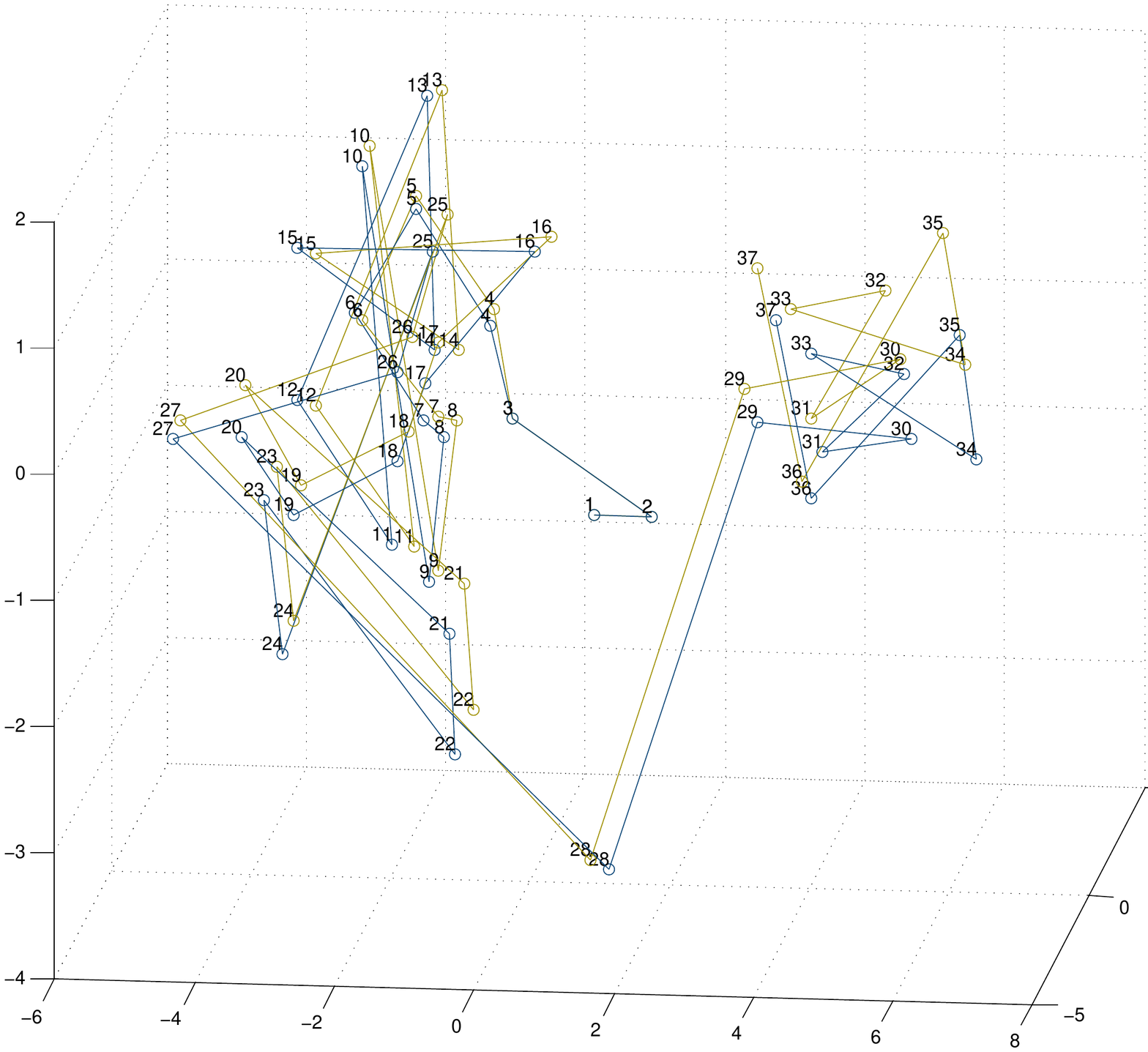} 
\end{minipage}} &
\multicolumn{3}{c|}{% 
\begin{minipage}{4cm}
  \includegraphics[width=3.5cm]{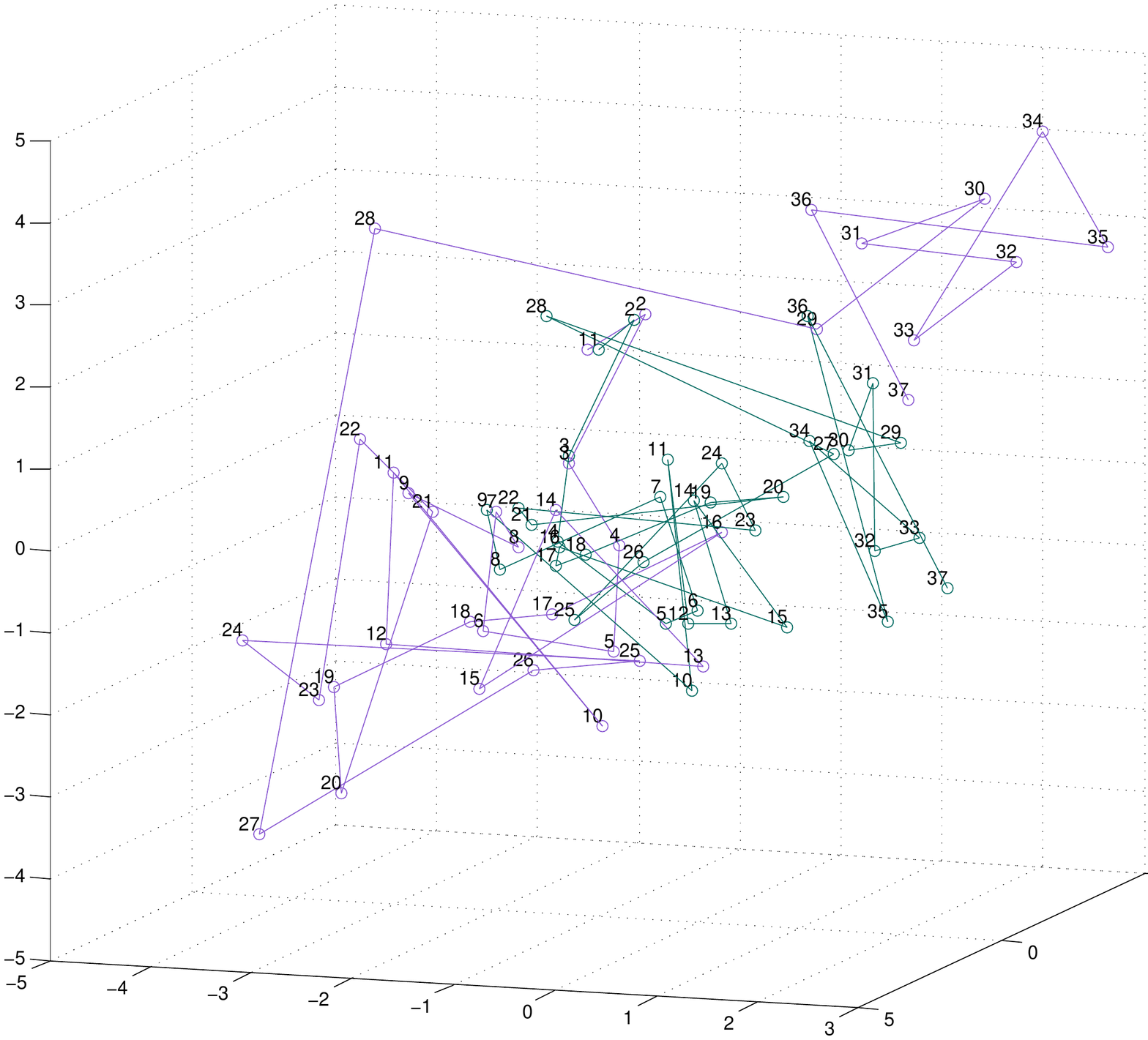} 
\end{minipage}} \\ \hline
$\partial$ &\multicolumn{3}{c|}{2.3972} &\multicolumn{3}{c|}{15.6098} \\ \hline \hline
\end{tabular}
}
\end{center}
\caption{Performance of {\it semidefinite programming} on {\tt
    tiny}. For each formulation (variant), we report the
  edge errors $\Phi,\Psi$, the CPU time, a 3D plot of the solution
  $x_{\mbox{\tt\scriptsize tiny}}$ given in the PDB file versus the
  solution $x_{\mbox{\scriptsize DEMI}}$ found by solving the DEMI
  instance with $x=x_{\mbox{\tt\scriptsize tiny}}$ and $y$ given by
  the solution of the solver, and the corresponding DEMI measure
  $\partial(x,y)=\min\limits_{g,\rho}\|x-g\rho(y)\|$.}
\label{t:sdp}
\end{table}

\subsubsection{Yajima's SDP relaxation}

This formulation was proposed in \cite{yajima}. The term
$2\sum_{\{u,v\}\in E} X_{uv}$ added to the objective function is equal
to $\mbox{Tr}(\mathbf{1}X)$ (where $\mathbf{1}$ is the all-one matrix)
and has a regularization purpose, ensuring that
$\mbox{Tr}(\mathbf{1}X)=0$ and hence that $\mbox{rk}(X)\le n-1$.
\begin{equation}
  \left. \begin{array}{rrcl}
    \min\limits_{s\in\mathbf{M}^+,X\succeq 0} & \sum\limits_{\{u,v\}\in E}
       (s_{uv} - (X_{uu}+X_{vv}-2X_{uv}) + L^2_{uv}) &+&
    2\sum\limits_{\{u,v\}\in E} X_{uv}  \\ 
    \forall \{u,v\}\in E & (X_{uu} + X_{vv} - 2X_{uv}) - L^2_{uv}  &\le& s_{uv} \\
    \forall \{u,v\}\in E & 2(X_{uu} + X_{vv} - 2X_{uv}) - L^2_{uv}-
      U^2_{uv} &\le& s_{uv} 
  \end{array} \right\} \label{eq:dgp6}
\end{equation}

We propose no variants for this formulation. The performance of
Yajima's SDP relaxation on the {\tt tiny} instance is shown in Table
\ref{t:yajima}.

\begin{table}
\begin{center}
{\small
\begin{tabular}{|l||ccc|ccc|} \hline 
& \multicolumn{3}{c|}{Original} \\
{\it Solver} & $\Phi$ & $\Psi$ & {\it CPU}  \\ \hline \hline
{\sc Mosek} & 0.3864 & 2.6086 & 1.65 \\ \hline
\begin{minipage}{1cm}
$x_{\mbox{\tiny\tt tiny}}$ and $x_{\mbox{\tiny DEMI}}$
\end{minipage} &
\multicolumn{3}{c|}{% 
\begin{minipage}{5cm}
  \includegraphics[width=5cm]{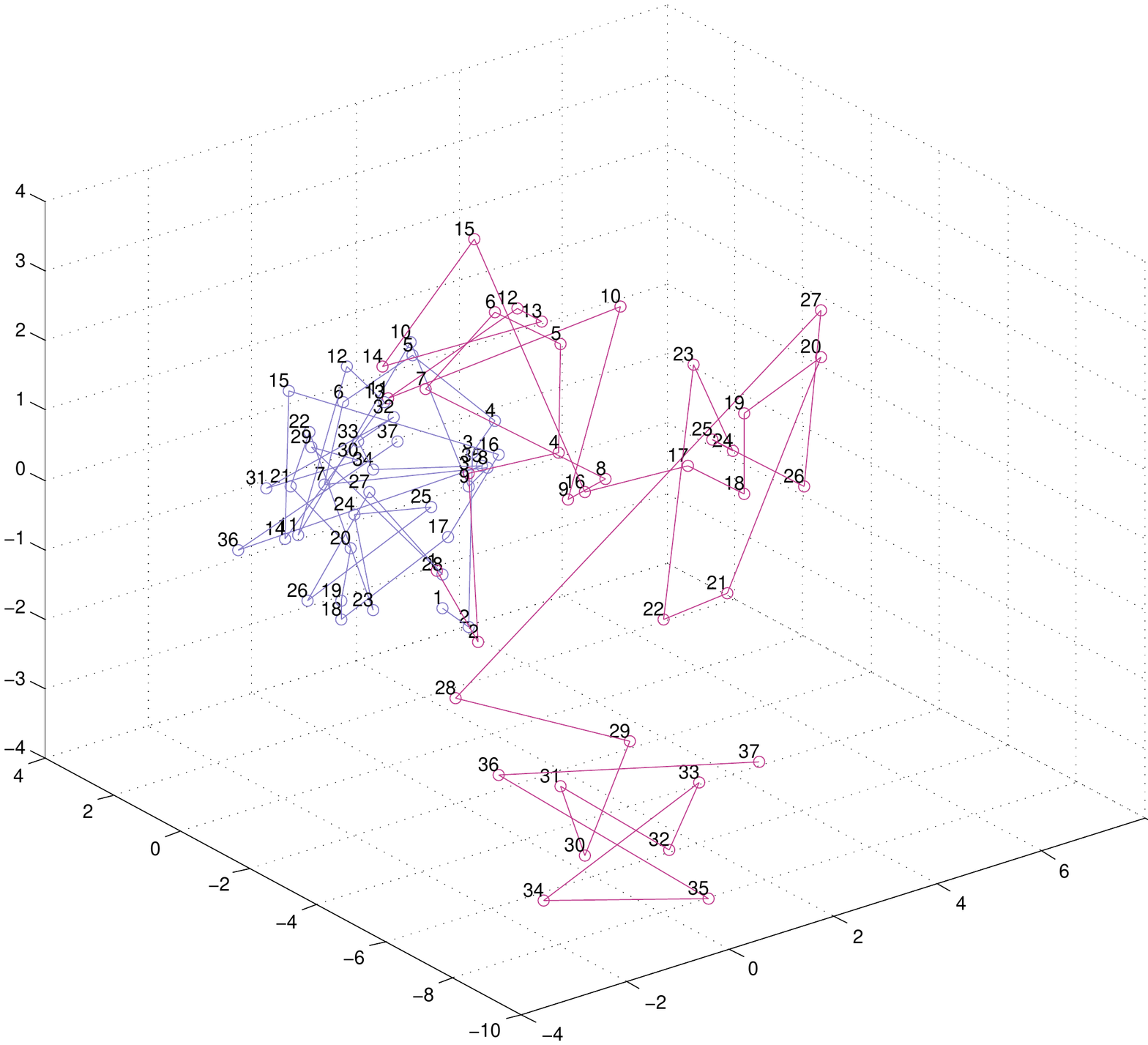} 
\end{minipage}} \\ \hline
$\partial$ &\multicolumn{3}{c|}{27.3586}  \\ \hline \hline
\end{tabular}
}
\end{center}
\caption{Performance of {\it Yajima's SDP} on {\tt tiny}. We
  report the edge errors $\Phi,\Psi$, the CPU time, a 3D plot of the
  solution $x_{\mbox{\tt\scriptsize tiny}}$ given in the PDB file
  versus the solution $x_{\mbox{\scriptsize DEMI}}$ found by solving
  the DEMI instance with $x=x_{\mbox{\tt\scriptsize tiny}}$ and $y$
  given by the solution of the solver, and the corresponding DEMI
  error $\partial(x,y)=\min\limits_{g,\rho}\|x-g\rho(y)\|$.}
\label{t:yajima}
\end{table}

\subsection{A pointwise reformulation}
\label{s:pointwise}
Pointwise reformulations are only exact for a specific set of values
assigned to certain parameters. Typically, replacing variables or
entire terms by parameters makes it possible to obtain formulations
for which there exist very efficient solution methods. This
reformulation will be used in a stochastic search setting (see
Sect.~\ref{s:mwu} below) where the global search phase occurs over the
parameter values.

We replace the term $(x_{uk}-x_{vk})^2=(x_{uk}-x_{vk})(x_{uk}-x_{vk})$
by a linear term $\theta_{uvk}(x_{uk}-x_{vk})$ whenever it occurs in
Eq.~\eqref{eq:Idgp2} and \eqref{eq:dgp4} in a nonconvex way:
\begin{equation}
  \left. \begin{array}{rrcl}
     \max\limits_{x} & \sum\limits_{\{u,v\}\in E}\sum\limits_{k\le K}
     \theta_{uvk}(x_{uk}-x_{vk}) & & \\  
     \forall\{u,v\}\in E & \|x_u-x_v\|_2^2 &\le & U_{uv}^2 \\
     \forall\{u,v\}\in E & \sum\limits_{k\le K}
       \theta_{uvk}(x_{uk}-x_{vk}) &\ge & L_{uv}^2.
  \end{array} \right\} \label{eq:dgp7}
\end{equation}
It should be clear that for each solution $x^\ast$ of
Eq.~\eqref{eq:Idgp2}, there is a parameter matrix
$\theta^\ast\in\mathbb{R}^{mK}$ such that $x^\ast$ is a feasible
solution of Eq.~\eqref{eq:dgp7}: it suffices to choose
$\theta_{uvk}^\ast=(x^\ast_{uk}-x^\ast_{vk})$ for each $\{u,v\}\in E$
and $k\le K$. Note that Eq.~\eqref{eq:dgp7} is a convex MP, and can
therefore be solved efficiently. We let $\mbox{\sf PtwCvx}(\theta)$ be
the solution of Eq.~\eqref{eq:dgp7} with input parameters $\theta$. 

\section{A new \iDGP\ algorithm}
\label{s:mwu}
In this section we discuss an adaptation to the \iDGP\ of the
well-known MWU method \cite{kale}. As explained in \cite{kale}, the
MWU is in fact a meta-algorithm: it has been rediscovered along the
years applied to many different optimization problems. Differently
from most meta-heuristics, the MWU is as much a theoretical tool as a
practical method, insofar as it provides a ``generic'' asymptotic
performance guarantee which works for all problems where the MWU
applies. The performance guarantee proof can be modified according to
the specific features of the given problem to yield theoretical
results. Among the problems listed in \cite{kale}, possibly the most
interesting for the GO community are the Plotkin-Shmoys-Tardos LP
feasibility approximation algorithm \cite{plotkin} and the SDP
approximation algorithm in \cite{kalesdp}.

The MWU is applied to a multi-iteration setting over a given horizon
$\{1,\ldots,T\}$ where, at each iteration $t\le T$, $m$ ``advisors''
express an opinion about a certain decision. The advisors' opinion
yield a gain/loss vector $\psi^t=(\psi^t_i\;|\;i\le m)$ in
$[-1,1]^m$. The MWU method associates a discrete distribution
$\rho^t=(\rho^t_i\;|\;i\le m)$ on the advisors, which is updated using
the rule
\begin{equation}
  \omega^{t}_i=\omega^{t-1}_i(1-\eta\psi_i^{t-1})
  \label{eq:mwudist}
\end{equation}
for each $t>1$, where $\rho^t_i=\frac{\omega^t_i}{\sum_\ell
  \omega^t_i}$ and $\eta\le\frac{1}{2}$ is a user-defined
parameter. This distribution essentially measures the reliability of
each advisor. The method then stochastically takes the decision given
by advisor $i$ with probability $\rho^t_i$. The average gain/loss made
by MWU is therefore given by the weighted average
$\Omega^t=\psi^t\cdot\rho^t$. It is shown in \cite{kale} that the
following bound holds:
\begin{equation}
  \sum_{t\le T} \Omega^t \le \sum_{t\le T} \psi_{\ell}^t +
  \eta\sum_{t\le T} |\psi_{\ell}^t| + \frac{\ln
    m}{\eta}, \label{eq:mwubound}
\end{equation}
where $\ell$ is the index of the {\it best advisor} on average over
all iterations. For fixed $m$ and $T\to\infty$,
Eq.~\eqref{eq:mwubound} states that the cumulative gain/loss made by
the MWU method is bounded by a (piecewise) linear function of the
gain/loss made by the best advisor, which is somewhat
counterintuitive, given that $\ell$ is not known in advance.

\subsection{The MWU method in the \iDGP\ setting}
We now reinterpret the MWU method in the setting of the \iDGP, which
aims to solve the problem via the pointwise reformulation
Eq.~\eqref{eq:dgp7}. Consider a loop over $T$ iterations: the convex
pointwise reformulation Eq.~\eqref{eq:dgp7} is solved at each
iteration and efficiently yields a candidate realization $\bar{x}$.
This is then refined using $\bar{x}$ as a starting point to a local
Nonlinear Programming (NLP) solver applied to the penalty minimization
formulation of Eq.~\eqref{eq:dgp1}, which yields a current iterate
$x$.  

We now explain how $x$ is used to stochastically update $\theta$ at
iteration $t\le T$ along the lines of the MWU method (see the summary
in Fig.~\ref{f:update}):
\begin{itemize}
\item let $(D_{uv})=(\|x_u-x_v\| \;|\; u,v\in V)$ be the distance matrix
  corresponding to $x$;
\item for each $\{u,v\}\in E$ and $t\le T$, let:
\begin{equation}
   \psi_{uv}^t=\frac{\alpha_{uv}}{\max\limits_{\{w,z\}\in E}\alpha_{wz}}
   \label{eq:psi}
\end{equation}
be the relative error of $D$ with respect to $[L,U]$, where
$\alpha_{uv}$ is defined in Eq.~\eqref{eq:alpha} --- note that
$\psi^t$ is a scaled edge error vector with every component in
$[0,1]$;
\item for each $\{u,v\}\in E$ and $1<t\le T$ let
\begin{equation}
  \omega_{uv}^t = \omega_{uv}^{t-1}(1-\eta\psi_{uv}^{t-1});
  \label{eq:omega}
\end{equation}
\begin{figure}[!ht]
\begin{center}
\begin{tikzpicture}[scale=2.5]
\node (theta) at (-1,1) {$\theta$};
\node (x) at (1,1) {$x$};
%\node (start) at (1.3,1.3) {\scriptsize\color{violet} (start here)};
\node (D) at (1,-1) {$D$};
\node (psi) at (-1,-1) {$\psi^t$};
\draw [arrow] (theta) -- node [above] {\footnotesize $x=\mbox{\sf
    PtwCvx}(\theta)$} (x);  
\draw [arrow] (x) -- node [right] {\footnotesize 
\begin{minipage}{2.5cm}$D=\mbox{\sf EDM}(x)$\\error\end{minipage}} (D);
\draw [arrow] (D) -- node [below]
      {\footnotesize 
\begin{minipage}{3cm}\begin{center}
      $\psi_{uv}^t=\frac{\alpha_{uv}}{\max_{wz}\alpha_{wz}}$\\
      scaled error\end{center}\end{minipage}}
      (psi);  
\draw [arrow] (psi) -- node [left] {\footnotesize
\begin{minipage}{4cm}\begin{flushright}
      $\theta_{uvk}\sim[0,\omega_{uv}^t\psi_{uv}^t(x_{uk}-x_{vk})]$\\ $\omega^t$
    updated as per Eq.~\eqref{eq:omega}
\end{flushright}\end{minipage}}
(theta);
\end{tikzpicture}
\end{center}
\caption{The update of $\theta$ from a candidate realization $x$ at
  each iteration $t$ of the MWU method. The oracle $\mbox{\sf
    PtwCvx}(\theta)$ solves the pointwise reformulation
  Eq.~\eqref{eq:dgp7} parametriezd with $\theta$, and uses the
  solution as a starting point to a local NLP algorithm solving an
  exact formulation of the \iDGP, say Eq.~\eqref{eq:dgp1}.}
\label{f:update}
\end{figure}
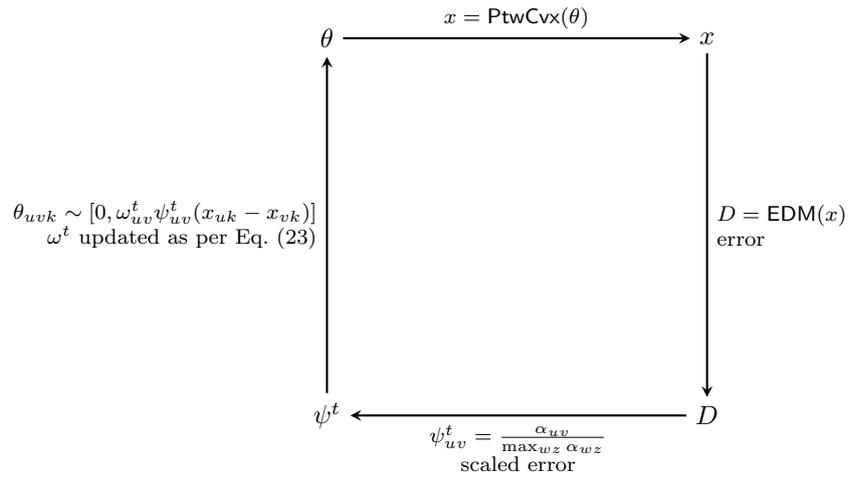
\item let $\theta_{uvk}$ be a random value sampled from the uniform
  distribution on $[0,\omega_{uv}(x_{uk}-x_{vk})]$.
\end{itemize}
We remark that the distribution $\rho^t$ is defined in terms of the
edge weights $\omega^t$:
\begin{equation}
  \rho_{uv}^t=\frac{\omega_{uv}^t}{\sum\limits_{\{w,z\}\in E}
  \omega_{wz}^t}.
  \label{eq:rho}
\end{equation}
The MWU method applied to the \iDGP\ is given as Alg.~\ref{a:mwu}.
\begin{algorithm}[!ht]
\begin{algorithmic}[1]
  \STATE let $\omega^0=1$
  \STATE let $x$ be the output of a local NLP solver applied to
         Eq.~\eqref{eq:dgp1} 
  \STATE let $x'=x$ be the best solution so far
  \FOR{$t\le T$}
    \STATE derive $\theta$ from $x$ as explained above
    \STATE compute a new candidate realization
    $\bar{x}=\mbox{\sf PtwCvx}(\theta)$
    \STATE let $x$ be the solution returned by a local NLP
    solver on Eq.~\eqref{eq:dgp1} with $\bar{x}$ as starting point
    \STATE if $x$ is an improvement with respect to $x'$ according to
           the average error $\Omega^t$, let $x=x'$ \label{edgerr}
  \ENDFOR 
\end{algorithmic}
\caption{$\mbox{\sc MultiplicativeWeightsUpdate}(\eta,T)$}
\label{a:mwu}
\end{algorithm}

%% \begin{enumerate}
%% \item choose a value for $\eta,T$;
%% \item let $\omega^0=1$ and $x$ be the result of applying the local NLP
%%   solver to Eq.~\eqref{eq:dgp1} with a random starting point;
%% \item let $x'=x$ be the best solution so far;
%% \item for each $t\le T$:
%% \begin{enumerate}
%% \item derive $\theta$ from $x$ as explained above;
%% \item compute a new candidate realization $\bar{x}=\mbox{\sf
%%   PtwCvx}(\theta)$;
%% \item if $\bar{x}$ is an improvement with respect to $x$ according to
%%   the average error $\Omega^t$, let $x'=\bar{x}$.
%% \end{enumerate}
%% \end{enumerate}

\subsection{The MWU approximation guarantee for the \iDGP}
One specific feature of the \iDGP\ is that the ``advisors'' never
yield gains but only a cost vector $\psi^t$ having components in
$[0,1]$. This allows us to prove the following result:

\ifspringer\begin{proposition}\else\begin{prop}\fi
After $T$ iterations of the MWU method, the following relationship holds:
\begin{equation}
  \min_{t\le T}\Omega^t \le
    \frac{1}{T}\left(\frac{\ln m}{\eta} +
    (1+\eta)\min_{\{u,v\}\in E}\sum_{t\le T} \psi_{uv}^t\right).
    \label{eq:mwudgp}
\end{equation}
\label{p:mwu}
\ifspringer\end{proposition}\else\end{prop}\fi
\begin{proof}
By Line \ref{edgerr} in Alg.~\ref{a:mwu}, $\min\limits_{t\le T}
\Omega^t$ is the per-edge error (weighted by the distribution $p^t$)
associated to $x'$. From Eq.~\eqref{eq:mwubound}, because
$\psi_{uv}^t\ge 0$ for all $\{u,v\}\in E, t\le T$, we get $\psi_{uv}^t
= |\psi_{uv}^t|$, whence, by definition of $\ell$ in
Eq.~\eqref{eq:mwubound}:
\begin{equation*}
  \sum_{t\le T} \Omega^t \le
    (1+\eta)\min_{\{u,v\}\in E}\sum_{t\le T} \psi_{uv}^t + \frac{\ln m}{\eta}.
\end{equation*}
Since $x'$ is the realization with lowest error over all $t\le T$,
then $T\min\limits_{t\le T} \Omega^t\le \sum\limits_{t\le T}\Omega^t$, which implies:
\begin{equation*}
  T\min_{t\le T} \Omega^t\le (1+\eta)\min_{\{u,v\}\in E}\sum_{t\le T}
    \psi_{uv}^t + \frac{\ln m}{\eta}.
\end{equation*}
Dividing through by $T$ yields the result. \qed
\end{proof}
We remark that the RHS of Eq.~\eqref{eq:mwubound} is the average
weighted error of the best realization found by the MWU in $T$
iterations. Prop.~\ref{p:mwu} states that this error is in the order of
a linear function of the smallest scaled error (see
Eq.~\eqref{eq:psi}) over all edges.

\subsection{Pointwise reformulation feasibility}
Although the pointwise reformulation is exact for a certain value of
$\theta$, it may fail to even be feasible for certain other values of
$\theta$. Since this would be an issue for the MWU method, we further
relax it to the following (always feasible) form:
\begin{equation}
  \left. \begin{array}{rrcl}
     \max\limits_{x,s} & \sum\limits_{\{u,v\}\in E}\left(\sum\limits_{k\le
       K} \theta_{uvk}(x_{uk}-x_{vk}) - s_{uv}\right) && \\  
     \forall\{u,v\}\in E & \|x_u-x_v\|_2^2 &\le & U_{uv}^2 \\
     \forall\{u,v\}\in E & \sum\limits_{k\le K}
     \theta_{uvk}(x_{uk}-x_{vk}) &\ge & L_{uv}^2 - s_{uv} \\
     & s &\ge& 0.
  \end{array} \right\} \label{eq:dgp8}
\end{equation}

\section{Computational assessment}
\label{s:compres}
The aim of this section is to present results obtained by four solvers
(MS, VNS, MWU, and {\sc Mosek}) over 19 different formulations, for each of
61 \iDGP\ instances. Since not every solver can be applied to every
formulation, and sometimes errors are generated for combinations of
solver$+$formulation with some of the instances, the number of measure
vectors is less than $4\times 19\times 61$.

\subsection{Solver$+$formulation combinations}
More precisely, we apply MS and VNS to Eq.~\eqref{eq:dgp1} and its 4
variants (the square root variant and 3 explicitly listed ones),
Eq.~\eqref{eq:dgp3} and its square root variant, Eq.~\eqref{eq:dgp4}
and its positively weighted objective function variant, for a total of
9 formulations. We apply MWU to Eq.~\eqref{eq:dgp7}, and {\sc Mosek}
to Eq.~\eqref{eq:dgp5} and its trace variant, and to
Eq.~\eqref{eq:dgp6}. We therefore consider 22 different
solver$+$formulation combinations.

Unlike in the validation experiments, we did not consider the sBB
solver as most instances are excessively difficult. The rest of the
solver set-up is the same. The solvers MS, VNS, MWU, which are all
implemented in AMPL, solve NLP subproblems at each iteration using the
local NLP solver {\sc Ipopt}. The SDP formulations were modelled using
YalMIP running under MATLAB and solved using {\sc Mosek}. Like the
validation experiments, all results were obtained on an Intel i7 CPU
running at 2.0GHz with 8GB RAM under the Darwin Kernel v.~13.3.0.

\subsection{User-configurable parameters}
Each of the MP solvers was given at most 20s of user CPU time,
excluding the time taken by {\sc Ipopt}. Each call to {\sc Ipopt} was
also limited to 20s; however, the {\sc Ipopt} documentation warns that
its stopwatch is not checked regularly, but only after certain
operations, which on certain instances appear to take place very
rarely. This is apparent in Table \ref{t:cpu}, where many solvers
exceed the 20s CPU time limit. {\sc Mosek} was given no time limit,
since we wanted to find the optimal solution of the SDP.

All tolerances in the AMPL code were set to $1\times 10^{-6}$. {\sc
  Ipopt} was used in its default configuration. The VNS maximum
neighbourhood radius and the maximum number of local searches deployed
in each neighbourhood were both set to $5$.  The $\eta$ parameter in
MWU was set to $0.5$ (its maximum value) after some preliminary
testing. {\sc Mosek} was used in its default configuration.

\subsection{Instances}
Instances were obtained from a selection of PDB files by extracting
all the atomic coordinates, computing all of the inter-atomic
distances, and discarding all those distances exceeding $5${\AA} (so
as to mimic NMR data). More precisely, covalent bonds and angles are
known fairly precisely; since each covalent angle is incident to two
covalent bonds, the remaining side of the triangle they define can
also be computed precisely. Other known distances can be found through
NMR experiments, which yield an interval measurement. We extracted the
protein backbone from each considered PDB dataset, computed all
precise distances, and then we replace all other distances $d_{uv}$
smaller than $5${\AA} by the interval
$[d_{uv}-0.1d_{uv},d_{uv}+0.1d_{uv}]$.

\revision{The mean pruning group generator size $|Z|$ over the test instances is 1.78 and the standard deviation is 4.92, but this is due to a single outlier with $|Z|=34$. Removing the outlier, we have mean $|Z|$ 1.04 and standard deviation 0.30, consistent with Sect.~\ref{s:demisizeofZ}. The sparsity of the pruning edges over the test instances is 0.14.}

Table \ref{t:stats} reports the instance names,
their sizes, and whether they are classified as easy or hard (last
column), see Sect.~\ref{s:easyhard}.
\begin{table}[!ht]
{\small  \begin{center}
  \begin{tabular}{|l|rr|l|} \hline
{\it Instance} & $|V|$ & $|E|$ & Hard? \\ \hline
{\tt 100d} & 489 & 5741 & H\\
{\tt 1guu-1} & 150 & 959 & H\\
{\tt 1guu-4000} & 150 & 968 & H \\
{\tt 1guu} & 150 & 955 & H \\
{\tt 1PPT} & 302 & 3102 & H \\
{\tt 2erl-frag-bp1} & 39 & 406 &  \\
{\tt 2kxa} & 177 & 2711 & H \\
{\tt C0020pdb} & 107 & 999 & H \\
{\tt C0030pkl} & 198 & 3247 & H \\
{\tt C0080create.1} & 60 & 681 & H \\
{\tt C0080create.2} & 60 & 681 & H \\
{\tt C0150alter.1} & 37 & 335 & H$^\ast$ \\
{\tt C0700odd.1} & 18 & 39 &  \\
{\tt C0700odd.2} & 18 & 39 &  \\
{\tt C0700odd.3} & 18 & 39 &  \\
{\tt C0700odd.4} & 18 & 39 &  \\
{\tt C0700odd.5} & 18 & 39 &  \\
{\tt C0700odd.6} & 18 & 39 &  \\
{\tt C0700odd.7} & 18 & 39 &  \\
{\tt C0700odd.8} & 18 & 39 &  \\
{\tt C0700odd.9} & 18 & 39 &  \\
{\tt C0700odd.A} & 18 & 39 &  \\
{\tt C0700odd.B} & 18 & 39 &  \\
{\tt C0700odd.C} & 36 & 242 &  \\
{\tt C0700odd.D} & 36 & 242 &  \\
{\tt C0700odd.E} & 36 & 242 &  \\
{\tt C0700odd.F} & 18 & 39 &  \\
{\tt C0700.odd.G} & 36 & 308 &  \\
{\tt C0700.odd.H} & 36 & 308 &  \\
{\tt cassioli-protein-130731} & 281 & 4871 & H \\ %\hline
{\tt GM1\_sugar} & 68 & 610 & H \\
\hline
\end{tabular}
\begin{tabular}{|l|rr|l|} \hline
{\it Instance} & $|V|$ & $|E|$ & Hard? \\ \hline
{\tt helix\_amber} & 392 & 6265 & H \\
{\tt labelplot} & 37 & 49 & E \\
{\tt lavor11\_7-1} & 11 & 47 &  \\
{\tt lavor11\_7-2} & 11 & 47 &  \\
{\tt lavor11\_7-b} & 11 & 47 &  \\
{\tt lavor11\_7} & 11 & 47 &  \\
{\tt lavor11} & 11 & 40 &  \\
{\tt lavor30\_6-1} & 30 & 192 &  \\
{\tt lavor30\_6-2} & 30 & 202 & H$^\ast$ \\
{\tt lavor30\_6-3} & 30 & 195 & H$^\ast$ \\
{\tt lavor30\_6-4} & 30 & 191 & H$^\ast$ \\
{\tt lavor30\_6-5} & 30 & 195 &  \\
{\tt lavor30\_6-6} & 30 & 195 &  \\
{\tt lavor30\_6-7} & 30 & 195 &  \\
{\tt lavor30\_6-8} & 30 & 193 &  \\
{\tt mdgp4-heuristic} & 4 & 6 &  \\
{\tt mdgp4-optimal} & 4 & 6 &  \\
{\tt names} & 86 & 849 & H \\
{\tt odd01} & 18 & 39 &  \\
{\tt odd02} & 36 & 308 &  \\
{\tt pept} & 107 & 999 & H \\
{\tt res\_0} & 108 & 1410 & H \\
{\tt res\_1000} & 108 & 1506 & H \\
{\tt res\_2000} & 108 & 1404 & H \\
{\tt res\_2kxa} & 177 & 2627 & H \\
{\tt res\_3000} & 108 & 1487 & H \\
{\tt res\_5000} & 108 & 1392 & H \\
{\tt small02} & 36 & 242 &  \\
{\tt tiny} & 37 & 335 & H{}$^\ast$ \\
{\tt water} & 648 & 11939 & H \\ 
 & & & \\
\hline
\end{tabular}
\end{center} }
  \caption{The test set: 61 instances, from the PDB and \cite{Lav05}, their sizes, and the estimated difficulty of solution.}
  \label{t:stats}
\end{table}

\subsection{Weeding out obvious losers}
\label{s:losers}
Not every combination of solver and formulation variant is worth
considering. Those which find a solution with high average edge error
$\Phi$ and/or maximum edge error $\Psi$ should be excluded. We
proceeded to record $\Phi,\Psi$, and seconds of user CPU time for every
combination on every instance, and we computed the average values
(over all instances) of $\Phi,\Psi$, and CPU time.

\begin{figure}[!ht]
  \begin{center}
    \includegraphics[width=10cm]{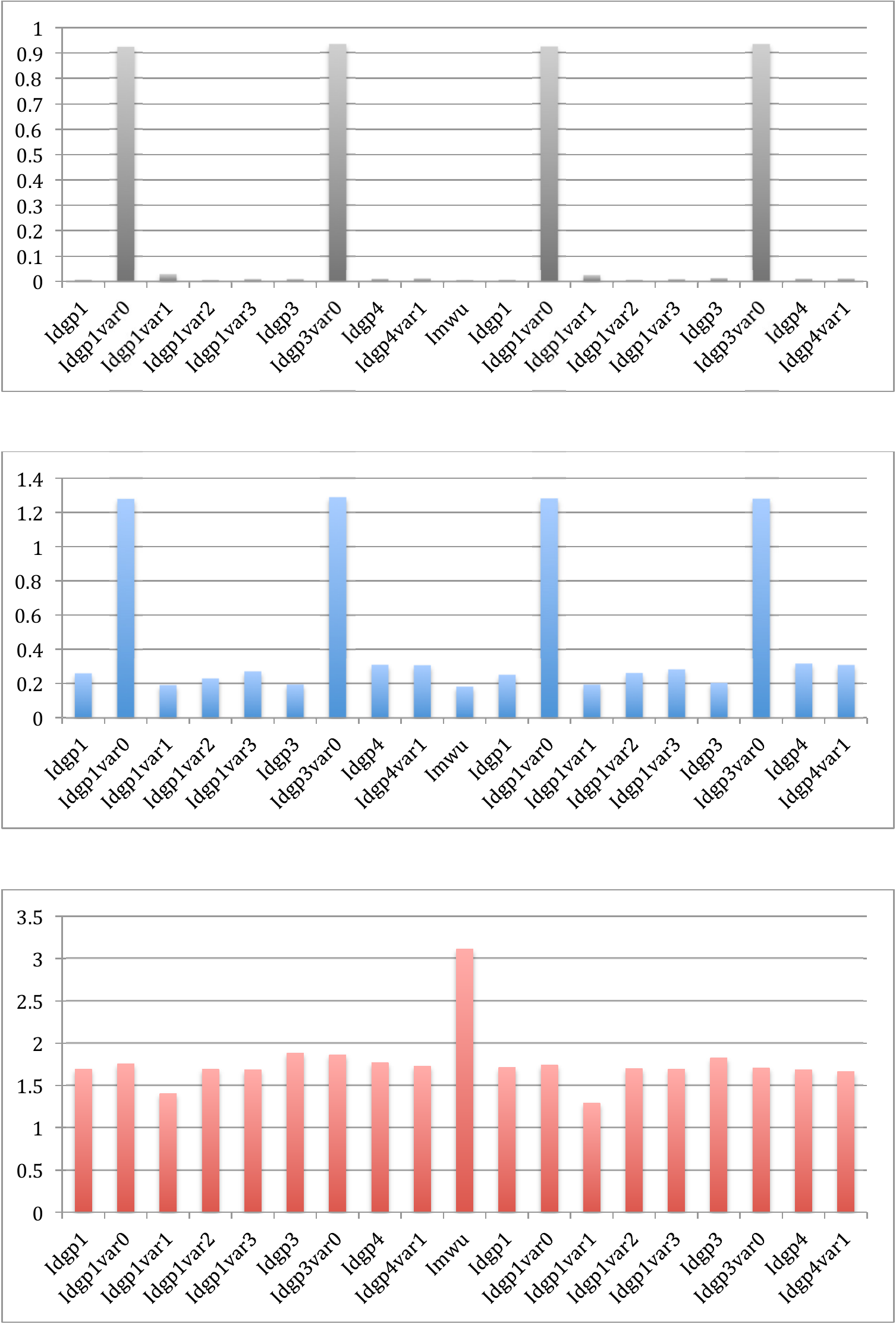}
  \end{center}
  \caption{Histogram plots of the statistic $\log(1+\mu)$ whenever
    $\mu$ is the average of $\Phi$ (top), $\Psi$ (middle), and CPU
    time (bottom) over all instances, for each relevant combination of
    solver$+$formulation, with ``solver'' in MS (left), MWU (middle)
    and VNS (right).}
  \label{f:avgstats}
\end{figure}

The statistics for the MS, MWU, and VNS solvers are shown in
Fig.~\ref{f:avgstats} (more precisely, if $\mu$ is an average, we
plotted $\log(1+\mu)$). All variants involving square roots perform
really poorly in terms of edge errors.
The statistics for the {\sc Mosek} solver, limited to instances where
$n\le 200$ because of RAM limitations, are given below.
\begin{center}
  \begin{tabular}{|l||r|r|r|} \hline
    {\it Formulation} & $\Phi$ & $\Psi$ & CPU \\ \hline
    {\tt sdprel} & {\bf 0.037} & {\bf 0.516} & 51 \\
    {\tt sdprel1} & 0.123 & 0.678 & {\bf 45} \\
    {\tt yajima} & 0.113 & 0.717 & {\bf 45} \\ \hline
  \end{tabular}
\end{center}
The relevant figure in this table is that the SDP relaxation {\tt
  sdprel} has much lower average edge error than the other
formulations, lower maximum error, and slightly higher CPU time. 
  
These tests show that, on average, the SDP trace variant, Yajima's
relaxation, and all square root variants are not worth
considering. The reason why introducing square roots results in
performance losses may be related to the use of the same local
subsolver ({\sc Ipopt}) within all global optimization solver, since
it carries out most floating point computations.

A remark about Yajima's relaxation: although it was introduced
specifically for the \iDGP, it was originally solved using an ad-hoc
interior point method. Even though our results show it underperforms
on average with respect to {\sc Mosek}, this does not negate the
(good) results reported in \cite{yajima}.

We call {\it bad} the solver$+$formulation combinations we excluded,
and {\it good} the rest. The good combinations are shown below, marked
by a ``1'' in the corresponding entry.
\begin{center}
  \begin{tabular}{|l|l|l||c|c|c|c|} \hline 
    \multicolumn{3}{|c||}{Formulation} & \multicolumn{4}{c|}{Solver}\\ \hline
    {\it Description} & {\it Notation} & {\it Name}
    & {\it MS} & {\it MWU} & {\it VNS} & {\it Mosek} \\ \hline
    \eqref{eq:dgp1}        & \eqref{eq:dgp1} & {\tt Idgp1} & 1 &   & 1 &  \\
    \eqref{eq:dgp1} variant (i) & \eqref{eq:dgp1}.1 & {\tt Idgp1var1} & 1 & & 1 & \\
    \eqref{eq:dgp1} variant (ii) & \eqref{eq:dgp1}.2 & {\tt Idgp1var2} & 1 & & 1 & \\
    \eqref{eq:dgp1} variant (iii) & \eqref{eq:dgp1}.3 & {\tt Idgp1var3} & 1 & & 1 & \\
    \eqref{eq:dgp3}        & \eqref{eq:dgp3} & {\tt Idgp3} & 1 &   & 1 &  \\
    \eqref{eq:dgp4}        & \eqref{eq:dgp4} & {\tt Idgp4} & 1 &   & 1 &  \\
    \eqref{eq:dgp4} variant (i) & \eqref{eq:dgp4}.1 & {\tt Idgp4var1} & 1 & & 1 & \\
    \eqref{eq:dgp7}        & \eqref{eq:dgp7} & {\tt Imwu} &  & 1 &   &  \\
    \eqref{eq:dgp5}        & \eqref{eq:dgp5} & {\tt sdprel} &  &  & & 1 \\
\hline
  \end{tabular}
  \end{center}

\subsection{Focusing on the hard instances}
\label{s:easyhard}
We also make a qualitative distinction between easy and hard
instances. We call an instance {\it easy} if at least one third of the
good combinations find a solution with $\Phi,\Psi$ approximately zero
within 1s of user CPU time, and {\it hard} the rest. The
classification is reported in the last column of Table
\ref{f:avgstats}: hard instances are marked ``H''. We marked
H${}^\ast$ the instances which are ``borderline hard'', i.e.,~there is
at least one good combination which finds a solution with $\Phi,\Psi$
approximately zero within 1s of user CPU time.

The computational results below will focus on the hard instances for
$\Phi,\Psi$; because of excessive computational requirements, however,
we shall relax this constraint for the results on the DEMI measure
$\partial$.

\subsection{Testing heuristics without averages}
\label{s:noaverage}
When benchmarking heuristic algorithms, such as MS, VNS, or MWU, it is
customary to present results based on a number of runs (the higher the
better) of the same instance. Because of the complexity of this
computational comparison, and the absolute time taken to perform it,
it was ungainly for us to multiply this effort by a significant factor
(say 10 or 100). Does this mean that our results are unreliable?
Though it could be argued that the instance-by-instance results are in
fact unreliable, as can be gleaned by the difference in $\partial$
measure for the {\tt tiny} instance in Sect.~\ref{s:formulations} and
those in Table \ref{t:demi}, we think the averages (reported in Tables
\ref{t:avg}-\ref{t:demi}) are not. Since we never claim in our
computational comparisons that one method is best for a certain
instance, but only suggest first and second best over all tested
instances, we think our computational benchmark is significant.

\subsection{Comparative results on edge errors and CPU}
\label{s:comparison}
In this section we discuss an overarching comparison yielding an
overall ``winner''. Our most meaningful measure, if $\Phi$ and $\Psi$
are nonzero, is the DEMI measure $\partial(\cdot,y)$, where $y$ is a
given solution of the \kDMDGP\ instance being solved. By
Sect.~\ref{s:demipractice}, however, we are not able to compute it for
every instance, and hence we focus on $\Phi,\Psi$ for our global
comparison, and only look at $\delta$ on a subset of instances
(Sect.~\ref{s:demires}).

Tables \ref{t:avg}-\ref{t:cpu} report the average edge
errors $\Phi$, the maximum edge error $\Psi$, and the CPU time taken
by the good combinations when solving hard instances.
\begin{sidewaystable}[!ht]
  {\tiny
   %\scriptsize
    \begin{center}
      \ifspringer
      \vspace*{10cm}
      \fi
      \input{avg.tex}
    \end{center}
  }
\caption{$\Phi$ statistics: blue indicates best.}
\label{t:avg}
\end{sidewaystable}
We remark that the MWU algorithm is best with respect to the edge
errors $\Phi$ and $\Psi$, and the worst with respect to CPU
time. However, since CPU time is of least consequence in protein
conformation computations, CPU time information has a much lower
priority than solution quality.
\begin{sidewaystable}[!ht]
  {\tiny
   %\scriptsize
    \begin{center}
      \ifspringer
      \vspace*{10cm}
      \hspace*{-4cm}
      \fi
      \input{max.tex}
    \end{center}
  }
\caption{$\Psi$ statistics: blue indicates best.}
\label{t:max}
\end{sidewaystable}
We can therefore make the following claim.
\begin{center}
  {\bf The MWU algorithm is the best solver on average}.
\end{center}
\begin{sidewaystable}[!ht]
  {\tiny
   %\scriptsize
    \begin{center}
      \ifspringer
      \vspace*{10cm}
      \fi
      \input{cpu.tex}
    \end{center}
  }
\caption{CPU time statistics: blue indicates best.}
\label{t:cpu}
\end{sidewaystable}
Given the consequential CPU time difference between MWU and the other
solvers, it is worth ranking the solver$+$formulation combinations by
$\Phi,\Psi$, and CPU time (see below).
{\small
  \begin{center}
  \begin{tabular}{|r||ll|ll|ll|} \hline
{\it Rank} & \multicolumn{2}{c|}{$\Phi$}&\multicolumn{2}{c|}{$\Psi$} &\multicolumn{2}{c|}{CPU}\\ \hline
    1  & {\tt mwu+Imwu}      & 0.029 & {\tt mwu+Imwu}      & 1.111 & {\tt vns+Idgp1var1} & 41.53  \\
    2  & {\tt ms+Idgp1var2}  & 0.031 & {\tt ms+Idgp1var1}  & 1.237 & {\tt ms+Idgp1var1}  & 55.11  \\
    3  & {\tt vns+Idgp1}     & 0.032 & {\tt vns+Idgp1var1} & 1.259 & {\tt ms+Idgp4var1}  & 96.85  \\
    4  & {\tt vns+Idgp1var2} & 0.032 & {\tt ms+Idgp3}      & 1.265 & {\tt vns+Idgp4var1} & 99.05  \\
    5  & {\tt ms+Idgp1}      & 0.033 & {\tt mosek+sdprel}  & 1.267 & {\tt vns+Idgp4}     & 105.70 \\
    6  & {\tt vns+Idgp1var3} & 0.045 & {\tt vns+Idgp3}     & 1.281 & {\tt ms+Idgp1var3}  & 107.41 \\
    7  & {\tt ms+Idgp4}      & 0.046 & {\tt ms+Idgp1var2}  & 1.560 & {\tt vns+Idgp1var3} & 108.15 \\
    8  & {\tt ms+Idgp1var3}  & 0.047 & {\tt vns+Idgp1}     & 1.752 & {\tt ms+Idgp1var2}  & 108.63 \\
    9  & {\tt vns+Idgp4}     & 0.047 & {\tt ms+Idgp1}      & 1.828 & {\tt ms+Idgp1}      & 109.26 \\
    10 & {\tt ms+Idgp3}      & 0.048 & {\tt vns+Idgp1var2} & 1.849 & {\tt vns+Idgp1var2} & 109.83 \\
    11 & {\tt vns+Idgp4var1} & 0.049 & {\tt ms+Idgp1var3}  & 1.939 & {\tt ms+Idgp4}      & 111.16 \\
    12 & {\tt ms+Idgp4var1}  & 0.052 & {\tt vns+Idgp1var3} & 2.007 & {\tt vns+Idgp1}     & 113.35 \\
    13 & {\tt vns+Idgp3}     & 0.064 & {\tt vns+Idgp4var1} & 2.027 & {\tt vns+Idgp3}     & 146.98 \\
    14 & {\tt mosek+sdprel}  & 0.078 & {\tt ms+Idgp4var1}  & 2.028 & {\tt ms+Idgp3}      & 170.79 \\
    15 & {\tt vns+Idgp1var1} & 0.129 & {\tt ms+Idgp4}      & 2.064 & {\tt mosek+sdrel}   & 472.82 \\
    16 & {\tt ms+Idgp1var1}  & 0.147 & {\tt vns+Idgp4}     & 2.140 & {\tt mwu+Imwu}      & 27669  \\
    \hline
  \end{tabular}
  \end{center}
}%
%%(see Table \ref{t:ranking}).
%%\begin{table}[!ht]
%%  \caption{Ranking solver$+$formulation combinations according to
%%    $\Phi,\Psi$ and CPU time.}
%%  \label{t:ranking}
%%\end{table}
This ranking shows that {\tt mwu+Imwu} has the only consistent ranking
in both $\Phi$ and $\Psi$.  It also shows that no other
solver$+$formulation combination has the same desirable property of
approximately equal rank w.r.t.~both $\Phi$ and $\Psi$. The issue is
not only relative: values of $\Phi$ higher than $0.1$ and of $\Psi$
higher than $1.5$ may well imply that the realization is fundamentally
wrong, and the only combinations with $\Phi<0.1$ and $\Psi<1.5$ are
{\tt ms+Idgp3}, {\tt vns+Idgp3}, {\tt mosek+sdprel}. However, the
statistics for the latter were computed on a subset of instances (all
those with $n\le 200$) due to the high RAM requirements of {\sc Mosek}
when applied to large instances (see Sect.~\ref{s:losers}). Based on
these observations, we claim that:
\begin{quote}
{\bf the formulation {\tt Idgp3} in Eq.~\eqref{eq:dgp3}, when used
  with MS or VNS, is second best}.
\end{quote}
We observe that the usual trade-off between quality and efficiency is
also at play: solving Eq.~\eqref{eq:dgp3} takes longest over all
formulations solved by both MS and VNS.

\subsection{Results on DEMI}
\label{s:demires}
Table \ref{t:demi} reports the results on the DEMI measure. Note that
the instances in the test set are not the same as for the tests on
$\Phi,\Psi$, and CPU (Tables \ref{t:avg}-\ref{t:cpu}). As mentioned in
Sect.~\ref{s:demipractice}, it is not always possible to determine a
cTOP order automatically (or disprove that one exists) in acceptable
amounts of CPU time, which is a requirement for computing the DEMI
measure. Table \ref{t:demi} includes all instances for which this task
could be carried out within 150s of CPU time.

Although it is clear that the SDP relaxation Eq.~\eqref{eq:dgp5}
scores the best performance in terms of the DEMI measure, we mentioned
above that the {\sc Mosek} solver is unable to scale up to desired
sizes. We must therefore resort to the second best, which happens to
be the MWU algorithm, consistently with Sect.~\ref{s:comparison}. We
also observe that VNS attains lower average DEMI measure values more
often than MS.

We recall that the DEMI measure values for {\tt tiny} differ from
those given in Sect.~\ref{s:formulations} for the reasons given in
Sect.~\ref{s:noaverage}.

\section{Conclusion}
Our main aim is to find the best general-purpose continuous search
methods for solving \iDGP\ instances. To answer this question, we
need: (i) a set of benchmarking measures; (ii) a set of
\iDGP\ formulations; (iii) a set of methods; (iv) extensive
computational results. Since a preliminary study \cite{ios14slides}
showed that two standard metaheuristics and the existing benchmark
measures were insufficient, we decided to introduce a new measure and
a new method.

Accordingly, this paper presents several notions: (a) a coordinate
root mean square deviation modulo partial reflections (called DEMI
measure), for benchmarking the performance of \iDGP\ algorithms on
protein isomers; (b) a zoo of mathematical programming formulations
for the \iDGP; (c) a new method for solving the \iDGP, based on the
well-known Multiplicative Weights Update (MWU) algorithm; (d) a
complex computational benchmark for the best formulation-based methods
on the hardest instances.

Our study shows that, on average:
\begin{itemize}
  \item the new MWU-based heuristic yields
    \iDGP\ solutions of highest quality with respect to existing measures;
  \item the Square Factoring formulation in Eq.~\eqref{eq:dgp3} is second best;
  \item as concerns the new DEMI measure, the SDP relaxation in
    Eq.~\eqref{eq:dgp5} is best, but only on a limited set of
    instances, whereas the MWU-based heuristic is second best.
\end{itemize}

\revision{Future research directions for the topics presented in this paper include: (i) the algorithmic exploitation of the DEMI measure for more effective pruning within the Branch-and-Prune algorithm; (ii) the insertion of a limited diving device within the Branch-and-Prune: instead of branching in order to find possible positions of the next atom in the order, it would be desirable to realize a considerable number of successive atoms by means of one of the continuous methods presented in this paper.}

\begin{sidewaystable}[!ht]
  {\tiny
   %\scriptsize
    \begin{center}
      \ifspringer
      \vspace*{10cm}
      \hspace*{-4cm}
      \fi
      \input{demi.tex}
    \end{center}
  }
\caption{DEMI measure statistics: blue indicates best. Although
  the best on average is {\sc Mosek} on {\tt sdprel}, this combination
  was unable to scale up to desired sizes. We therefore also
  emphasized the second best: MWU on {\tt Imwu}.}
\label{t:demi}
\end{sidewaystable}

\section*{Acknowledgments}
The second author (VKK) is supported by a Microsoft Research
PhD Fellowship. The third author (CL) is grateful to the Brazilian
funding agencies FAPESP and CNPq for financial support. The fourth
author (LL) is partly supported by the ANR grant ``Bip:Bip'' under
contract ANR-10-BINF-0003. The fifth author (NM) is grateful to the
Brazilian funding agencies FAPERJ and CNPq for financial support.

% end paper

\bibliographystyle{plain}
\bibliography{ibmer}

\end{document}

%% file: avg.tex
\begin{tabular}{|l||r|r|r|r|r|r|r|r|r|r|r|r|r|r|r|r|r|r|}\hline
{\it Instance} & $\mbox{\tt mosek}\atop\mbox{\tt sdprel}$ & $\mbox{\tt ms}\atop\mbox{\tt Idgp1}$ & $\mbox{\tt ms}\atop\mbox{\tt Idgp1var1}$ & $\mbox{\tt ms}\atop\mbox{\tt Idgp1var2}$ & $\mbox{\tt ms}\atop\mbox{\tt Idgp1var3}$ & $\mbox{\tt ms}\atop\mbox{\tt Idgp3}$ & $\mbox{\tt ms}\atop\mbox{\tt Idgp4}$ & $\mbox{\tt ms}\atop\mbox{\tt Idgp4var1}$ & $\mbox{\tt mwu}\atop\mbox{\tt Imwu}$ & $\mbox{\tt vns}\atop\mbox{\tt Idgp1}$ & $\mbox{\tt vns}\atop\mbox{\tt Idgp1var1}$ & $\mbox{\tt vns}\atop\mbox{\tt Idgp1var2}$ & $\mbox{\tt vns}\atop\mbox{\tt Idgp1var3}$ & $\mbox{\tt vns}\atop\mbox{\tt Idgp3}$ & $\mbox{\tt vns}\atop\mbox{\tt Idgp4}$ & $\mbox{\tt vns}\atop\mbox{\tt Idgp4var1}$ \\ \hline
{\tt 100d} & NA & 0.076 & 0.208 & {\color{blue}\bf 0.07} & 0.104 & 0.124 & 0.127 & 0.125 & 0.0932 & 0.124 & 0.222 & 0.111 & 0.138 & 0.124 & 0.129 & 0.143 \\
{\tt 1PPT} & NA & 0.0925 & 0.209 & 0.0969 & 0.131 & 0.0742 & 0.0994 & 0.107 & 0.0787 & 0.0769 & 0.227 & {\color{blue}\bf 0.07} & 0.0891 & 0.0974 & 0.13 & 0.164 \\
{\tt 1guu} & NA & 0.000348 & 0.000408 & 0.000106 & 0.00242 & 0.00141 & 0.0132 & 0.0188 & 0.00033 & 0.00144 & 0.00129 & 0.000792 & 8.33e-05 & {\color{blue}\bf 0.00} & 0.0134 & 0.0181 \\
{\tt 1guu-1} & 0.297 & 0.00745 & 0.012 & 0.00467 & 0.00716 & {\color{blue}\bf 0.00} & 0.0178 & 0.0102 & 0.0103 & 0.00345 & 0.00682 & 0.00574 & 0.0118 & 0.00319 & 0.0177 & 0.0214 \\
{\tt 1guu-4000} & 0.285 & 0.00654 & 0.0213 & 0.00401 & 0.0101 & 0.00387 & 0.0201 & 0.0315 & 0.0149 & 0.00704 & 0.0171 & 0.0077 & 0.0216 & {\color{blue}\bf 0.00} & 0.0158 & 0.0242 \\
{\tt 2kxa} & NA & 0.063 & 0.432 & 0.0479 & 0.045 & 0.136 & 0.0711 & 0.0312 & {\color{blue}\bf 0.01} & 0.0285 & 0.34 & 0.085 & 0.0676 & 0.137 & 0.122 & 0.0689 \\
{\tt C0020pdb} & {\color{blue}\bf 0.02} & 0.0376 & 0.148 & 0.0388 & 0.0776 & 0.0464 & 0.0702 & 0.062 & 0.0435 & 0.0248 & 0.144 & 0.0438 & 0.0452 & 0.0527 & 0.079 & 0.0354 \\
{\tt C0030pkl} & 0.0245 & 0.0509 & 0.46 & 0.042 & 0.159 & 0.0374 & 0.104 & 0.0802 & {\color{blue}\bf 0.02} & 0.0456 & 0.25 & 0.0329 & 0.075 & 0.141 & 0.0568 & 0.108 \\
{\tt C0080create.1} & 0.0248 & {\color{blue}\bf 0.00} & {\color{blue}\bf 0.00} & {\color{blue}\bf 0.00} & {\color{blue}\bf 0.00} & 0.00416 & 0.00523 & 0.0049 & {\color{blue}\bf 0.00} & {\color{blue}\bf 0.00} & 0.121 & {\color{blue}\bf 0.00} & {\color{blue}\bf 0.00} & 0.034 & 0.00523 & 0.00516 \\
{\tt C0080create.2} & 0.0248 & {\color{blue}\bf 0.00} & 0.0169 & {\color{blue}\bf 0.00} & {\color{blue}\bf 0.00} & {\color{blue}\bf 0.00} & 0.00523 & 0.00551 & {\color{blue}\bf 0.00} & {\color{blue}\bf 0.00} & 0.0306 & {\color{blue}\bf 0.00} & 0.00387 & 0.0321 & 0.00523 & 0.00486 \\
{\tt C0150alter.1} & 0.0169 & {\color{blue}\bf 0.00} & {\color{blue}\bf 0.00} & {\color{blue}\bf 0.00} & {\color{blue}\bf 0.00} & {\color{blue}\bf 0.00} & 0.00159 & 0.00169 & {\color{blue}\bf 0.00} & {\color{blue}\bf 0.00} & {\color{blue}\bf 0.00} & {\color{blue}\bf 0.00} & {\color{blue}\bf 0.00} & {\color{blue}\bf 0.00} & 0.00159 & 0.0021 \\
{\tt GM1\_sugar} & 0.0187 & {\color{blue}\bf 0.00} & 0.0525 & 0.000893 & 0.015 & {\color{blue}\bf 0.00} & 0.00504 & 0.00552 & 1.92e-06 & 0.0111 & 0.0246 & {\color{blue}\bf 0.00} & 0.00962 & 0.0279 & 0.00507 & 0.00459 \\
{\tt cassioli\dots} & NA & 0.0345 & 0.357 & {\color{blue}\bf 0.00} & 0.0423 & 0.0485 & 0.0553 & 0.0938 & 0.00685 & 0.0383 & 0.191 & 0.0584 & 0.0762 & 0.0797 & 0.0784 & 0.0757 \\
{\tt helix\_amber} & NA & 0.0451 & 0.418 & 0.0516 & 0.0872 & 0.0889 & 0.0523 & 0.12 & 0.0453 & 0.0461 & 0.324 & {\color{blue}\bf 0.01} & 0.0574 & 0.135 & 0.0848 & 0.104 \\
{\tt lavor30\_6-2} & 0.0643 & {\color{blue}\bf 0.00} & {\color{blue}\bf 0.00} & {\color{blue}\bf 0.00} & {\color{blue}\bf 0.00} & {\color{blue}\bf 0.00} & 0.0108 & 0.0157 & {\color{blue}\bf 0.00} & {\color{blue}\bf 0.00} & {\color{blue}\bf 0.00} & {\color{blue}\bf 0.00} & {\color{blue}\bf 0.00} & {\color{blue}\bf 0.00} & 0.011 & 0.0134 \\
{\tt lavor30\_6-3} & 0.0902 & {\color{blue}\bf 0.00} & {\color{blue}\bf 0.00} & {\color{blue}\bf 0.00} & {\color{blue}\bf 0.00} & {\color{blue}\bf 0.00} & 0.0101 & 0.0113 & 0.00312 & {\color{blue}\bf 0.00} & {\color{blue}\bf 0.00} & {\color{blue}\bf 0.00} & {\color{blue}\bf 0.00} & {\color{blue}\bf 0.00} & 0.0101 & 0.00926 \\
{\tt lavor30\_6-4} & 0.139 & {\color{blue}\bf 0.00} & {\color{blue}\bf 0.00} & {\color{blue}\bf 0.00} & {\color{blue}\bf 0.00} & {\color{blue}\bf 0.00} & 0.0103 & 0.0114 & {\color{blue}\bf 0.00} & {\color{blue}\bf 0.00} & {\color{blue}\bf 0.00} & 0.00319 & {\color{blue}\bf 0.00} & {\color{blue}\bf 0.00} & 0.0103 & 0.013 \\
{\tt names} & 0.0238 & 0.00658 & 0.148 & 0.015 & 0.0172 & 0.0566 & 0.0115 & 0.00434 & 0.0142 & {\color{blue}\bf 0.00} & 0.0877 & 0.0212 & 0.0315 & 0.0688 & 0.0349 & 0.00768 \\
{\tt pept} & 0.0419 & 0.0575 & 0.104 & 0.0256 & 0.0685 & 0.0339 & 0.0567 & 0.0629 & 0.0755 & 0.0647 & 0.114 & 0.0175 & 0.0496 & 0.0517 & 0.0329 & {\color{blue}\bf 0.01} \\
{\tt res\_0} & NA & 0.0318 & 0.119 & 0.0248 & 0.0483 & 0.0247 & 0.0727 & 0.0446 & 0.0604 & 0.026 & 0.163 & {\color{blue}\bf 0.01} & 0.0552 & 0.0227 & 0.0482 & 0.0478 \\
{\tt res\_1000} & NA & {\color{blue}\bf 0.02} & 0.109 & 0.0373 & 0.0191 & 0.0602 & 0.0811 & 0.0855 & 0.0424 & 0.018 & 0.191 & 0.0216 & 0.0282 & 0.081 & 0.0666 & 0.0526 \\
{\tt res\_2000} & NA & 0.0119 & 0.177 & 0.0236 & 0.0397 & 0.0669 & 0.0472 & 0.0524 & {\color{blue}\bf 0.00} & 0.0496 & 0.0884 & 0.0322 & 0.0215 & 0.129 & 0.0688 & 0.0472 \\
{\tt res\_2kxa} & NA & 0.0487 & 0.14 & 0.0693 & 0.051 & 0.0448 & 0.0558 & 0.0792 & 0.0377 & 0.0563 & 0.303 & 0.0578 & 0.0963 & 0.0508 & {\color{blue}\bf 0.02} & 0.0202 \\
{\tt res\_3000} & NA & 0.0347 & 0.191 & 0.00338 & 0.0506 & 0.0905 & 0.0539 & 0.0316 & {\color{blue}\bf 0.00} & 0.0335 & 0.0616 & 0.0256 & 0.046 & 0.0471 & 0.0345 & 0.0508 \\
{\tt res\_5000} & NA & 0.0504 & 0.158 & 0.0395 & 0.0316 & 0.0231 & 0.0446 & 0.0582 & {\color{blue}\bf 0.00} & 7.91e-05 & 0.132 & 0.0355 & 0.0676 & 0.121 & 0.0313 & 0.0294 \\
{\tt tiny} & 0.0132 & {\color{blue}\bf 0.00} & {\color{blue}\bf 0.00} & {\color{blue}\bf 0.00} & {\color{blue}\bf 0.00} & {\color{blue}\bf 0.00} & 0.00475 & 0.0047 & {\color{blue}\bf 0.00} & {\color{blue}\bf 0.00} & {\color{blue}\bf 0.00} & {\color{blue}\bf 0.00} & {\color{blue}\bf 0.00} & {\color{blue}\bf 0.00} & 0.00475 & 0.00493 \\
{\tt water} & NA & 0.23 & 0.492 & 0.229 & 0.272 & 0.336 & {\color{blue}\bf 0.13} & 0.232 & 0.227 & 0.215 & 0.438 & 0.212 & 0.215 & 0.291 & 0.163 & 0.242 \\
\hline
{\it Average} & 0.078 & 0.033 & 0.147 & 0.031 & 0.047 & 0.048 & 0.046 & 0.052 & {\color{blue}\bf 0.029} & 0.032 & 0.129 & 0.032 & 0.045 & 0.064 & 0.047 & 0.049 \\ \hline
\end{tabular}

%% file: max.tex
\begin{tabular}{|l||r|r|r|r|r|r|r|r|r|r|r|r|r|r|r|r|r|r|}\hline
{\it Instance} & $\mbox{\tt mosek}\atop\mbox{\tt sdprel}$ & $\mbox{\tt ms}\atop\mbox{\tt Idgp1}$ & $\mbox{\tt ms}\atop\mbox{\tt Idgp1var1}$ & $\mbox{\tt ms}\atop\mbox{\tt Idgp1var2}$ & $\mbox{\tt ms}\atop\mbox{\tt Idgp1var3}$ & $\mbox{\tt ms}\atop\mbox{\tt Idgp3}$ & $\mbox{\tt ms}\atop\mbox{\tt Idgp4}$ & $\mbox{\tt ms}\atop\mbox{\tt Idgp4var1}$ & $\mbox{\tt mwu}\atop\mbox{\tt Imwu}$ & $\mbox{\tt vns}\atop\mbox{\tt Idgp1}$ & $\mbox{\tt vns}\atop\mbox{\tt Idgp1var1}$ & $\mbox{\tt vns}\atop\mbox{\tt Idgp1var2}$ & $\mbox{\tt vns}\atop\mbox{\tt Idgp1var3}$ & $\mbox{\tt vns}\atop\mbox{\tt Idgp3}$ & $\mbox{\tt vns}\atop\mbox{\tt Idgp4}$ & $\mbox{\tt vns}\atop\mbox{\tt Idgp4var1}$ \\ \hline
{\tt 100d} & NA & 3.21 & 2.09 & 3.45 & 3.91 & 2.47 & 4.05 & 3.85 & 3.05 & 4.29 & {\color{blue}\bf 2.05} & 3.9 & 3.7 & 2.58 & 3.62 & 3.81 \\
{\tt 1PPT} & NA & 3.31 & 2.14 & 2.87 & 3.56 & 2.12 & 3.53 & 3.76 & 2.66 & 3.28 & {\color{blue}\bf 1.93} & 4.01 & 3.0 & 2.79 & 3.5 & 3.71 \\
{\tt 1guu} & NA & 0.332 & 0.0167 & 0.0667 & 0.517 & 0.0603 & 0.965 & 1.36 & 0.143 & 0.36 & 0.053 & 0.345 & 0.0629 & {\color{blue}\bf 0.00} & 1.33 & 1.19 \\
{\tt 1guu-1} & 3.2 & 0.394 & 0.233 & 0.45 & 0.813 & {\color{blue}\bf 0.06} & 1.23 & 0.677 & 0.453 & 0.38 & 0.0846 & 0.44 & 1.22 & 0.101 & 0.948 & 1.0 \\
{\tt 1guu-4000} & 2.97 & 1.48 & 0.564 & 0.912 & 1.29 & 0.151 & 1.49 & 1.58 & 0.95 & 1.27 & 0.256 & 1.62 & 1.75 & {\color{blue}\bf 0.08} & 1.0 & 3.03 \\
{\tt 2kxa} & NA & 3.62 & 2.13 & 3.57 & 2.7 & 2.59 & 3.32 & 2.38 & {\color{blue}\bf 1.21} & 3.37 & 2.01 & 2.76 & 3.51 & 3.04 & 3.76 & 3.59 \\
{\tt C0020pdb} & 0.898 & 1.99 & 1.86 & 2.1 & 2.01 & 0.89 & 3.35 & 2.07 & 1.45 & 1.93 & 1.4 & 2.48 & 2.48 & {\color{blue}\bf 0.81} & 3.22 & 2.57 \\
{\tt C0030pkl} & {\color{blue}\bf 1.27} & 3.83 & 2.35 & 2.75 & 3.74 & 1.35 & 3.59 & 4.0 & 1.62 & 3.51 & 2.55 & 3.42 & 3.72 & 1.78 & 3.42 & 3.58 \\
{\tt C0080create.1} & 0.546 & {\color{blue}\bf 0.00} & {\color{blue}\bf 0.00} & {\color{blue}\bf 0.00} & {\color{blue}\bf 0.00} & 0.304 & 0.272 & 0.254 & {\color{blue}\bf 0.00} & {\color{blue}\bf 0.00} & 1.71 & {\color{blue}\bf 0.00} & {\color{blue}\bf 0.00} & 0.953 & 0.272 & 0.314 \\
{\tt C0080create.2} & 0.546 & {\color{blue}\bf 0.00} & 0.406 & {\color{blue}\bf 0.00} & {\color{blue}\bf 0.00} & {\color{blue}\bf 0.00} & 0.272 & 0.211 & {\color{blue}\bf 0.00} & {\color{blue}\bf 0.00} & 0.618 & {\color{blue}\bf 0.00} & 0.911 & 1.04 & 0.272 & 0.313 \\
{\tt C0150alter.1} & 0.443 & {\color{blue}\bf 0.00} & {\color{blue}\bf 0.00} & {\color{blue}\bf 0.00} & {\color{blue}\bf 0.00} & {\color{blue}\bf 0.00} & 0.0856 & 0.176 & {\color{blue}\bf 0.00} & {\color{blue}\bf 0.00} & {\color{blue}\bf 0.00} & {\color{blue}\bf 0.00} & {\color{blue}\bf 0.00} & {\color{blue}\bf 0.00} & 0.0856 & 0.142 \\
{\tt GM1\_sugar} & 0.474 & {\color{blue}\bf 0.00} & 0.505 & 0.32 & 1.54 & {\color{blue}\bf 0.00} & 0.31 & 0.312 & 0.00117 & 1.13 & 0.309 & {\color{blue}\bf 0.00} & 1.19 & 0.635 & 0.55 & 0.419 \\
{\tt cassioli\dots} & NA & 3.16 & 2.59 & {\color{blue}\bf 0.00} & 3.57 & 2.64 & 3.27 & 3.84 & 1.67 & 3.3 & 2.36 & 3.39 & 3.65 & 1.76 & 3.54 & 3.85 \\
{\tt helix\_amber} & NA & 3.39 & 2.69 & 2.99 & 3.57 & 3.65 & 3.76 & 3.56 & 2.77 & 3.52 & {\color{blue}\bf 2.12} & 3.39 & 3.88 & 2.15 & 3.84 & 3.52 \\
{\tt lavor30\_6-2} & 0.959 & {\color{blue}\bf 0.00} & {\color{blue}\bf 0.00} & {\color{blue}\bf 0.00} & {\color{blue}\bf 0.00} & {\color{blue}\bf 0.00} & 0.414 & 0.418 & {\color{blue}\bf 0.00} & {\color{blue}\bf 0.00} & {\color{blue}\bf 0.00} & {\color{blue}\bf 0.00} & {\color{blue}\bf 0.00} & {\color{blue}\bf 0.00} & 0.374 & 0.369 \\
{\tt lavor30\_6-3} & 1.42 & {\color{blue}\bf 0.00} & {\color{blue}\bf 0.00} & {\color{blue}\bf 0.00} & {\color{blue}\bf 0.00} & {\color{blue}\bf 0.00} & 0.4 & 0.307 & 0.314 & {\color{blue}\bf 0.00} & {\color{blue}\bf 0.00} & {\color{blue}\bf 0.00} & {\color{blue}\bf 0.00} & {\color{blue}\bf 0.00} & 0.4 & 0.332 \\
{\tt lavor30\_6-4} & 2.1 & {\color{blue}\bf 0.00} & {\color{blue}\bf 0.00} & {\color{blue}\bf 0.00} & {\color{blue}\bf 0.00} & {\color{blue}\bf 0.00} & 0.362 & 0.368 & {\color{blue}\bf 0.00} & {\color{blue}\bf 0.00} & {\color{blue}\bf 0.00} & 0.314 & {\color{blue}\bf 0.00} & {\color{blue}\bf 0.00} & 0.362 & 0.508 \\
{\tt names} & 0.951 & 0.845 & 1.86 & 0.807 & 1.77 & 2.26 & 1.03 & {\color{blue}\bf 0.29} & 0.928 & 0.295 & 1.28 & 1.7 & 1.44 & 1.51 & 3.09 & 0.978 \\
{\tt pept} & 1.66 & 3.43 & 0.699 & 1.47 & 2.59 & 1.62 & 2.02 & 3.02 & 1.99 & 2.39 & 1.96 & 2.66 & 3.56 & 0.887 & 2.72 & {\color{blue}\bf 0.35} \\
{\tt res\_0} & NA & 2.85 & 1.03 & 2.89 & 3.36 & {\color{blue}\bf 0.82} & 3.02 & 3.2 & 2.19 & 2.47 & 2.16 & 2.47 & 2.1 & 0.824 & 3.0 & 3.11 \\
{\tt res\_1000} & NA & 2.49 & {\color{blue}\bf 1.83} & 2.68 & 2.59 & 1.86 & 2.82 & 3.21 & 2.3 & 2.75 & 2.04 & 2.67 & 2.94 & 2.3 & 3.24 & 2.63 \\
{\tt res\_2000} & NA & 1.42 & 2.26 & 2.79 & 2.16 & 1.2 & 3.19 & 2.58 & {\color{blue}\bf 0.00} & 3.21 & 1.25 & 2.07 & 2.12 & 2.08 & 2.62 & 2.22 \\
{\tt res\_2kxa} & NA & 3.29 & 2.07 & 3.3 & 2.81 & 2.26 & 2.91 & 3.17 & 2.53 & 3.19 & 2.42 & 3.35 & 3.68 & {\color{blue}\bf 1.65} & 2.63 & 2.72 \\
{\tt res\_3000} & NA & 2.8 & 1.35 & 0.867 & 3.28 & 1.78 & 2.91 & 3.17 & {\color{blue}\bf 0.00} & 2.73 & 1.08 & 1.87 & 2.9 & 1.8 & 3.09 & 3.55 \\
{\tt res\_5000} & NA & 3.39 & 2.06 & 3.66 & 2.1 & 1.16 & 2.8 & 2.81 & {\color{blue}\bf 0.00} & 0.11 & 1.78 & 2.77 & 2.55 & 1.72 & 2.32 & 2.68 \\
{\tt tiny} & 0.306 & {\color{blue}\bf 0.00} & {\color{blue}\bf 0.00} & {\color{blue}\bf 0.00} & {\color{blue}\bf 0.00} & {\color{blue}\bf 0.00} & 0.334 & 0.318 & {\color{blue}\bf 0.00} & {\color{blue}\bf 0.00} & {\color{blue}\bf 0.00} & {\color{blue}\bf 0.00} & {\color{blue}\bf 0.00} & {\color{blue}\bf 0.00} & 0.334 & 0.385 \\
{\tt water} & NA & 4.13 & 2.66 & 4.17 & 4.46 & 4.9 & 4.03 & 3.86 & 3.77 & 3.82 & {\color{blue}\bf 2.57} & 4.3 & 3.82 & 4.1 & 4.23 & 3.86 \\
\hline
{\it Average} & 1.267 & 1.828 & 1.237 & 1.560 & 1.939 & 1.265 & 2.064 & 2.028 & {\color{blue}\bf 1.111} & 1.752 & 1.259 & 1.849 & 2.007 & 1.281 & 2.140 & 2.027 \\ \hline
\end{tabular}

%% file: cpu.tex
\begin{tabular}{|l||r|r|r|r|r|r|r|r|r|r|r|r|r|r|r|r|r|r|}\hline
{\it Instance} & $\mbox{\tt mosek}\atop\mbox{\tt sdprel}$ & $\mbox{\tt ms}\atop\mbox{\tt Idgp1}$ & $\mbox{\tt ms}\atop\mbox{\tt Idgp1var1}$ & $\mbox{\tt ms}\atop\mbox{\tt Idgp1var2}$ & $\mbox{\tt ms}\atop\mbox{\tt Idgp1var3}$ & $\mbox{\tt ms}\atop\mbox{\tt Idgp3}$ & $\mbox{\tt ms}\atop\mbox{\tt Idgp4}$ & $\mbox{\tt ms}\atop\mbox{\tt Idgp4var1}$ & $\mbox{\tt mwu}\atop\mbox{\tt Imwu}$ & $\mbox{\tt vns}\atop\mbox{\tt Idgp1}$ & $\mbox{\tt vns}\atop\mbox{\tt Idgp1var1}$ & $\mbox{\tt vns}\atop\mbox{\tt Idgp1var2}$ & $\mbox{\tt vns}\atop\mbox{\tt Idgp1var3}$ & $\mbox{\tt vns}\atop\mbox{\tt Idgp3}$ & $\mbox{\tt vns}\atop\mbox{\tt Idgp4}$ & $\mbox{\tt vns}\atop\mbox{\tt Idgp4var1}$ \\ \hline
{\tt 100d}         & NA & 612.0 & {\color{blue}\bf 63.30} & 312.0 & 375.0 & 345.0 & 307.0 & 319.0 & 8230.0 & 316.0 & 65.2 & 359.0 & 290.0 & 1010.0 & 441.0 & 188.0 \\
{\tt 1PPT}         & NA & 87.1 & {\color{blue}\bf 25.90} & 52.4 & 70.8 & 926.0 & 55.8 & 61.0 & 2292.1 & 54.9 & 31.7 & 62.0 & 77.6 & 262.0 & 98.3 & 95.7 \\
{\tt 1guu}         & NA & 21.1 & 20.1 & 22.4 & 20.6 & 26.6 & 20.6 & 20.3 & 64.3 & 10.9 & 20.4 & 20.9 & 20.5 & {\color{blue}\bf 6.03} & 21.5 & 21.3 \\
{\tt 1guu-1}       & 1020.0 & 22.1 & {\color{blue}\bf 20.20} & 20.7 & 20.6 & 21.6 & 20.7 & 20.6 & 70.63 & 21.1 & 20.3 & 22.2 & 20.5 & {\color{blue}\bf 20.20} & {\color{blue}\bf 20.20} & {\color{blue}\bf 20.20} \\
{\tt 1guu-4000}    & 816.0 & 21.1 & 20.4 & 22.5 & 21.9 & 21.9 & 22.4 & 20.6 & 84.66 & 21.1 & 20.4 & 22.7 & 21.0 & 22.9 & 20.9 & {\color{blue}\bf 20.10} \\
{\tt 2kxa}         & NA & 26.6 & {\color{blue}\bf 21.90} & 29.7 & 31.7 & 227.0 & 36.2 & 32.1 & 2018.66 & 101.0 & 28.6 & 29.8 & 30.5 & 74.3 & 36.2 & 50.3 \\
{\tt C0020pdb}     & 135.0 & 22.0 & 20.2 & 23.8 & {\color{blue}\bf 20.00} & 25.3 & 24.8 & 22.0 & 198.66 & 22.0 & 21.2 & 20.8 & 20.4 & 20.4 & 20.8 & 24.8 \\
{\tt C0030pkl}     & 4440.0 & 60.8 & 26.1 & 51.4 & 47.3 & 69.8 & {\color{blue}\bf 23.60} & 54.0 & 3108.14 & 51.4 & 25.0 & 57.7 & 85.5 & 72.8 & 43.5 & 27.0 \\
{\tt C0080create.1}& 12.7 & 4.52 & {\color{blue}\bf 1.22} & 2.28 & 2.91 & 21.4 & 20.2 & 20.4 & 55.31 & 17.1 & 2.99 & 2.81 & 3.35 & 21.1 & 20.2 & 20.1 \\
{\tt C0080create.2}& 12.7 & 3.95 & 20.1 & 8.1 & 4.39 & 3.92 & 20.5 & 20.3 & 51.90 & 18.3 & 20.2 & {\color{blue}\bf 2.14} & 7.24 & 20.6 & 20.4 & 20.1 \\
{\tt C0150alter.1} & 4.32 & 2.76 & 1.75 & 1.28 & {\color{blue}\bf 0.67} & 1.85 & 20.2 & 20.2 & 17.31 & 7.72 & 1.47 & 0.888 & 7.66 & 12.3 & 7.78 & 5.57 \\
{\tt GM1\_sugar}   & 15.6 & 13.8 & 20.4 & 20.7 & 21.7 & 9.15 & 21.3 & 20.1 & 53.28 & {\color{blue}\bf 7.83} & 20.1 & 20.8 & 20.0 & 20.1 & 20.0 & 20.4 \\
{\tt cassioli\dots}& NA & 120.0 & {\color{blue}\bf 33.60} & 171.0 & 121.0 & 572.0 & 175.0 & 88.3 & 7842.0 & 186.0 & 34.2 & 104.0 & 100.0 & 67.5 & 87.5 & 81.1 \\
{\tt helix\_amber} & NA & 210.0 & {\color{blue}\bf 80.30} & 193.0 & 147.0 & 267.0 & 168.0 & 112.0 & 12628.8 & 222.0 & 101.0 & 259.0 & 209.0 & 157.0 & 221.0 & 91.7 \\
{\tt lavor30\_6-2} & 3.04 & 1.39 & 1.36 & 2.12 & 1.36 & 1.51 & 20.0 & 20.0 & 7.82 & 0.444 & 2.32 & {\color{blue}\bf 0.34} & 0.427 & 9.98 & 4.08 & 2.52 \\
{\tt lavor30\_6-3} & 3.11 & 2.87 & 0.8 & 1.7 & 0.962 & {\color{blue}\bf 0.68} & 20.0 & 20.0 & 7.18 & 3.62 & 1.84 & 3.14 & 2.6 & 11.9 & 2.61 & 2.08 \\
{\tt lavor30\_6-4} & 3.05 & 1.34 & 0.653 & 1.45 & {\color{blue}\bf 0.31} & 1.94 & 20.0 & 20.0 & 7.46 & 1.59 & 2.67 & 1.55 & 6.88 & 0.332 & 0.412 & 0.37 \\
{\tt names}        & 44.1 & 20.4 & 21.9 & 21.7 & 21.4 & 23.5 & 21.7 & 20.5 & 104.0 & 22.9 & {\color{blue}\bf 4.59} & 15.4 & 22.1 & 37.4 & 20.3 & 20.9 \\
{\tt pept}         & 105.0 & 22.7 & 21.0 & 24.9 & 22.9 & 22.2 & 22.7 & 20.5 & 171.2 & {\color{blue}\bf 17.40} & 21.0 & 23.1 & 20.2 & 20.1 & 20.4 & 20.3 \\
{\tt res\_0}       & NA & 20.3 & 21.6 & 23.0 & 22.0 & 27.1 & 26.4 & 20.4 & 263.51 & 22.6 & 21.0 & 20.7 & 21.8 & 24.0 & {\color{blue}\bf 10.10} & 21.5 \\
{\tt res\_1000}    & NA & 23.6 & 23.8 & 22.6 & 23.0 & 21.0 & 21.1 & 25.3 & 342.46 & 23.3 & 21.2 & 22.4 & 21.5 & 23.6 & 21.7 & {\color{blue}\bf 20.60} \\
{\tt res\_2000}    & NA & 23.0 & 20.8 & 26.3 & 20.6 & 77.6 & 21.3 & 20.7 & 245.08 & 22.5 & 20.5 & 22.4 & 20.5 & 90.8 & {\color{blue}\bf 20.10} & 21.4 \\
{\tt res\_2kxa}    & NA & 28.1 & {\color{blue}\bf 20.30} & 30.5 & 33.0 & 43.9 & 30.2 & 26.4 & 1614.6 & 42.6 & 24.5 & 28.5 & 37.0 & 31.3 & 26.7 & 36.4 \\
{\tt res\_3000}    & NA & 24.0 & 20.7 & 24.0 & 23.5 & 28.4 & 20.6 & 29.2 & 270.56 & 20.7 & 22.4 & 22.2 & 26.5 & 21.0 & {\color{blue}\bf 20.00} & {\color{blue}\bf 20.00} \\
{\tt res\_5000}    & NA & 22.0 & 20.4 & 22.8 & 21.0 & 21.8 & 21.0 & 21.0 & 224.83 & 24.7 & 20.6 & {\color{blue}\bf 20.20} & 22.0 & 20.4 & 22.8 & 21.1 \\
{\tt tiny}         & 4.89 & 2.51 & 2.11 & 0.813 & 4.45 & 3.19 & 20.0 & 20.1 & 18.16 & 0.772 & 3.9 & 0.686 & 5.4 & {\color{blue}\bf 0.50} & 5.51 & 0.821 \\
{\tt water}        & NA & 1530.0 & 937.0 & 1800.0 & 1800.0 & 1800.0 & 1800.0 & 1520.0 & 38597.0 & 1800.0 & {\color{blue}\bf 542.00} & 1800.0 & 1800.0 & 1890.0 & 1600.0 & 1800.0 \\
\hline
{\it Average}      & 472.822 & 109.261 & 55.107 & 108.635 & 107.409 & 170.790 & 111.159 & 96.852 & 2910.0 & 113.351 & {\color{blue}\bf 41.529} & 109.828 & 108.154 & 146.983 & 105.703 & 99.050 \\ \hline
\end{tabular}

%% file: demi.tex
\begin{tabular}{|l||r|r|r|r|r|r|r|r|r|r|r|r|r|r|r|r|r|}\hline
{\it Instance} & $\mbox{\tt mosek}\atop\mbox{\tt sdprel}$ & $\mbox{\tt ms}\atop\mbox{\tt Idgp1}$ & $\mbox{\tt ms}\atop\mbox{\tt Idgp1var1}$ & $\mbox{\tt ms}\atop\mbox{\tt Idgp1var2}$ & $\mbox{\tt ms}\atop\mbox{\tt Idgp1var3}$ & $\mbox{\tt ms}\atop\mbox{\tt Idgp3}$ & $\mbox{\tt ms}\atop\mbox{\tt Idgp4}$ & $\mbox{\tt ms}\atop\mbox{\tt Idgp4var1}$ & $\mbox{\tt mwu}\atop\mbox{\tt Imwu}$ & $\mbox{\tt vns}\atop\mbox{\tt Idgp1}$ & $\mbox{\tt vns}\atop\mbox{\tt Idgp1var1}$ & $\mbox{\tt vns}\atop\mbox{\tt Idgp1var2}$ & $\mbox{\tt vns}\atop\mbox{\tt Idgp1var3}$ & $\mbox{\tt vns}\atop\mbox{\tt Idgp3}$ & $\mbox{\tt vns}\atop\mbox{\tt Idgp4}$ & $\mbox{\tt vns}\atop\mbox{\tt Idgp4var1}$ \\ \hline
{\tt 1guu-1} & 132.083 & 116.038 & 163.411 & 145.938 & 158.154 & 256.755 & 127.979 & 196.529 & 132.709 & 121.270 & 112.595 & 248.200 & 242.633 & 313.398 & 182.382 & {\color{blue}\bf 98.386} \\
{\tt 1guu} & 228.757 & 202.573 & 210.966 & 181.700 & 138.726 & 209.438 & 184.057 & 161.057 & 153.964 & 135.907 & 218.604 & 172.713 & {\color{blue}\bf 126.893} & 215.186 & 201.001 & 242.788 \\
{\tt 2erl-frag-bp1} & 4.203 & {\color{blue}\bf 0.205} & 26.066 & 26.118 & 26.119 & 26.119 & 2.841 & 2.302 & 0.208 & 26.119 & 0.210 & 0.221 & 10.312 & 16.053 & 2.841 & 3.403 \\
{\tt 2kxa} & {\color{blue}\bf 7.954} & 104.331 & 109.387 & 59.612 & 107.329 & 74.355 & 138.700 & 64.190 & 15.169 & 51.538 & 98.227 & 152.011 & 104.147 & 82.440 & 101.206 & 106.075 \\
{\tt C0150alter.1} & 2.105 & 0.511 & 0.403 & 3.412 & 0.610 & 0.522 & 3.028 & 2.158 & 0.564 & 2.598 & 0.246 & 5.464 & {\color{blue}\bf 0.240} & 0.250 & 3.028 & 2.910 \\
{\tt C0700.odd.G} & 5.690 & 38.839 & 15.052 & 27.131 & 15.474 & 15.159 & 9.485 & 7.726 & 35.427 & 23.890 & {\color{blue}\bf 0.848} & 7.142 & 15.183 & 2.549 & 9.485 & 4.876 \\
{\tt C0700.odd.H} & 5.690 & 0.520 & 7.389 & 7.809 & 7.391 & 15.156 & 9.485 & 29.485 & 9.005 & 27.733 & {\color{blue}\bf 0.422} & 25.227 & 0.462 & 17.623 & 9.485 & 10.781 \\
{\tt C0700odd.C} & 5.566 & 38.284 & 26.107 & 49.372 & 26.033 & {\color{blue}\bf 2.191} & 20.498 & 28.260 & 18.467 & 25.222 & 19.999 & 58.306 & 24.686 & 24.708 & 20.498 & 27.363 \\
{\tt C0700odd.D} & 5.566 & {\color{blue}\bf 1.795} & 24.602 & 40.674 & 20.131 & 21.033 & 30.645 & 33.841 & 26.072 & 21.885 & 19.998 & 36.257 & 19.982 & 1.799 & 30.645 & 26.527 \\
{\tt C0700odd.E} & 5.566 & 40.742 & {\color{blue}\bf 3.314} & 57.546 & 19.257 & 26.192 & 30.645 & 31.231 & 21.049 & 19.770 & 4.451 & 23.424 & 24.777 & 24.647 & 20.498 & 13.691 \\
{\tt lavor11\_7-1} & 5.970 & 4.326 & 0.067 & 6.800 & 3.813 & 0.057 & 3.232 & 1.886 & 4.274 & 2.203 & {\color{blue}\bf 0.047} & 4.267 & 0.109 & 0.056 & 3.232 & 2.752 \\
{\tt lavor11\_7-2} & 6.509 & {\color{blue}\bf 0.066} & 0.076 & 0.649 & 1.154 & 0.079 & 2.924 & 2.901 & 7.366 & 0.084 & 0.080 & 0.485 & 0.348 & 0.075 & 2.924 & 1.048 \\
{\tt lavor11\_7-b} & 5.970 & 4.274 & 5.905 & 3.730 & 4.273 & {\color{blue}\bf 0.058} & 3.232 & 3.237 & 4.277 & 0.379 & 0.681 & 1.487 & 0.603 & 0.096 & 3.232 & 2.874 \\
{\tt lavor11\_7} & 5.970 & 0.194 & 0.072 & 0.064 & 0.215 & 2.130 & 3.232 & 3.538 & 6.555 & 0.076 & {\color{blue}\bf 0.053} & 4.028 & 0.064 & 0.056 & 3.232 & 3.653 \\
{\tt lavor11} & 5.103 & 0.135 & 0.128 & 7.131 & 4.966 & 0.942 & 2.960 & 7.190 & 6.642 & {\color{blue}\bf 0.077} & 6.634 & 5.603 & 6.638 & 0.136 & 2.960 & 5.932 \\
{\tt lavor30\_6-1} & 32.441 & 32.157 & 13.647 & 30.050 & 13.649 & 14.926 & 21.203 & 24.563 & 26.190 & 14.949 & {\color{blue}\bf 2.286} & 35.156 & 16.135 & 15.448 & 21.203 & 22.334 \\
{\tt lavor30\_6-2} & 28.787 & 8.366 & 10.833 & 1.565 & 0.253 & 11.120 & 25.772 & 24.558 & 29.555 & 28.764 & 10.835 & {\color{blue}\bf 0.241} & 17.135 & 0.243 & 12.306 & 8.709 \\
{\tt lavor30\_6-3} & 37.932 & 42.267 & 16.883 & 34.249 & 30.305 & 34.248 & 41.663 & 7.342 & 55.839 & 41.242 & 16.878 & 38.034 & {\color{blue}\bf 2.153} & 16.082 & 41.663 & 34.898 \\
{\tt lavor30\_6-4} & 16.717 & 5.142 & 18.835 & 12.490 & 4.558 & 7.403 & 5.436 & 4.149 & 18.840 & {\color{blue}\bf 1.045} & 15.509 & 16.838 & 15.494 & 10.353 & 5.436 & 19.329 \\
{\tt lavor30\_6-5} & 22.715 & 31.712 & 15.159 & 25.267 & 1.847 & 26.029 & 26.834 & 26.800 & 44.336 & 23.498 & {\color{blue}\bf 1.838} & 15.800 & 13.133 & 16.565 & 26.834 & 28.453 \\
{\tt lavor30\_6-7} & 32.540 & 41.196 & {\color{blue}\bf 2.035} & 41.597 & 2.443 & 32.357 & 29.950 & 17.784 & 52.537 & 42.098 & 40.102 & 41.153 & {\color{blue}\bf 2.035} & 20.193 & 29.950 & 43.134 \\
{\tt lavor30\_6-8} & 10.389 & 44.385 & {\color{blue}\bf 2.041} & 42.796 & 44.901 & 12.187 & 45.894 & 48.735 & 44.294 & 44.276 & 2.136 & 24.175 & 2.086 & 29.897 & 45.894 & 46.677 \\
{\tt mdgp4-heuristic} & 0.460 & 0.894 & 0.190 & 0.120 & 0.557 & 0.116 & 0.460 & 0.460 & 0.445 & 0.338 & 0.345 & {\color{blue}\bf 0.037} & 0.655 & 0.119 & 0.460 & 0.460 \\
{\tt mdgp4-optimal} & 1.423 & 0.054 & 0.066 & 0.037 & 0.038 & 0.026 & 0.660 & 0.660 & 0.707 & 0.463 & 0.224 & 0.033 & 0.033 & {\color{blue}\bf 0.025} & 0.660 & 0.660 \\
{\tt res\_0} & {\color{blue}\bf 7.933} & 206.560 & 149.633 & 120.930 & 157.351 & 46.055 & 101.755 & 92.769 & 75.056 & 136.851 & 137.643 & 149.057 & 131.249 & 196.041 & 79.722 & 136.753 \\
{\tt res\_1000} & {\color{blue}\bf 8.151} & 24.677 & 82.179 & 99.962 & 170.710 & 122.709 & 114.645 & 142.391 & 99.926 & 46.920 & 119.482 & 85.413 & 38.340 & 77.778 & 121.924 & 60.485 \\
{\tt res\_2000} & 13.866 & 202.930 & 125.860 & 130.671 & 176.975 & 97.516 & 139.965 & 111.010 & {\color{blue}\bf 0.732} & 41.158 & 115.147 & 95.563 & 175.189 & 86.838 & 134.238 & 118.871 \\
{\tt res\_2kxa} & {\color{blue}\bf 11.008} & 106.131 & 72.845 & 94.618 & 82.791 & 145.808 & 83.657 & 90.800 & 149.216 & 110.327 & 90.640 & 64.137 & 118.709 & 104.359 & 38.168 & 35.297 \\
{\tt res\_3000} & {\color{blue}\bf 9.294} & 144.285 & 106.614 & 48.903 & 93.328 & 82.502 & 80.058 & 38.559 & 9.557 & 117.402 & 44.590 & 134.721 & 70.533 & 121.108 & 36.592 & 165.878 \\
{\tt res\_5000} & 7.633 & 101.283 & 122.206 & 105.803 & 51.874 & 100.349 & 99.724 & 102.613 & {\color{blue}\bf 6.202} & 44.859 & 125.495 & 99.729 & 61.313 & 103.984 & 44.655 & 82.487 \\
{\tt small02} & 5.566 & 31.324 & 14.138 & 23.329 & 23.036 & 25.506 & 20.498 & 28.181 & {\color{blue}\bf 2.307} & 57.688 & 24.687 & 62.995 & 24.689 & 23.558 & 30.645 & 17.314 \\
{\tt tiny} & 2.397 & 17.715 & 5.758 & 18.174 & 2.189 & 18.246 & 2.651 & 3.720 & 21.402 & 4.406 & 8.322 & 11.334 & 7.218 & {\color{blue}\bf 1.864} & 2.651 & 7.964 \\
\hline
{\it Averages} & {\color{blue}\bf 681.954} & 1593.911 & 1351.867 & 1448.247 & 1390.450 & 1427.289 & 1413.768 & 1340.625 & {\color{purple}\bf 1078.889} & 1215.035 & 1239.254 & 1619.251 & 1273.176 & 1523.527 & 1269.650 & 1382.762 \\ \hline
\end{tabular}

%% file: zoo-techrep.bbl
\begin{thebibliography}{10}

\bibitem{wolkowicz}
A.~Alfakih, A.~Khandani, and H.~Wolkowicz.
\newblock Solving {E}uclidean distance matrix completion problems via
  semidefinite programming.
\newblock {\em Computational Optimization and Applications}, 12:13--30, 1999.

\bibitem{altmehl}
H.~Alt, K.~Mehlhorn, H.~Wagener, and E.~Welzl.
\newblock Congruence, similarity and symmetries of geometric objects.
\newblock {\em Discrete Computational Geometry}, 3:237--256, 1988.

\bibitem{kalesdp}
S.~Arora, E.~Hazan, and S.~Kale.
\newblock Fast algorithms for approximate semidefinite programming using the
  multiplicative weights update method.
\newblock In {\em Foundations of Computer Science}, volume~46 of {\em FOCS},
  pages 339--348. IEEE, 2005.

\bibitem{kale}
S.~Arora, E.~Hazan, and S.~Kale.
\newblock The multiplicative weights update method: a meta-algorithm and
  applications.
\newblock {\em Theory of Computing}, 8:121--164, 2012.

\bibitem{atkinson}
M.~Atkinson.
\newblock An optimal algorithm for geometrical congruence.
\newblock {\em Journal of Algorithms}, 8:159--172, 1987.

\bibitem{bahr}
A.~Bahr, J.~Leonard, and M.~Fallon.
\newblock Cooperative localization for autonomous underwater vehicles.
\newblock {\em International Journal of Robotics Research}, 28(6):714--728,
  2009.

\bibitem{pollack}
S.~Basu, R.~Pollack, and M.-F. Roy.
\newblock {\em Algorithms in real algebraic geometry}.
\newblock Springer, New York, 2006.

\bibitem{dgpinnp}
N.~Beeker, S.~Gaubert, C.~Glusa, and L.~Liberti.
\newblock Is the distance geometry problem in {{\bf NP}}?
\newblock In Mucherino et~al. \cite{dgpbook}.

\bibitem{couenne}
P.~Belotti, J.~Lee, L.~Liberti, F.~Margot, and A.~W\"achter.
\newblock Branching and bounds tightening techniques for non-convex {MINLP}.
\newblock {\em Optimization Methods and Software}, 24(4):597--634, 2009.

\bibitem{benedetti}
R.~Benedetti and J.-J. Risler.
\newblock {\em Real algebraic and semi-algebraic sets}.
\newblock Hermann, Paris, 1990.

\bibitem{pdb}
H.~Berman, J.~Westbrook, Z.~Feng, G.~Gilliland, T.~Bhat, H.~Weissig, I.N.
  Shindyalov, and P.~Bourne.
\newblock The protein data bank.
\newblock {\em Nucleic Acid Research}, 28:235--242, 2000.

\bibitem{biswasphd}
P.~Biswas.
\newblock {\em Semidefinite programming approaches to distance geometry
  problems}.
\newblock PhD thesis, Stanford University, 2007.

\bibitem{biswasacm}
P.~Biswas, T.~Lian, T.~Wang, and Y.~Ye.
\newblock Semidefinite programming based algorithms for sensor network
  localization.
\newblock {\em ACM Transactions in Sensor Networks}, 2:188--220, 2006.

\bibitem{biswas2006ieee}
P.~Biswas, T.-C. Liang, K.-C. Toh, T.-C. Wang, and Y.~Ye.
\newblock Semidefinite programming approaches for sensor network localization
  with noisy distance measurements.
\newblock {\em IEEE Transactions on Automation Science and Engineering},
  3:360--371, 2006.

\bibitem{phaselift}
E.~Cand\`es, T.~Strohmer, and V.~Voroninski.
\newblock Phase{L}ift: {E}xact and stable signal recovery from magniture
  measurements via convex programming.
\newblock {\em Communications on Pure and Applied Mathematics},
  66(8):1241--1274, 2012.

\bibitem{bipbip}
A.~Cassioli, B.~Bordeaux, G.~Bouvier, A.~Mucherino, R.~Alves, L.~Liberti,
  M.~Nilges, C.~Lavor, and T.~Malliavin.
\newblock An algorithm to enumerate all possible protein conformations
  verifying a set of distance constraints.
\newblock {\em BMC Bioinformatics}, page 16:23, 2015.

\bibitem{orders-dam}
A.~Cassioli, O.~G\"unl\"uk, C.~Lavor, and L.~Liberti.
\newblock Discretization vertex orders for distance geometry.
\newblock {\em Discrete Applied Mathematics}, 197:27--41, 2015.

\bibitem{ipopt}
COIN-OR.
\newblock {\em Introduction to IPOPT: A tutorial for downloading, installing,
  and using IPOPT}, 2006.

\bibitem{coutsias}
E.~Coutsias, C.~Seok, and K.~Dill.
\newblock Using quaternions to calculate rmsd.
\newblock {\em Journal of Computational Chemistry}, 25(15):1849--1857, 2004.

\bibitem{mago14}
C.~D'Ambrosio, Vu~Khac Ky, C.~Lavor, L.~Liberti, and N.~Maculan.
\newblock Computational experience on distance geometry problems 2.0.
\newblock In L.~Casado, I.~Garcia, and E.~Hendrix, editors, {\em Mathematical
  and applied Global Optimization}, volume XII of {\em Global Optimization
  Workshop}, pages 97--100, Malaga, 2014. University of Malaga.

\bibitem{snl}
Y.~Ding, N.~Krislock, J.~Qian, and H.~Wolkowicz.
\newblock Sensor network localization, {E}uclidean distance matrix completions,
  and graph realization.
\newblock {\em Optimization and Engineering}, 11:45--66, 2010.

\bibitem{splogic}
H.~Du, N.~Alechina, K.~Stock, and M.~Jackson.
\newblock The logic of {NEAR} and {FAR}.
\newblock In T.~Tenbrink et~al., editor, {\em COSIT}, volume 8116 of {\em
  LNCS}, pages 475--494, Switzerland, 2013. Springer.

\bibitem{ampl}
R.~Fourer and D.~Gay.
\newblock {\em The {AMPL} Book}.
\newblock Duxbury Press, Pacific Grove, 2002.

\bibitem{goodall}
C.~Goodall.
\newblock Procrustes methods in the statistical analysis of shape.
\newblock {\em Journal of the Royal Statistical Society B}, 53(2):285--339,
  1991.

\bibitem{henneberg1911}
L.~Henneberg.
\newblock {\em Die {G}raphische {S}tatik der starren {S}ysteme}.
\newblock Teubner, Leipzig, 1911.

\bibitem{Lav05}
C.~Lavor.
\newblock On generating instances for the molecular distance geometry problem.
\newblock In L.~Liberti and N.~Maculan, editors, {\em Global Optimization: from
  Theory to Implementation}, pages 405--414. Springer, Berlin, 2006.

\bibitem{cliffordalgebra}
C.~Lavor, R.~Alves, W.~Figuereido, A.~Petraglia, and N.~Maculan.
\newblock Clifford algebra and the discretizable molecular distance geometry
  problem.
\newblock {\em Advances in Applied Clifford Algebras}, 25:925--942, 2015.

\bibitem{dvop}
C.~Lavor, J.~Lee, A.~{Lee-St.~John}, L.~Liberti, A.~Mucherino, and
  M.~Sviridenko.
\newblock Discretization orders for distance geometry problems.
\newblock {\em Optimization Letters}, 6:783--796, 2012.

\bibitem{dmdgp}
C.~Lavor, L.~Liberti, N.~Maculan, and A.~Mucherino.
\newblock The discretizable molecular distance geometry problem.
\newblock {\em Computational Optimization and Applications}, 52:115--146, 2012.

\bibitem{bpinterval}
C.~Lavor, L.~Liberti, and A.~Mucherino.
\newblock The {\it interval} {Branch-and-Prune} algorithm for the discretizable
  molecular distance geometry problem with inexact distances.
\newblock {\em Journal of Global Optimization}, 56:855--871, 2013.

\bibitem{refmathprog}
L.~Liberti.
\newblock Reformulations in mathematical programming: Definitions and
  systematics.
\newblock {\em RAIRO-RO}, 43(1):55--86, 2009.

\bibitem{vnssolver}
L.~Liberti and M.~Dra\v{z}ic.
\newblock Variable neighbourhood search for the global optimization of
  constrained {NLP}s.
\newblock In {\em Proceedings of GO Workshop, Almeria, Spain}, 2005.

\bibitem{ios14slides}
L.~Liberti and C.~Lavor.
\newblock Solving large-scale distance geometry problems exactly versus
  approximately.
\newblock In {\em Optimization Society}, Proceedings of the Annual Conference,
  Houston, 2014. INFORMS.

\bibitem{liberti-gsi13}
L.~Liberti, C.~Lavor, J.~Alencar, and G.~Abud.
\newblock Counting the number of solutions of ${}^k${DMDGP} instances.
\newblock In F.~Nielsen and F.~Barbaresco, editors, {\em Geometric Science of
  Information}, volume 8085 of {\em LNCS}, pages 224--230, New York, 2013.
  Springer.

\bibitem{lln5}
L.~Liberti, C.~Lavor, and N.~Maculan.
\newblock A branch-and-prune algorithm for the molecular distance geometry
  problem.
\newblock {\em International Transactions in Operational Research}, 15:1--17,
  2008.

\bibitem{dgp-sirev}
L.~Liberti, C.~Lavor, N.~Maculan, and A.~Mucherino.
\newblock Euclidean distance geometry and applications.
\newblock {\em SIAM Review}, 56(1):3--69, 2014.

\bibitem{bppolybook}
L.~Liberti, C.~Lavor, and A.~Mucherino.
\newblock The discretizable molecular distance geometry problem seems easier on
  proteins.
\newblock In Mucherino et~al. \cite{dgpbook}.

\bibitem{mdgpsurvey}
L.~Liberti, C.~Lavor, A.~Mucherino, and N.~Maculan.
\newblock Molecular distance geometry methods: from continuous to discrete.
\newblock {\em International Transactions in Operational Research}, 18:33--51,
  2010.

\bibitem{powerof2}
L.~Liberti, B.~Masson, C.~Lavor, J.~Lee, and A.~Mucherino.
\newblock On the number of realizations of certain {H}enneberg graphs arising
  in protein conformation.
\newblock {\em Discrete Applied Mathematics}, 165:213--232, 2014.

\bibitem{mwa_minlp_working}
L.~Liberti and L.~Mencarelli.
\newblock A multiplicative weights update algorithm for {MINLP}, 2014.
\newblock Working paper.

\bibitem{recipe}
L.~Liberti, N.~Mladenovi\'c, and G.~Nannicini.
\newblock A recipe for finding good solutions to {MINLP}s.
\newblock {\em Mathematical Programming Computation}, 3:349--390, 2011.

\bibitem{yalmip}
J.~L\"ofberg.
\newblock {YALMIP}: {A} toolbox for modeling and optimization in {MATLAB}.
\newblock In {\em Proceedings of the International Symposium of Computer-Aided
  Control Systems Design}, volume~1 of {\em CACSD}, Taipei, 2004. IEEE.

\bibitem{crippen-rmsd}
V.~Maiorov and G.~Crippen.
\newblock Significance of root-mean-square deviation in comparing
  three-dimensional structures of globular proteins.
\newblock {\em Journal of Molecular Biology}, 235:625--634, 1994.

\bibitem{muchbook}
T.~Malliavin, A.~Mucherino, and M.~Nilges.
\newblock Distance geometry in structural biology.
\newblock In Mucherino et~al. \cite{dgpbook}.

\bibitem{matlab}
The MathWorks, Inc., Natick, MA.
\newblock {\em MATLAB R2014a}, 2014.

\bibitem{milnor}
J.~Milnor.
\newblock {\em Topology from the differentiable viewpoint}.
\newblock University Press of Virginia, Charlottesville, 1969.

\bibitem{morewu2}
J.~Mor\'e and Z.~Wu.
\newblock Distance geometry optimization for protein structures.
\newblock {\em Journal of Global Optimization}, 15:219--234, 1999.

\bibitem{mosek7}
Mosek ApS.
\newblock {\em The {\tt mosek} manual, Version 7 (Revision 114)}, 2014.
\newblock \verb+(www.mosek.com)+.

\bibitem{dgpbook}
A.~Mucherino, C.~Lavor, L.~Liberti, and N.~Maculan, editors.
\newblock {\em Distance Geometry: Theory, Methods, and Applications}.
\newblock Springer, New York, 2013.

\bibitem{plotkin}
S.~Plotkin, D.~Shmoys, and \'E. Tardos.
\newblock Fast approximation algorithm for fractional packing and covering
  problems.
\newblock {\em Mathematics of Operations Research}, 20:257--301, 1995.

\bibitem{saxe79}
J.~Saxe.
\newblock Embeddability of weighted graphs in $k$-space is strongly {{\bf
  NP}}-hard.
\newblock {\em Proceedings of 17th Allerton Conference in Communications,
  Control and Computing}, pages 480--489, 1979.

\bibitem{singer4}
A.~Singer.
\newblock Angular synchronization by eigenvectors and semidefinite programming.
\newblock {\em Applied and Computational Harmonic Analysis}, 30:20--36, 2011.

\bibitem{tay-whiteley}
T.-S. Tay and W.~Whiteley.
\newblock Generating isostatic frameworks.
\newblock {\em Structural Topology}, 11:21--69, 1985.

\bibitem{waechter}
A.~W\"achter and L.~Biegler.
\newblock On the implementation of an interior-point filter line-search
  algorithm for large-scale nonlinear programming.
\newblock {\em Mathematical Programming}, 106(1):25--57, 2006.

\bibitem{wikisumcorr}
Wikipedia.
\newblock Variance, {S}um of correlated variables, 2016.
\newblock [Online; accessed 160622].

\bibitem{yajima}
Y.~Yajima.
\newblock Positive semidefinite relaxations for distance geometry problems.
\newblock {\em Japan Journal of Industrial and Applied Mathematics},
  19:87--112, 2002.

\end{thebibliography}
